\documentclass[11pt]{article}

\usepackage{amsmath}
\usepackage{amsfonts}
\usepackage{amssymb}
\usepackage{graphicx}
\usepackage{bm}
\usepackage{hyperref}

\newcommand{\nol}{:\!}
\newcommand{\nor}{\!:}

\begin{document}


\title
{Stochastic geometry of critical curves, Schramm-Loewner
evolutions, and conformal field theory}

\author
{Ilya A. Gruzberg
\\ \\
The James Franck Institute, The University of Chicago\\
5640 S. Ellis Avenue, Chicago, Il 60637 USA }

\date{August 9, 2006}

\maketitle

\begin{abstract}

Conformally-invariant curves that appear at critical points in
two-dimensional statistical mechanics systems, and their fractal
geometry have received a lot of attention in recent years. On the
one hand, Schramm \cite{Schramm} has invented a new rigorous as
well as practical calculational approach to critical curves, based
on a beautiful unification of conformal maps and stochastic
processes, and by now known as Schramm-Loewner evolution (SLE). On
the other hand, Duplantier \cite{Duplantier-PRL, Duplantier} has
applied boundary quantum gravity methods to calculate exact
multifractal exponents associated with critical curves.

In the first part of this paper I provide a pedagogical
introduction to SLE. I present mathematical facts from the theory
of conformal maps and stochastic processes related to SLE. Then I
review basic properties of SLE and provide practical derivation of
various interesting quantities related to critical curves,
including fractal dimensions and crossing probabilities.

The second part of the paper is devoted to a way of describing
critical curves using boundary conformal field theory (CFT) in the
so-called Coulomb gas formalism. This description provides an
alternative (to quantum gravity) way of obtaining the multifractal
spectrum of critical curves using only traditional methods of CFT
based on free bosonic fields.

\end{abstract}


\newpage

\tableofcontents

\newpage


\section{Introduction}

The area of two-dimensional (2D) critical phenomena has enjoyed a
recent breakthrough. A radically new development, referred to as
the Schramm (or stochastic) Loewner evolution (SLE)
\cite{Schramm}, has given new tools to study criticality in 2D,
and also provided us with a new interpretation of the traditional
conformal field theory (CFT) and Coulomb gas approaches. Examples
of systems described by SLE include familiar statistical models
--- Ising, Potts, O$(n)$ model, polymers, --- as well as
``geometric'' critical phenomena like percolation, self-avoiding
random walks, spanning trees and others. The new description
focuses directly on non-local structures that characterize a given
system, be it a boundary of an Ising or percolation cluster, or
loops in the O$(n)$ model. This description uses the fact that all
these non-local objects become random curves at a critical point,
and may be precisely characterized by stochastic dynamics of
certain conformal maps.

The SLE approach is complementary to that of CFT, and the new
description has not only reproduced many of the known results from
previous approaches, but also gave new results, either conjectured
before or unknown altogether. It appears that questions that are
difficult to pose and/or answer within CFT are easy and natural in
the SLE framework, and vice versa.

The SLE approach is very intuitive and transparent using
traditional pa\-ra\-digms of stochastic processes --- Brownian
motion, diffusion, and the like. In spite of all this, SLE is not
yet widely known in the physics community and deserves more
attention and study. One goal of this paper is to give a brief
introduction to this burgeoning field.

Another important recent advance (actually preceding the invention
of SLE) in the study of critical 2D systems, has been the
calculation of exact multifractal spectrum of critical clusters by
Duplantier \cite{Duplantier-PRL, Duplantier}, who has ingeniously
applied methods of boundary quantum gravity (the KPZ formula of
Ref. \cite{Knizhnik:1988ak}). The second goal of this paper is to
review Duplantier's results and rederive them using traditional
methods of CFT. To this end I to connect SLE with CFT in the so
called Coulomb gas formulation. In this formulation the curves
produced by SLE can be viewed as level lines of a height function
(bosonic field) that fluctuates and is described by a simple
Gaussain action with some extra terms. This formalism allows to
perform a very transparent translation between SLE and CFT, and
between geometric object (curves) and operators and states in the
CFT.

Many reviews of SLE and its applications in physics already exist.
They are listed in the references section in the end. Refs.
\cite{Bauer-Bernard}--\cite{Kager-Nienhuis} are geared for
physicists, and Refs. \cite{Lawler-book}--\cite{Werner2} for
mathematicians. My presentation is for physicists who may want to
read original mathematical papers on SLE. Therefore, I use
mathematical language and notation, explaining and illustrating
all important terms and ideas with plausible arguments and simple
calculations. The presentation is not rigorous, but I try to
formulate all important statements precisely. Another feature of
this paper (mainly in its second part) is that I assume that
readers are familiar with statistical mechanics and methods of
CFT, some of which are briefly summarized in appropriate sections
(see Ref. \cite{cft} for a thorough introduction).

The structure of the paper is as follows. In  Section
\ref{sec:critical curves} I first describe microscopic origins of
critical curves, as they appear in models of statistical mechanics
defined on 2D lattices. Then I introduce various quantities of
interest related to critical curves.

The next Section \ref{sec:Loewner} presents some properties of
conformal maps, especially those that map the complement of a
curve in the upper half plane (UHP) to the UHP. These maps can be
obtained as solutions of a simple differential equation introduced
by Loewner. I provide a heuristic derivation of this equation and
give a few example of explicit solutions.

In Section \ref{sec:Chordal SLE} I introduce SLE and describe its
basic properties (some of which are actually derived later) and
relation to particular statistical mechanics models.

Section \ref{sec:stochastic calculus} provides a quick
introduction to tools of stochastic analysis, the main ones being
It\^o formula and its consequences.

Basic properties of SLE, including its phases, locality and
restriction, are considered and derived using stochastic calculus
in Section \ref{sec:basic properties}.

In Section \ref{sec:SLE calculations} I give several examples of
non-trivial calculations within SLE, whose results provide
probabilistic and geometric properties of critical curves. This
section closes the part of the paper devoted to SLE.

The remaining sections, based on Refs.
\cite{BRGW-harmonic-measure-PRL, RBGW-long-paper}, develop an
alternative way of analyzing critical curves based on CFT methods
in the Coulomb gas formalism.

The name ``Coulomb gas'' refers to a group of techniques that have
been very fruitfully used to obtain exact critical exponents of
various lattice models of statistical mechanics (see Ref.
\cite{nienhuis}) for a review). A similar method was introduced by
Dotsenko and Fateev \cite{DF} to reproduce correlations of the
minimal models of CFT. The basic ingredient of all these methods
is a bosonic action for a Gaussian free field. In Section
\ref{sec:Coulomb gas} I show how lattice models can be related to
the bosonic action, and how critical curves can be created by
certain vertex operators.

Section \ref{sec:harmonic measure} begins with definitions and
properties of harmonic measure and its multifractal spectrum. The
identification of curve-creating operators as vertex operators of
Coulomb gas is used then to derive Duplantier's results for many
multifractal exponents characterizing stochastic geometry of
critical curves.

The final Section \ref{sec:omitted topics} briefly lists topics
related to SLE and its connection with CFT, that had to be
omitted. This section may serve as a (necessarily incomplete)
guide to the SLE literature.

\begin{figure}[t]
\centering
\includegraphics[width=0.47\textwidth]{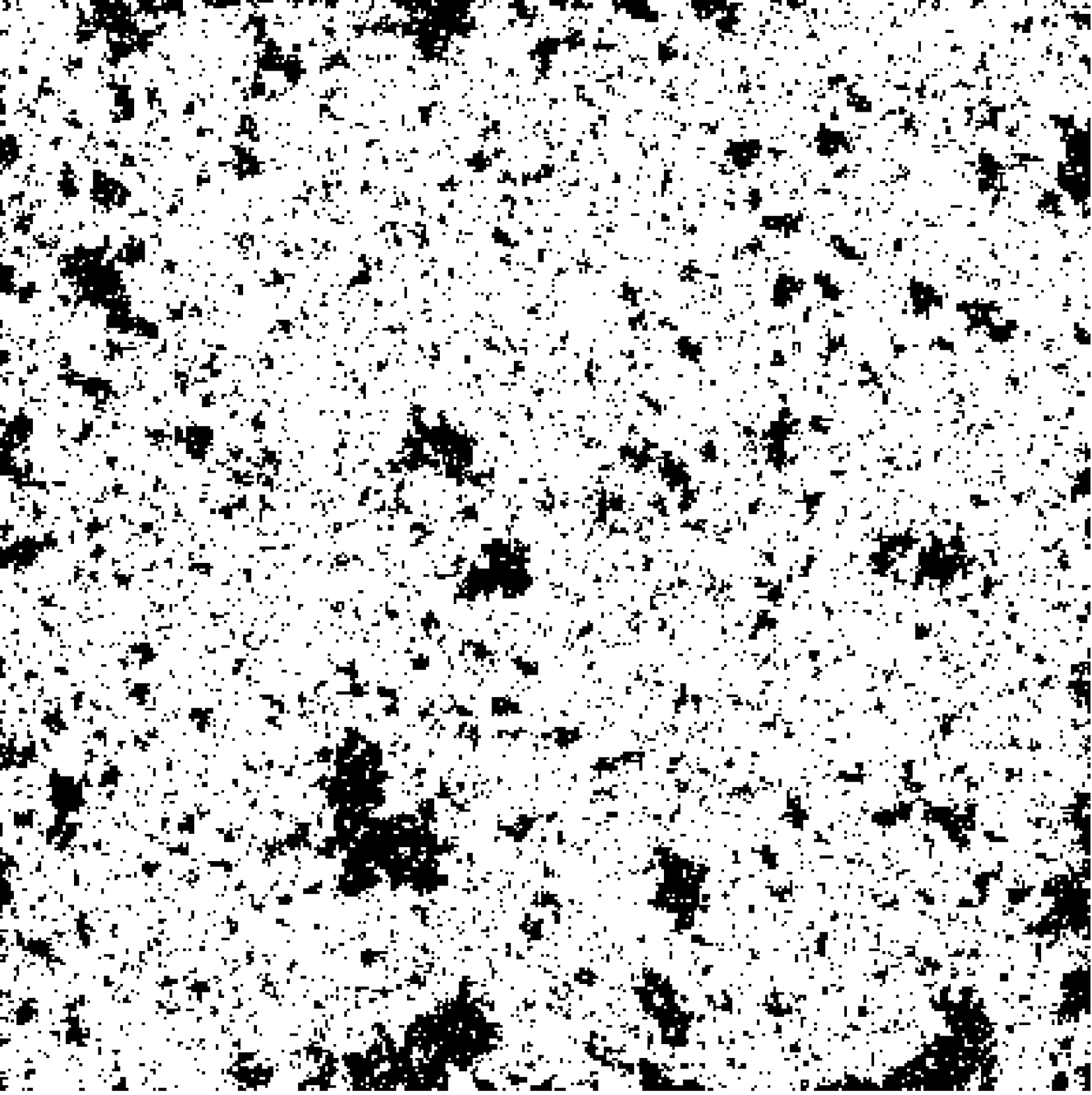}
\hfill
\includegraphics[width=0.47\textwidth]{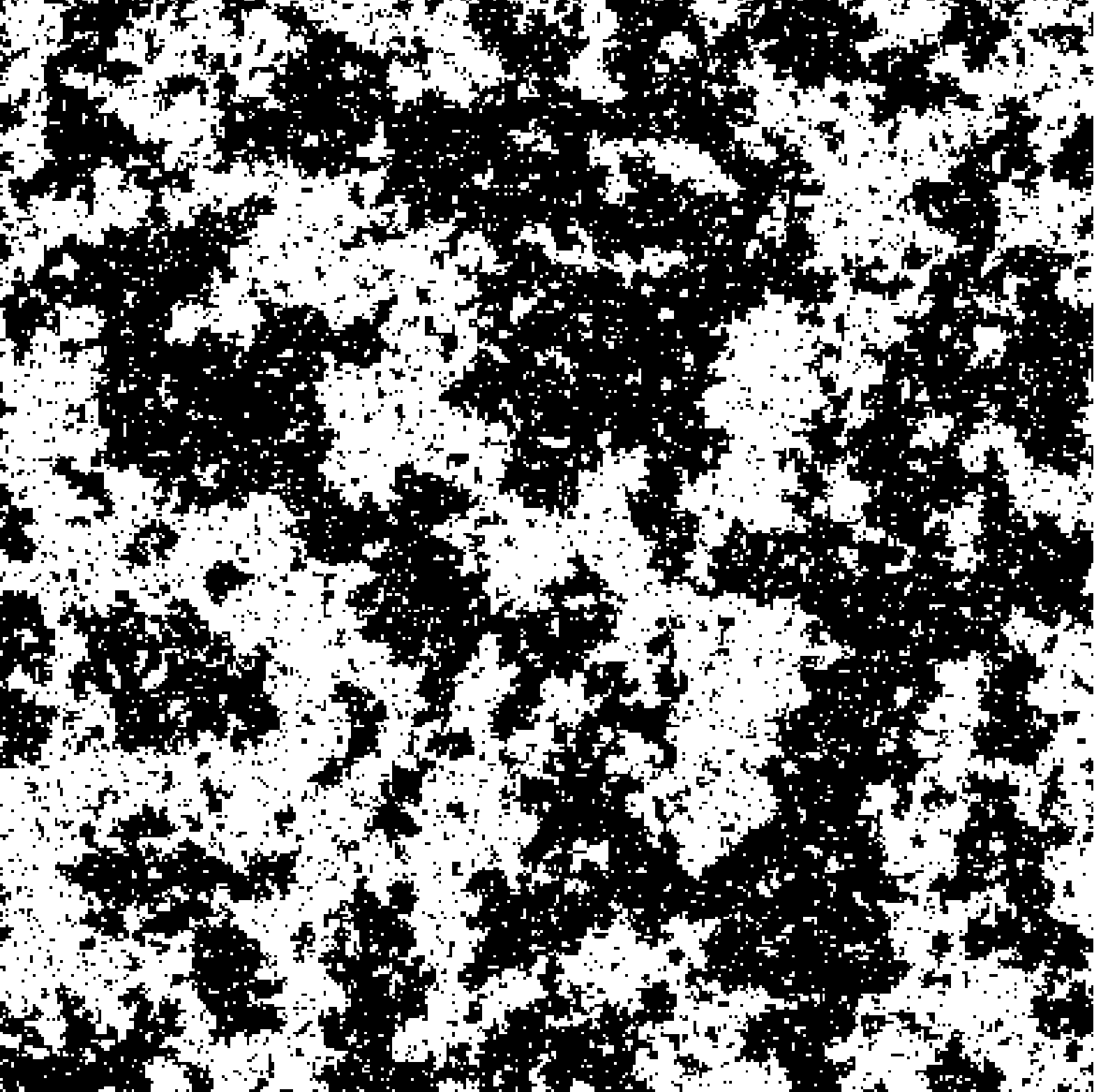}
\caption{Ising clusters at low (left) and critical (right)
temperature.} \label{fig:Ising}
\end{figure}

\section{Critical 2D systems and critical curves}
\label{sec:critical curves}

Many very simple lattice models of statistical mechanics exhibit
critical phenomena characteristic of continuous phase transitions.
The prototype of all such models is the Ising model which describes
the behavior of a collection of ``spin'' variables $S_i$ located on
sites of a lattice labeled by the index $i$, and taking values $\pm
1$. In this paper we will only consider two-dimensional models. In
the Ising model the energy of the system is given by $H = - J \sum
S_i S_j$, where the sum is over all nearest neighbor pairs of sites
$i j$.

At a finite temperature $T$ various possible spin configurations
$\{S\}$ of the system have probabilities given by Gibbs
distribution $e^{-H/T}/Z$, where the partition function is
obtained by summing over all possible configurations: $Z =
\sum_{\{S\}}e^{-H/T}$. Qualitative picture of this model is that
at low temperatures the $\mathbb{Z}_2$ symmetry between the up
($S_i = 1$) and down ($S_i = -1$) directions of the spins is
spontaneously broken, and the majority of spins points, say, up.
As the temperature is increased, typical configurations involve
small domains or connected clusters of down spins in the sea of up
spins. The typical size of such clusters --- the correlation
length --- increases indefinitely, as the temperature approaches a
specific critical value $T_c$. At this temperature the clusters of
up and down spins of all possible sizes are mixed together, and
the whole picture is scale-invariant, see Fig. \ref{fig:Ising}
\footnote{The
applet that produced these pictures is available at \\
http://www.ibiblio.org/e-notes/Perc/contents.htm}. The cluster
boundaries or domain walls at the critical point are fractal
curves that the SLE focuses on.

\begin{figure}[t]
\centering
\hbox{\includegraphics[width=0.5\textwidth]{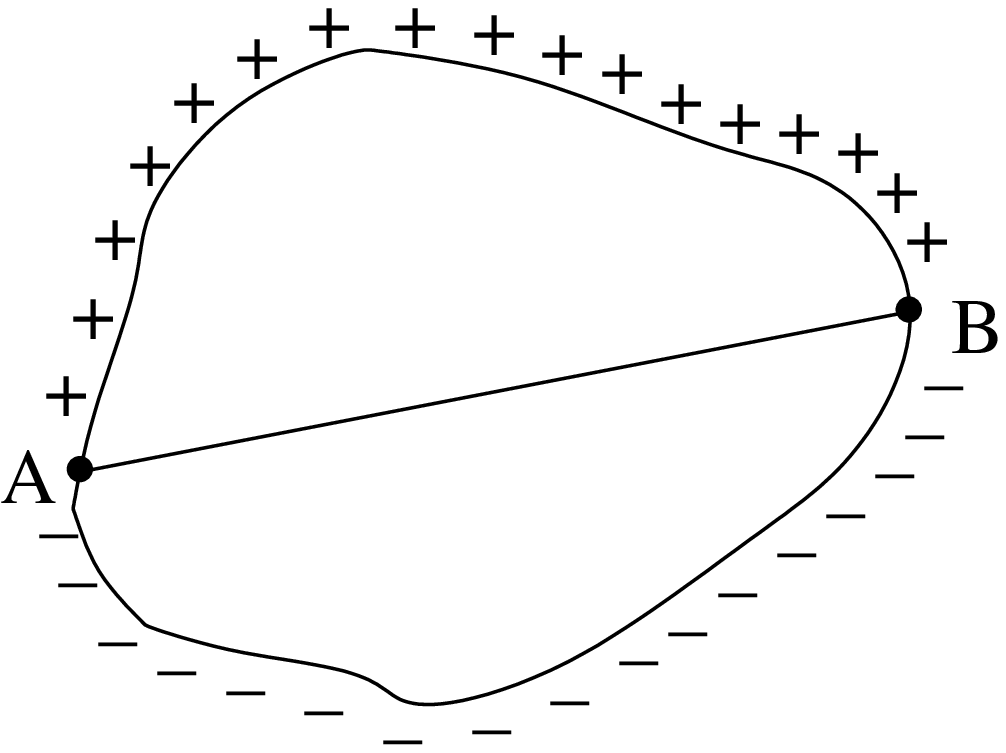}
\includegraphics[width=0.5\textwidth]{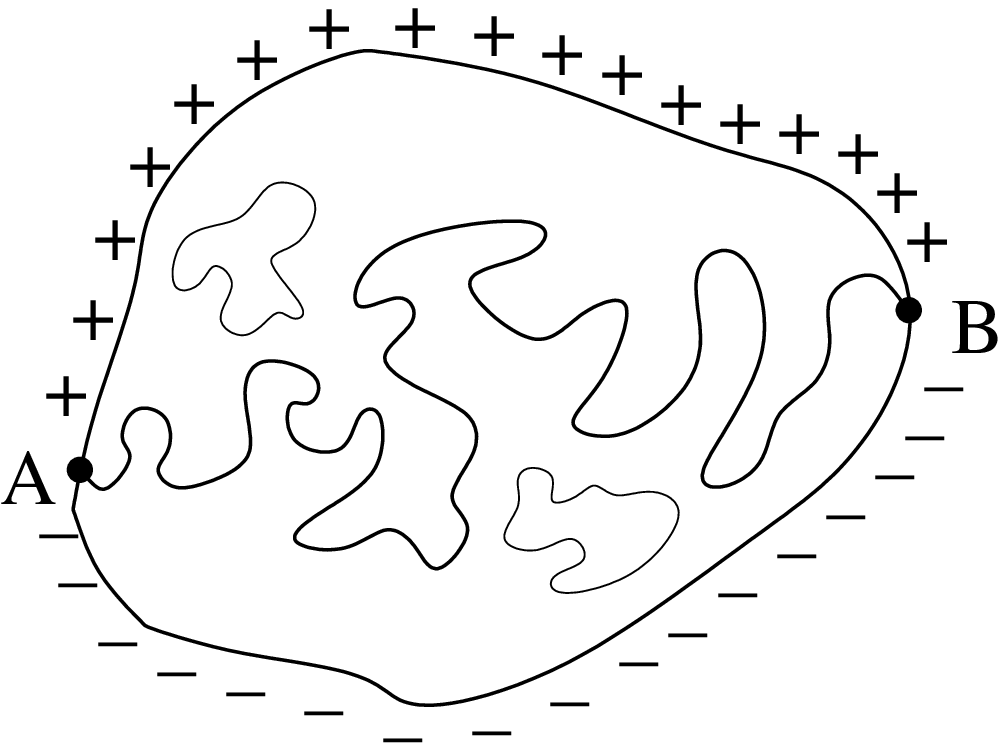}}
\caption{Domain walls in a finite Ising system. The boundary
conditions change at points $A$ and $B$, forcing a domain wall to
go between these points. Left figure: zero temperature. Right
figure: critical temperature.} \label{fig:Ising-domain-wall}
\end{figure}

To be slightly more precise, let us consider a system in a simply
connected region $D$ with the boundary $\partial D$, with a very
fine lattice inside (essentially, we want the lattice spacing to
be much smaller than the system size and the correlation length at
a given temperature), see Fig. \ref{fig:Ising-domain-wall}. We can
force a domain wall to go between two points A and B on the
boundary $\partial D$. To this purpose let me impose the following
boundary conditions. On the upper portion of the boundary between
A and B we force the spins to be up, and on the lower portion
--- to be down. Then at zero temperature there will be exactly one
straight domain wall between the points A and B. As the
temperature increases, the domain wall will wander off the
straight line, and eventually, at the critical temperature will
become a complicated fractal curve. These curves will differ
between the members of the statistical thermal ensemble, and will
have particular weights or distribution within the ensemble.

\begin{figure}[t]
\centering
\includegraphics[width=0.8\textwidth]{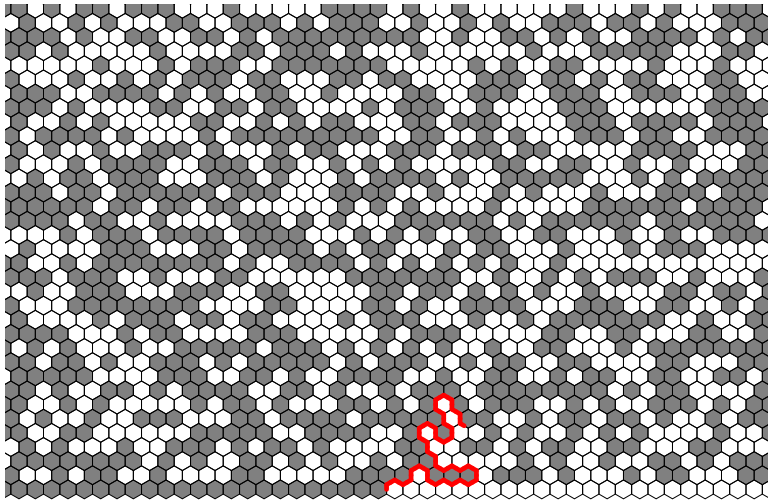}
\caption{Critical percolation on a triangular lattice with a
boundary condition that forces a domain wall (percolation hull)
between the origin and infinity. The figure is borrowed from Ref.
\cite{Werner1}.} \label{fig:percolation-1}
\end{figure}

\begin{figure}[t]
\centering
\includegraphics[width=0.8\textwidth]{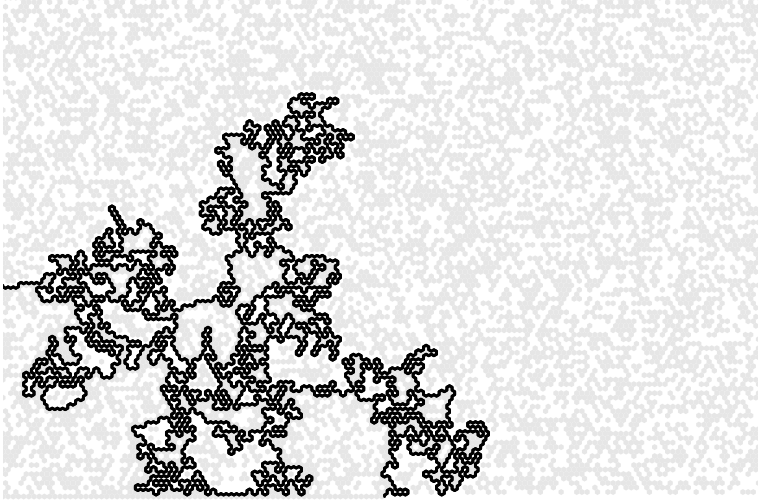}
\caption{A long percolation hull. The figure is borrowed from Ref.
\cite{Werner1}.} \label{fig:percolation-2}
\end{figure}

Another prototypical example of a model that exhibits critical
behavior is the site percolation. In this model each site on a
lattice is independently colored grey with probability $p$ or
white with probability $1-p$. For each lattice there is a critical
value $p = p_c$ such that an infinite connected cluster of grey
sites appears in the system (that is, for $p < p_c$ all the gray
clusters are finite). For a triangular lattice $p_c$ is known to
be exactly $1/2$. A good graphical representation of the critical
site percolation on a triangular lattice is obtained if we replace
every lattice point by a hexagon whose vertices lie on the dual
honeycomb lattice. Then we can again force a domain wall into the
system by making hexagons grey and white on two adjacent portions
of the boundary. The Fig. \ref{fig:percolation-1} (borrowed from
Ref. \cite{Werner1}) shows the upper half plane tiled with such
hexagons. All the hexagons to the left of the origin on the
horizontal axes (this is the boundary) are colored grey, and all
the hexagons to the right are colored white. This produces a
domain wall separating grey and white hexagons the beginning of
which is shown in Fig. \ref{fig:percolation-1}. Fig.
\ref{fig:percolation-2} shows a much bigger system with the same
boundary conditions.

Both the Ising and percolation models can be included in a larger
class of models, loosely called the loop models, since their
partition functions can be written as sums over loop configurations
$\mathcal L$, either on the original or some related (the dual or
the surrounding) lattice. Mappings between specific lattice models
and loop models are described in detail in many reviews
\cite{nienhuis, loop models-1, loop models-2, loop models-3}. Here I
only mention the O$(n)$ model because of the richness of its phase
diagram. The model is defined in the simplest way on the honeycomb
lattice directly in terms of closed loops:
\begin{align}
Z_{\text{O}(n)} &= \sum_{\mathcal{L}} x^L n^N. \label{O(n)
partition}
\end{align}
Here $x$ is the variable related to the temperature, $L$ is the
total length of all loops and $N$ is their number in the
configuration $\mathcal L$.

It is known from various approaches that the O$(n)$ model has a
critical point at some value $x_c(n)$ for all $n$ in the range $-2
\leqslant n \leqslant 2$. At the critical point the mean length of
a loop diverges, but loops are dilute in the sense that the
fraction of the vertices visited by the loops is zero. For $x >
x_c(n)$ the loops are still critical but now visit a finite
fraction of the sites. This is called the dense phase of the loop
model. Finally, at zero temperature ($x = \infty$) the loops go
through every point on the lattice, and this is called the fully
packed phase.

For some values of the parameter $n$ the O$(n)$ model is related
to other known statistical mechanics models: $n=2$ corresponds to
the XY model, the limit $n=0$ describes self-avoiding walks or
polymers, and $n=-2$ corresponds to the so-called loop-erased
random walk. The dense phase of the O$(n)$ model is also related
to the critical point of the $q$-states Potts model. The Potts
critical point exists for all $0 \leqslant q \leqslant 4$, and at
that point the boundaries of the so-called Fortuin-Kasteleyn
clusters that appear in the high-temperature expansion of the
$q$-states Potts model, are essentially the same as the loops in
the dense phase of the O$(n)$ model with $n = \sqrt{q}$.

In all the models mentioned above, one can choose boundary
conditions so as to introduce an open curve starting at one point
on the boundary of a domain and ending at another boundary point.
The continuum limit of these open curves with fixed ends on the
boundary is exactly what is being studied using SLE.

One crucial paradigm in the study of critical phenomena is that of
conformal invariance \cite{Polyakov:1970xd, Belavin:1984vu}. For
the critical curve described above this means that if we map
conformally the region $D$ in which our system is defined into
another region $D'$, then the statistical weights or the
distribution of the critical curves will be invariant under such
mapping. In other words, any two curves that map into each other
will have the same weight in the corresponding thermal ensembles.
Then we can study these curves in a standard simple region, which
we choose here to be the upper half plane $\mathbb{H}$ with point
A at the origin and point B at infinity.

In this setup we may ask various questions about the critical
curves. Some of them are geometric. For example, we may want to
know the fractal dimension of a critical curve. More generally, we
can imagine that the cluster surrounded by a critical curve is
charged, and then the charge distribution on the domain boundary
will be very uneven or ``lumpy''. This lumpiness is characterized
by what is known as the spectrum of multifractal exponents. More
precise definition uses the notion of the harmonic measure of the
cluster boundary, and is explained in Section \ref{sec:harmonic
measure}.

Another class of possible questions is probabilistic. We may ask
about the probability that the critical curve between the origin
and the infinity in $\mathbb{H}$ passes to the left of a given
point. Another question asks for the probability of the critical
curve to touch the boundary at certain places in certain order.
This is related to the so-called crossing probability in
percolation that is defined as the probability for a connected
cluster to span the critical system between two disjoint segments
of the boundary. Sometimes we are interested only in the
asymptotic behavior of probabilities of such events for long times
or large spatial distances. These asymptotic probabilities behave
in a power law fashion with some universal exponents that need to
be found.

SLE provides an easy way of answering the above geometric and
probabilistic questions and computing the corresponding
quantities. In the following sections I will introduce the
necessary tools from the theory of conformal maps and stochastic
processes and will describe some calculations with SLE.

\section{Conformal maps and Loewner equation}
\label{sec:Loewner}

Consider the upper half plane $\mathbb H = \{z: \mathrm{Im} z >
0\}$, with a curve $\gamma$ starting at the origin on the real
axis such that $\gamma \in {\mathbb H}\cup \{0\}$. We parametrize
the curve by a real variable $t \in [0,\infty)$, and denote a
point on $\gamma$ as $\gamma(t)$ and (closed) segments as
$\gamma[t_1,t_2]$.

A segment $\gamma[0,t]$ is an example of the so-called hull. A
{\bf hull} $K \subset \overline{\mathbb H}$ is a bounded subset of
$\overline{\mathbb H}$ such that ${\mathbb H}\setminus K$ is
simply connected and $K = \overline{K \cap {\mathbb H}}$. So, a
hull is, essentially, a bounded (but not necessarily connected)
set bordering on the real line $\mathbb R$. By Riemann's mapping
theorem (see Ref. \cite{Ahlfors1}), for each such hull there is a
conformal map $g_K$ that maps ${\mathbb H}\setminus K$ to
${\mathbb H}$. Since conformal automorphisms of ${\mathbb H}$ are
M\"obius transformations with real coefficients,
we can make $g_K$
unique by fixing three real parameters. A conventional
``hydrodynamic'' normalization is such that
\begin{align}
\lim_{z \to \infty} (g_K(z) - z) = 0, \label{normalization}
\end{align}
or, equivalently, that near $z = \infty$ the map has the form
\begin{align}
g_K(z) = z + \sum_{n=1}^\infty \frac{a_n}{z^n}. \label{Laurent}
\end{align}
If the hull $K$ is located a finite distance away from the origin,
then $g_K(z)$ is regular at $z=0$. In this situation it is more
convenient for some purposes (see Sections \ref{subsec:locality},
\ref{subsec:restriction} below) to consider the map $\Phi_K(z) =
g_K(z) - g_k(0)$, normalized as
\begin{align}
\Phi_K(0) &= 0, & \Phi_K(\infty) &= \infty, & \Phi_K'(\infty) &=
1, \label{mapPhi}
\end{align}

Since the function $g_K(z)$ takes real values on the boundary of
${\mathbb H}\setminus K$, the coefficients $a_n \in {\mathbb R}$.
The coefficient $a_1 = a(K)$ is called the half-plane capacity (or
simply capacity) of the hull $K$. For any $r > 0$ the map $g_K(z)$
satisfies the scaling relation $g_{rK}(z) = r g_K(z/r)$, which
implies the scaling for the capacity
\begin{align}
a(r K) &= r^2 a(K), & \forall r > 0. \label{scaling}
\end{align}
Thus the capacity has the dimension of area. Geometrically, it is
bounded above by $R^2$, where $R$ is the radius of the smallest
semicircle that completely encloses the hull $K$.

\begin{figure}[t]
\centering
\includegraphics[width=0.7\textwidth]{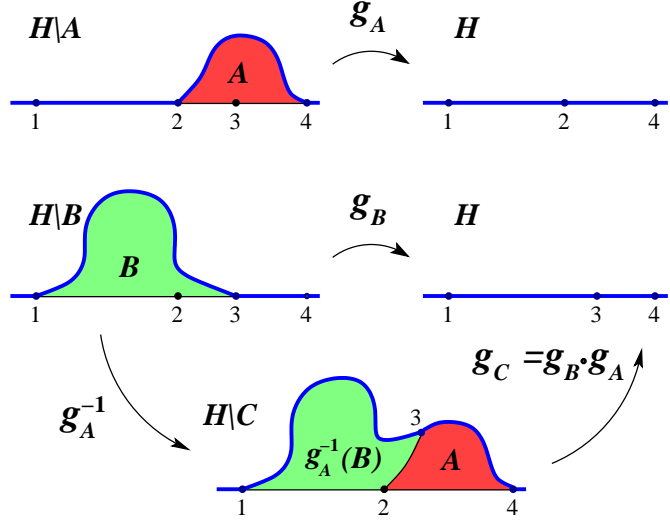}
\caption{The composition of conformal maps. Here we denote $C = A
\cup g_A^{-1}(B)$.} \label{fig:composition}
\end{figure}
Conformal maps for hulls can be composed as shown in Figure
\ref{fig:composition}. Note that the mapping region monotonically
shrinks under such a composition, and the hulls grow. Their
capacities satisfy another important property (additivity or
composition rule):
\begin{align}
a(A \cup g_A^{-1}(B)) = a(A) + a(B), \label{additivity}
\end{align}
which can be easily checked by composing the conformal maps $g_A$
and $g_B$ in the form of Laurent expansions (\ref{Laurent}) and
finding the coefficient $a_1$ of the composed map.

Consider now the map $g_{\gamma[0,t]} = g_t$ for the hull that is
a segment of a curve, as in the beginning of this section. We can
always choose the parametrization for the curve in such a way that
\begin{align}
a(\gamma[0,t]) = 2t. \label{parameter}
\end{align}
We will call the parameter $t$ ``time'', since the evolution of
$g_t$ in this variable will be of importance. Then it can be shown
that the map $g_t$ satisfies a very simple differential equation
called Loewner equation \cite{Loewner}\footnote{This equation has
a fascinating history. It was invented in 1923 by Karl L\"owner
(who later changed his name to Charles Loewner, see more about him
at
http://www-gap.dcs.st-and.ac.uk/\~{}history/Mathemati\-cians/Loewner.html)
to partially solve a famous conjecture from the theory of
univalent functions proposed by Bieberbach in 1916. After many
partial successes, the conjecture was finally proved by de Branges
in 1985. The key element of the proof was the same Loewner
equation! A very readable account of this story and the proof is
given in Ref. \cite{Gong}.}:
\begin{align}
\partial_t g_t(z) &= \frac{2}{g_t(z) - \xi_t}, & g_0(z) &= z,
\label{LE}
\end{align}
where $\xi_t$ is a real function that is the image of the tip of
the cut $\gamma(t)$ under the map $g_t$:
\begin{align}
\xi_t = g_t(\gamma(t)). \nonumber
\end{align}

\begin{figure}[t]
\centering
\includegraphics[width=0.8\textwidth]{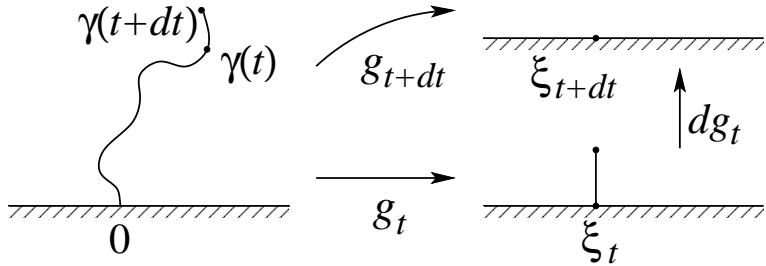}
\caption{Illustration for the derivation of Loewner equation.}
\label{fig:loewner-derivation}
\end{figure}

Let me give an intuitive derivation of this equation. Suppose that
we already know the map $g_t$ and want to find out what happens
during the time increment between $t$ and $t + dt$. Using the
composition of maps we write $g_{t+dt} = dg_t \circ g_t$. This
composition is illustrated in Fig. \ref{fig:loewner-derivation}.
Under the map $g_t$ the segment $\gamma[t,t+dt]$ is mapped to a
(almost) straight short vertical segment beginning at point $\xi_t
\in {\mathbb R}$. Using the additivity property, Eq.
(\ref{additivity}), the capacity of this little segment is
\begin{align}
a(g_t(\gamma[t,t+dt])) = 2dt. \label{capacity}
\end{align}
The corresponding conformal map $dg_t$ removing the segment is
elementary:
\begin{align}
dg_t(w) = \xi_t + \sqrt{(w - \xi_t)^2 + 4dt}.
\label{infinitesimal-map}
\end{align}
Composing this with $g_t$ and expanding in small $dt$ we get
\begin{align}
g_{t+dt}(z) &= dg_t(g_t(z)) = \xi_t + \sqrt{(g_t(z) - \xi_t)^2 +
4dt} \nonumber \\ & \approx g_t(z) + \frac{2dt}{g_t(z) - \xi_t}.
\nonumber
\end{align}
This immediately leads to Loewner equation (\ref{LE}) in the limit
$dt \to 0$.

There are two ways in which one can think about Loewner equation.
The first one was just presented: given a curve $\gamma$ in the
upper half plane, we can obtain, at least in principle, the real
function $\xi_t$ in the equation by constructing the corresponding
conformal maps. The second way is the opposite: given a real
continuous ``driving'' function $\xi_t$ we can plug it into
Loewner equation and solve it forward in time starting with the
initial condition $g_0(z) = z$. It is known that the solution
exists, but does not necessarily describe a map from $\mathbb{H}$
cut along a segment of a curve. In some cases the hull that
corresponds to the solution $g_t$ contains two-dimensional regions
of the upper half plane, as one of the examples below shows.

In general, the hull generated by the solution of Eq. (\ref{LE})
is defined as follows. For a given point $z \in \overline{\mathbb
H}$, the solution of Eq. (\ref{LE}) is well defined as long as
$g_t(z) - \xi_t \neq 0$. Thus, we define $\tau_z$ as the first
time $\tau$ such that $\lim_{t \nearrow \tau}(g_t(z) - \xi_t) =
0$. For some points in $\overline{\mathbb H}$ the time $\tau_z =
\infty$, meaning that at these points the Loewner map is defined
for all times. The union of all the points $z$ for which $\tau_z
\leqslant t$ is the hull corresponding to the map $g_t(z)$:
\begin{align}
K_t = \{z \in \overline{\mathbb H}: \tau_z \leqslant t\},
\nonumber
\end{align}
and its complement $H_t = \{z \in {\mathbb H}: \tau_z > t\} =
{\mathbb H} \setminus K_t$ is the domain of $g_t$, that is the set
of points for which $g_t(z)$ is still defined.

Another useful notion is that of the {\bf trace} $\gamma$ produced
by Loewner equation. This is defined as the union points
\begin{align}
\gamma(t) = \lim_{z \to 0} g_t^{-1}(z + \xi_t), \label{trace}
\end{align}
where the limit is taken within the upper half plane. Note that
the trace and the hull are not necessarily the same objects, as we
will see in a simple example below, and especially in the case of
SLE$_\kappa$ for some values of the parameter $\kappa$ (see
Section \ref{subsec:SLE phases}). The reason for this is that
points may enter the growing hull in two different ways. Some of
them are added to the trace itself, but others are swallowed, or
enclosed by the trace ``inside'' the hull, see examples below.

One can exhibit many explicit solutions of the Loewner equation
for several forms of the driving function $\xi_t$, see Ref.
\cite{Loewner-exact-solutions}. I will give here two of them as
illustrations. If $\xi_t = c$ is a constant, the solution of Eq.
(\ref{LE}) is simply
\begin{align}
g_t(z) = c + \sqrt{(z - c)^2 + 4t}. \nonumber
\end{align}
The corresponding hull is the vertical straight segment between
$c$ and $c + 2i\sqrt{t}$. In this case the map $g_t$ can be found
by elementary means.

\begin{figure}[t]
\centering
\includegraphics[width=0.8\textwidth]{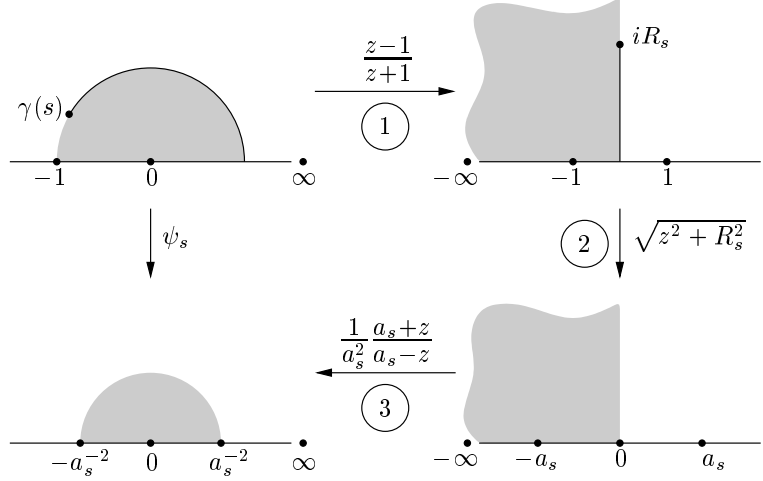}
\caption{The sequence of maps for the construction of the function
$g_s(z)$ in Eq. (\ref{half-circle-maps}). The figure is borrowed
from Ref. \cite{Loewner-exact-solutions}.} \label{fig:halfcricle}
\end{figure}

Another straightforward but instructive example described in detail
in Refs. \cite{Loewner-exact-solutions, Bauer-Bernard-1} deals with
a circular arc of radius $r$ growing in the complex $z$ plane from
the point $r$ on the real axis towards the point $-r$. The segment
of this arc spanning the angle $s \in [0,\pi)$ is mapped to an
interval on the imaginary axis $[0, i R_s]$, where $R_s = \tan
(s/2)$, by the M\"obius transformation $z_1 = (z - r)/(z + r)$, and
then removed by the transformation from the previous example: $z_2 =
\sqrt{R_s^2 + z_1^2}$. Further M\"obius transformations are
necessary to satisfy the hydrodynamic normalization
(\ref{normalization}). This leads to the mapping
\begin{align}
g_s(z) &= \frac{r}{a_s^2}\Big(\frac{a_s + z_2}{a_s - z_2} + 2 - 2
a_s^2 \Big), \label{half-circle-maps}
\end{align}
where $a_s^2 = 1 + R_s^2 = 1/\cos^2(s/2)$. The first three
conformal maps in this sequence for $r=1$ are illustrated in Fig.
\ref{fig:halfcricle}.

Expanding the function $g_s(z)$ near $z = \infty$ we find the
capacity of the arc to be $2t = r^2(1 - a_s^{-4})$. After the
reparametrization of the arc and the map $g_s$ in terms of $t$ we
get the solution of Loewner equation
\begin{align}
g_t(z) = \frac{(z-r)^2 + 2z\sqrt{r^2 - 2t} + (z+r)\sqrt{(z+r)^2 -
4z \sqrt{r^2 - 2t}}}{2z}, \label{half-circle-maps-1}
\end{align}
corresponding to the driving function $\xi_t = 3\sqrt{r^2 - 2t} -
2r$. The branches of the square roots in Eq.
(\ref{half-circle-maps-1}) have to be chosen in such a way that
\begin{align}
\lim_{t \to r^2/2} g_t(z) = \left\{\begin{array}{ll} z + r^2/z, &
\text{for  $|z| \geqslant r$}, \\
-2r, & \text{for  $|z| < r$}.
\end{array} \right. \nonumber
\end{align}

\begin{figure}[t]
\centering
\includegraphics[width=0.4\textwidth,angle=-90]{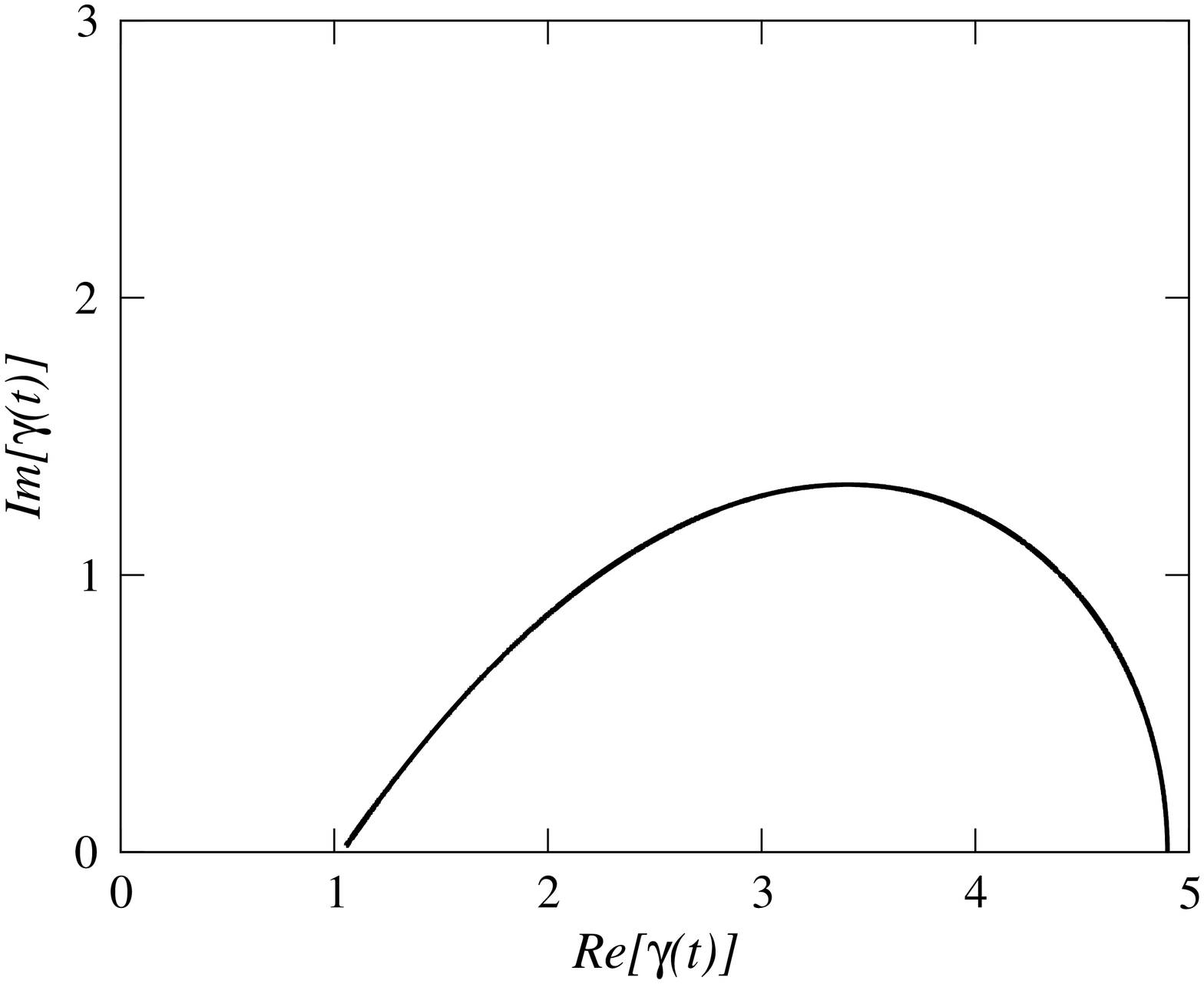}
\hfill
\includegraphics[width=0.4\textwidth,angle=-90]{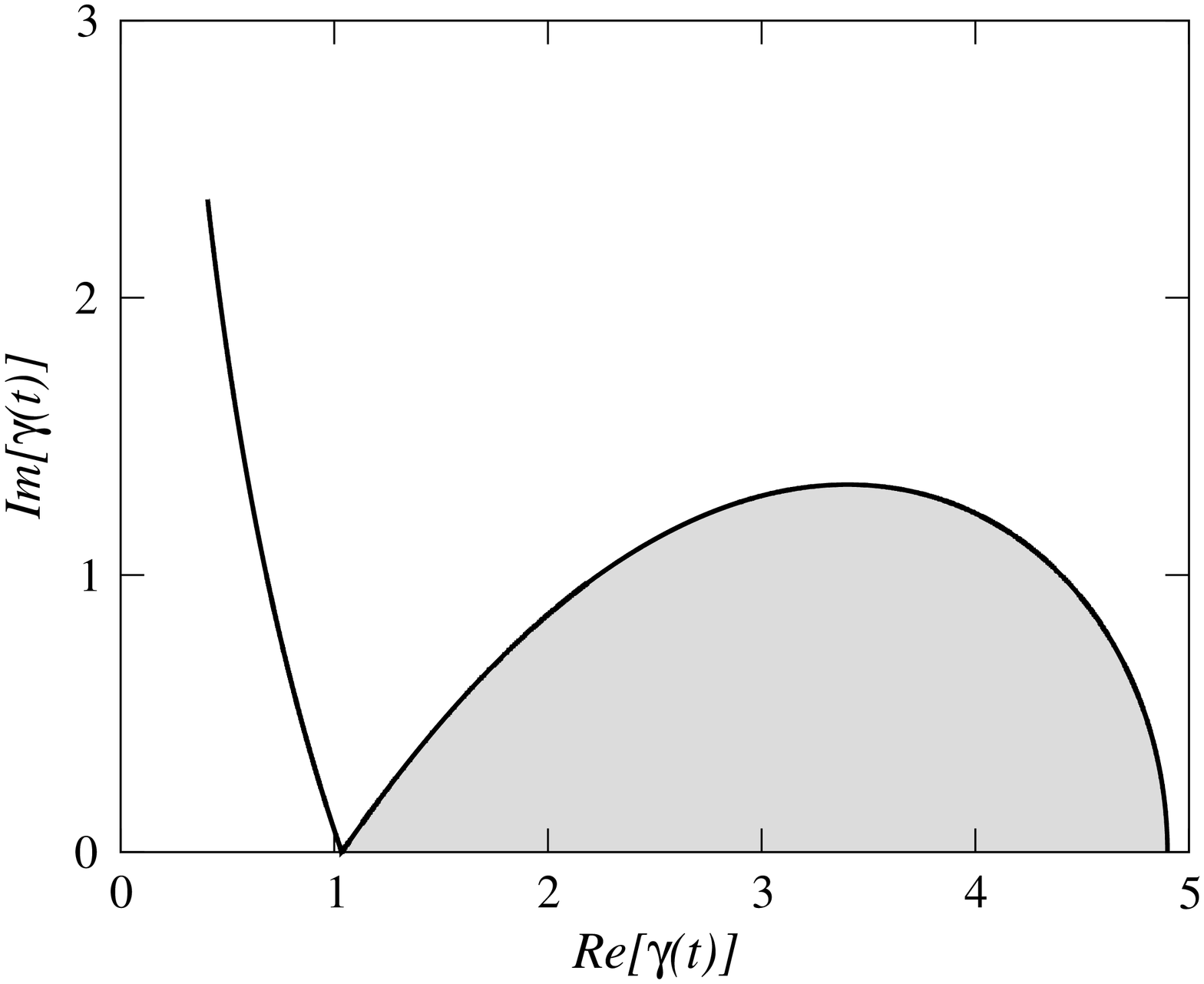}
\caption{The trace and the hull for a touching event. Here
$\xi_t=2\sqrt{6(1 - t)}$ for $t \in (0,1)$ and zero elsewhere.
Left shows the situation just before touching ($t \to 1^{-}$);
right shows the situation after ($t > 1$). The trace is the thick
dark line. The hull consists of that line plus the grey area. That
area is added to the hull at $t = 1$. Note that there is a
continuum of points added to the hull at the time of touching, but
only one of these, $\gamma(1)$, is on the trace and is not
swallowed. The figure is borrowed from Ref. \cite{ROKG-SLE-Levy}.}
\label{fig:swallowing}
\end{figure}
Note that at time $\tau = r^2/2$ the map $g_t$ changes
discontinuously. At any time before that the hull of the map is
the segment of the arc. But exactly at $t = \tau$ the whole region
$D = \{|z| < r, \mathop{\rm Im} z \geqslant 0 \}$ (the upper half
of the disc of radius $r$) is mapped to the point $-2r$: all the
points in this region are swallowed!

Let me define this notion more rigorously. We say that a point $z
\in \overline{\mathbb H}$ is swallowed if $z \notin
\gamma[0,\infty)$ but $z \in K_t$ for some $t$. In other words,
swallowed points do not lie on the trace, but get enclosed by the
trace in the ``interior'' portions of the hull. The time when a
point gets swallowed is called the swallowing time for this point.

The time $\tau = r^2/2$ is the swallowing time for the whole
region $D$. At this time the hull of the evolution $K_\tau$ is the
closed semi-disc $\overline D$, while the trace is still the
semi-circular arc. One may continue the evolution with the driving
function $\xi_t = 0$ for $t > r^2/2$, and the trace will continue
to grow as a simple curve from the point $-r$ on the real axis,
while the hull will be $K_t = {\overline D} \cup \gamma[r^2/2,
t]$, the union of the semi-disc and the portion of the trace grown
after time $\tau$. Similar swallowing of a region is illustrated
in Fig. \ref{fig:swallowing} for $\xi_t=2\sqrt{6(1 - t)}$ for $t
\in (0,1)$ and zero elsewhere.

\section{Schramm-Loewner evolution}
\label{sec:Chordal SLE}

The remarkable discovery of Schramm \cite{Schramm} was that one
can study Loewner equation (\ref{LE}) with random driving
functions and in this way obtain all possible ensembles of curves
with conformally invariant probabilities. Motivated by the
conformal invariance of interfaces in statistical mechanical
models, Schramm had argued that the driving function $\xi_t$ has
to be a continuous stationary stochastic process with independent
increments. This argument is well explained in the existing
reviews, here I simply indicate the basic idea. First, if we want
to produce a curve without branching or self-intersections, we
need to have a continuous input $\xi_t$. Next, for the curves to
possess a conformally-invariant distribution, the corresponding
maps have to be composed of statistically independent
infinitesimal maps of the form (\ref{infinitesimal-map}). Together
with the reflection symmetry this leads to essentially unique
choice of $\xi_t= \sqrt \kappa B_t$, where $\kappa > 0$, and $B_t$
is the standard Brownian motion started at $\xi(0) = 0$ (that is,
$W_t = dB_t/dt$ is the white noise with unit strength: $\langle
\dot W_t \dot W_s \rangle = \delta(t-s)$). The resulting
stochastic Loewner equation
\begin{align}
\partial_t g_t(z) &= \frac{2}{g_t(z) - \sqrt
\kappa B_t}, & g_0(z) &= z, \label{SLE}
\end{align}
and the sequence of conformal maps that it produces came to be
known as SLE$_\kappa$, where SLE stands for stochastic Loewner
evolution or Schramm-Loew\-ner evolution.

Notice that after assuming the hydrodynamic normalization
(\ref{normalization}) and the pa\-ra\-met\-rization in terms of
the capacity (\ref{parameter}), $\kappa$ is the only important
parameter of SLE. A we will see shortly, it completely determines
the properties of SLE, its hulls and traces.

Often a shifted version of $g_t(z)$ in introduced:
\begin{align}
w_t(z) &= g_t(z) - \xi_t, & w_0(z) &= z. \nonumber
\end{align}
This function satisfies the simple Langevin-type equation
\begin{align}
\partial_t w_t(z) &= \frac{2}{w_t(z)} -
\sqrt{\kappa} \dot B_t, & w_0(z) &= z. \label{shiftedLE}
\end{align}
It is the simplicity of this equation together with the powerful
methods of the theory of stochastic processes that makes SLE a
very versatile calculational tool. I will show how to do
computations with it in the following sections. But first let me
summarize the most important properties of SLE (some of them will
be derived later in Section \ref{sec:basic properties}). These
properties are quite non-trivial. Some of them have been
rigorously formulated and established in Refs.
\cite{Rohde-Schramm}.

\begin{figure}[t]
\centering
\includegraphics[width=\textwidth]{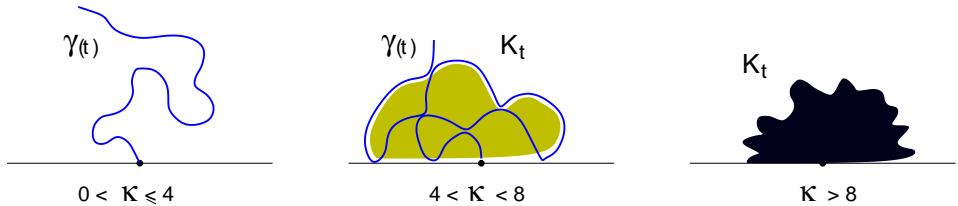}
\caption{The phases of SLE. The figure is borrowed from Ref.
\cite{Bauer-Bernard-1}.} \label{fig:phases}
\end{figure}

\begin{itemize}

\item

First of all, for all values of $\kappa$ one can still define the
trace of an SLE as the union of points $\gamma(t) = \lim\limits_{z
\to 0, z \in {\mathbb H}} w_t^{-1}(z)$ (and this limit exists).
Moreover, the trace is a continuous curve staring at $\gamma(0) =
0$, reaching infinity as $t \to \infty$ and never crossing itself
(self-avoiding).

\item

For $0 \leqslant \kappa \leqslant 4$ an SLE trace $\gamma$ is a
simple curve (does not have double points). In this case the SLE
hull coincides with the trace: $K_t = \gamma[0,t]$, and no point
in $\overline{\mathbb H}$ gets swallowed.

\item

For $4 < \kappa < 8$ an SLE trace has infinite number of double
points. The trace sort of ``touches'' itself and the real axis at
every scale. Every time such touching occurs, a whole finite
region of the plane gets swallowed. As time goes on, almost all
the points in $\overline{\mathbb H}$ (except the points on the
trace) get swallowed.

\item

For $\kappa \geqslant 8$ the trace is a space-filling curve (a
random analog of the Peano curve). This means that no point gets
swallowed, but all the points in $\overline{\mathbb H}$ lie on the
trace.

\item

The fractal dimension of the trace is (proven in Refs.
\cite{Beffara-1, Beffara-2})
\begin{align}
d_f(\kappa) &= \left\{
\begin{array}{ll}
1 + \dfrac{\kappa}{8} & \textrm{for $\kappa \leqslant 8$}, \\
2 & \textrm{for $\kappa \geqslant 8$}.
\end{array} \right.
\label{dimension}
\end{align}

\end{itemize}

The different behaviors of traces and hulls of SLE for different
values of $\kappa$ may be called phases in analogy with
statistical mechanics. These are schematically shown in Fig.
\ref{fig:phases}.

A heuristic derivation of some of these properties given below in
Section \ref{sec:basic properties} will not be rigorous but will
still require the use of probabilistic techniques. Therefore, in
Section \ref{sec:stochastic calculus} I briefly summarize relevant
results of stochastic calculus. Then in Section \ref{sec:SLE
calculations} I give examples of practical calculations with SLE,
deriving a number of non-trivial critical exponents and scaling
functions.

\begin{figure}[t]
\centering
\includegraphics[width=0.8\textwidth]{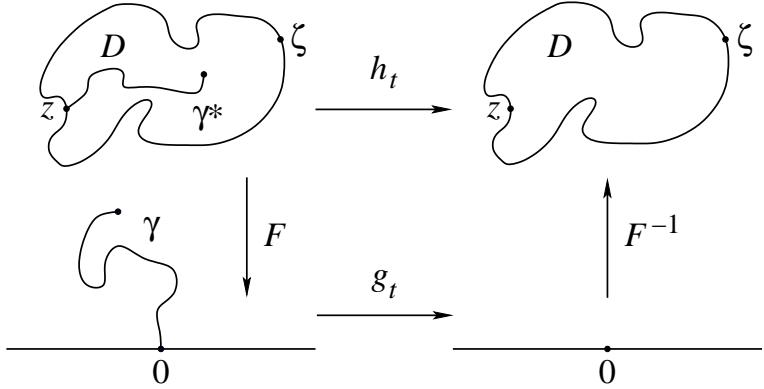}
\caption{Definition of SLE in an arbitrary simply-connected domain
$D$.} \label{fig:SLE-region-D}
\end{figure}
So far we have defined chordal SLE $g_t(z)$ and its traces and
hulls only in the upper half plane $\mathbb H$. We can now map any
simply-connected domain $D$ to $\mathbb H$  by a conformal
transformation $F(z)$. We fix this function uniquely by requiring
that
\begin{align}
F(z) &= 0, & F(\zeta) &= \infty, & F'(\zeta) &= 1, \label{mapF}
\end{align}
where $z$ and $\zeta$ are two distinct points on the boundary of
$D$. Then, by definition, the chordal SLE in $D$ from $z$ to
$\zeta$ is the family of maps $h_t(z) = F^{-1}(g_t(F(z)))$, with a
possible random time change, see Fig \ref{fig:SLE-region-D}. The
trace of the new SLE is $\gamma^* = F^{-1}(\gamma)$, and the hulls
are $K^*_t = F^{-1}(K_t)$.

Conformal invariance of SLE then means that, first of all, a trace
$\gamma^*$ in the domain $D$ locally looks the same as a trace
$\gamma$ in $\mathbb H$. In particular, it has the same fractal
dimension. Secondly, various random events (crossings,
swallowings, etc.) that correspond to each other under the map
$F$, have the same probabilities. The requirement of such
conformal invariance was crucial in the original definition of
SLE.

Most importantly for applications in statistical mechanics, SLE
produces conformally-invariant self-avoiding random traces that
are statistically equivalent to critical curves in statistical
mechanics models. It is actually very difficult to make this
statement precise, and a lot of efforts has gone and is going into
establishing the correspondence between SLE for different values
of $\kappa$ with critical points of various lattice models.

\begin{figure}[t]
\centering
\includegraphics[width=0.7\textwidth]{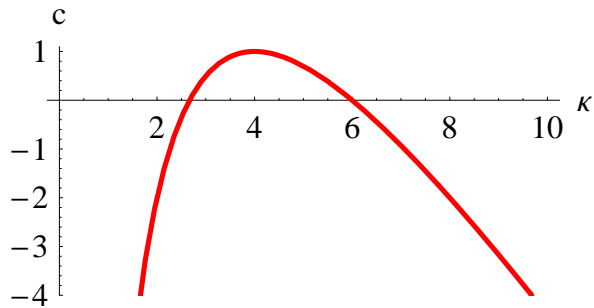}
\caption{Central charge as a function of $\kappa$. Notice that
$\kappa$ and $\kappa' = 16/\kappa$ correspond to the same $c$. For
example, for both $\kappa = 8/3$ and $\kappa = 6$ the central
charge is zero. These values correspond to self-avoiding walks and
boundaries of percolation clusters.} \label{fig:central-charge}
\end{figure}

Correlation functions of local quantities at these critical points
are described by CFTs with central charges $c \leqslant 1$. Bauer
and Bernard \cite{Bauer-Bernard-1} argued that SLE describes
critical curves in all these CFTs. The relation between the SLE
parameter $\kappa$ and the central charge happens to be
\begin{align}
c_\kappa = \frac{(8 - 3\kappa)(\kappa - 6)}{2\kappa} = 1 -
3\frac{(\kappa - 4)^2}{2\kappa}. \label{central charge 1}
\end{align}
This function is plotted in Fig. \ref{fig:central-charge}.
It possesses a remarkable duality, namely
\begin{align}
c_\kappa &= c_{\kappa'}, && {\rm where} \qquad \kappa' =
\frac{16}{\kappa}. \label{duality}
\end{align}
Duplantier \cite{Duplantier-PRL, Duplantier} has argued that this
duality has a geometric meaning. Namely, in the language of SLE,
for $\kappa > 4$ an SLE$_\kappa$ hull $K_t$ has the boundary
$\partial K_t$ (also called external perimeter or frontier), which
locally looks like an SLE$_{\kappa'}$ simple curve with fractal
dimension $d_f(\kappa') = 1 + 2/\kappa$.

Notice that the central charge vanishes for $\kappa = 8/3$ and
$\kappa = 6$. These values are special from SLE point of view,
since for these values the SLE hulls possess very special
properties called locality and restriction, correspondingly. We
will consider these properties in Sections \ref{subsec:locality},
\ref{subsec:restriction}.

Many different arguments, including comparison of critical
exponents, have lead to correspondences between SLE$_\kappa$ and
specific lattice models that we summarize in the Table
\ref{tab:lattice models}.
\begin{table}[t]
\centering
\begin{tabular}{|l|c|c|c|c|c|}
\hline Lattice model & $\kappa$ & $c_\kappa$ & $d_f(\kappa)$  &
$d_f(\kappa')$ & References
\\
\hline \hline
Loop-erased random walk & 2 & $-2$ & 5/4 & -- & \cite{Schramm, LSW-LERW} \\
\hline
Self-avoiding random walk & 8/3 & 0 & 4/3 & -- & \cite{LSW-SAW} \\
\hline Ising model & & & & & \\
spin cluster boundaries & 3 & 1/2 & 11/8 & -- & \cite{Rohde-Schramm} \\
\hline
Dimer tilings & 4 & 1 & 3/2 & -- & \cite{Rohde-Schramm, Kenyon} \\
\hline Harmonic explorer & 4 & 1 & 3/2 & -- &
\cite{Schramm-Sheffield-harmonic-explorer}
\\
\hline Level lines of Gaussian field & 4 & 1 & 3/2 & -- &
\cite{Schramm-Sheffield-level-lines}
\\
\hline
Ising model & & & & & \\
FK cluster boundaries & 16/3 & 1/2 & 5/3 & 11/8 & \cite{Rohde-Schramm} \\
\hline Percolation cluster boundaries & 6 & 0 & 7/4 & 4/3 &
\cite{Schramm, Smirnov, Smirnov-Werner} \\
\hline
Uniform spanning trees  & 8 & $-2$ & 2 & 5/4 & \cite{LSW-LERW} \\
\hline
\end{tabular}
\caption{Some lattice models for which a correspondence with SLE
has been conjectured or rigorously established. The dash in the
$d_f(\kappa')$ column means that the hull and the trace are the
same.} \label{tab:lattice models}
\end{table}
All these models can be related either to the critical point or
the dense phase of the O$(n)$ model, or to the critical point of
the $q$-states Potts model. The Coulomb gas methods map these
models to a Gaussian bosonic field theory with coupling constant
$g$ and with background and screening charges. Within this theory
one can identify the curve creating operators and establish a
relation between $g$ and $\kappa$. As a result, we get the
following relations between the parameters $n$, $q$, $g$ and
$\kappa$:
\begin{align}
n &= - 2 \cos \pi g, & g = \frac{4}{\kappa},
&& 2 &\leqslant \kappa \leqslant \infty,
\label{n g kappa}\\
q &= 2 + 2 \cos 2\pi g, & g = \frac{4}{\kappa}, && 4 & \leqslant
\kappa \leqslant 8. \label{q g kappa}
\end{align}
Mathematically rigorous formulation of conjectures related to
general Potts and O$(n)$ models can be found in Refs.
\cite{Schramm-overview, Rohde-Schramm}.

The critical points of O$(n)$ model correspond to the range $2
\leqslant \kappa \leqslant 4$ ($1 \leqslant g \leqslant 2$), while
the dense phase is described by $\kappa \geqslant 4$ ($0 \leqslant
g \leqslant 1$). Parts of these ranges correspond to negative $n$.
The critical point of the Potts model is the same as the dense
phase of the O$(n)$ model with $n = \sqrt{q}$ only for positive
$n$, which explains the restriction on $\kappa$ in Eq. (\ref{q g
kappa}). All these relations are reviewed below in Section
\ref{sec:Coulomb gas}.

Results presented in Section \ref{sec:Coulomb gas} allow us to
arrive at detailed geometric description of critical curves by
calculating the spectrum of multifractal exponents of the harmonic
measure. These exponents include and generalize the fractal
dimension $d_f(\kappa)$ (\ref{dimension}). This is done in Section
\ref{sec:harmonic measure}.

\section{Basic results from stochastic calculus}
\label{sec:stochastic calculus}

To analyze SLE and apply it to the study of critical curves, we
need to use stochastic calculus. This section provides a brief
summary of the necessary techniques.

Here we only consider one-dimensional stochastic processes. As an
example it is useful to keep in mind a simple diffusion of a
particle on an interval $(a,b)$ on the real line. Everything
trivially generalizes to higher dimensions.

All the material in this section and much more is very nicely
presented in Refs. {\cite{Oksendal, Klebaner}}.

\subsection{Stochastic differential equations, It\^o integrals
and martingales}
\label{subsec:SDE}

A stochastic differential equation (SDE) is, essentially, a
Langevin equation, which mathematicians like to write in terms of
differentials:
\begin{align}
\label{SDE} dx_t = u(x_t,t) \, dt + v(x_t,t) \, dB_t.
\end{align}
Here the first term in the right-hand side is called the drift
term, and in the second term $B_t$ is the standard Brownian motion
(BM) started at $B_0 = 0$ (that is, $W_t = dB_t/dt$ is the white
noise with unit strength). The process $x_t$ describes a random
``trajectory'' of a Brownian particle. The simplest example is
\begin{align}
dx_t = \sqrt\kappa \, dB_t. \nonumber
\end{align}
This describes a simple diffusion with the diffusion coefficient
$\kappa$.

BM is a Gaussian process with independent increments: for any set
of times $0 \leqslant t_1 < t_2 < \ldots < t_k$ the random
variables
\begin{align}
B_{t_1}, B_{t_2} - B_{t_1}, \ldots , B_{t_k} - B_{t_{k-1}}
\nonumber
\end{align}
are independent and normally distributed with zero means and
variances $t_1, t_2 - t_1$, etc. We note here that from this
definition it can be shown that all of the following are standard
BMs:
\begin{align}
&-B_t, &&& \text{reflection invariance}, \nonumber \\
&B_{s+t} - B_s,  &\forall s,t > 0, && \text{time homogeneity}, \nonumber \\
&a^{-1} B_{a^2 t},  &\forall a > 0, && \text{scaling}, \label{Brownian scaling} \\
&t B_{1/t}, &&& \text{time inversion}. \nonumber
\end{align}

One can write the solution of Eq. (\ref{SDE}) as
\begin{align}
x_t = x_0 + \int_0^t \!\! u(x_s,s) \, ds +  \int_0^t \!\! v(x_s,s)
\, dB_s. \nonumber
\end{align}
The last term here is an {\bf It\^o integral} defined as the limit
of finite sums
\begin{align}
\sum_i v(x_{s_i}, s_i) (B_{s_{i+1}} - B_{s_i}). \nonumber
\end{align}
Note the important property that the integrand is taken always at
the left end of the time interval. Then it is always independent
from the increment of the BM multiplying the integrand. This means
that upon averaging over $B_t$ all the terms in the above sum
vanish. The same is true then for the It\^o integral:
\begin{align}
\label{E-Ito-integral} {\mathbf E}^x \Bigl[ \int_0^t \!\! v(x_s,s)
\, dB_s \Bigr] = 0.
\end{align}
Here ${\mathbf E}^x[\ldots]$ stands for the expectation value, or
the average over the realizations of $B_t$, and the superscript
$x$ refers to the initial condition $x_0 = x$. Similarly, we have
the following It\^o isometry:
\begin{align}
{\mathbf E}^x \Bigl[ \Bigl( \int_0^t \!\! v(x_s,s) \, dB_s
\Bigr)^2 \Bigr] = {\mathbf E}^x \Bigl[ \int_0^t \!\! v^2(x_s,s) \,
ds \Bigr]. \nonumber
\end{align}

The previous two equations imply another very important property
of the It\^o integral, namely, that it is a martingale. {\bf
Martingale} is, essentially, a stochastic process $M_t$ which
satisfies the following properties:
\begin{align}
& {\mathbf E}[|M_t|]  < \infty, && \forall \, t, \nonumber \\
& {\mathbf E}[M_t \mid \text{history of } M \text{ up to }s] =
M_s, && \forall \, t \geqslant s. \nonumber
\end{align}
The second condition (which is formalized using the notion of
filtration of $\sigma$-algebras ${\mathcal M}_t$ to describe ``the
history of $M$'') contains a conditional expectation value (see
Section \ref{subsec:conditioning}), and it means that if we know
that at some time $s$ the process $M$ has the value $M_s$, then
the expectation value of this process in the future at any moment
$t \geqslant s$ is going to be the same $M_s$ independently of
time $t$. In particular, the unconditional expectation value
${\mathbf E}[M_t] = M_0$ is simply given by the initial value of
the martingale. Martingales necessarily satisfy stochastic
differential equations without drift terms:
\begin{align}
dM_t &= v(M_t,t) \, dB_t. \label{martingale SDE}
\end{align}

\subsection{It\^o formula}

Next we need the so called It\^o formula, which describes a change
of variables in stochastic calculus. Let me formulate it for 1D
processes first. Suppose that we have an It\^o stochastic process
$x_t$ which satisfies the SDE
\begin{align}
dx_t = u(x_t,t) \, dt + v(x_t,t) \, dB_t, \nonumber
\end{align}
where $B_t$ is the standard BM (that is, $W_t = dB_t/dt$ is the
white noise with unit strength). Now we take any ``reasonable''
function $f(x,t)$ (it should be twice continuously differentiable
in both arguments) and define $y_t = f(x_t,t)$. Then $y_t$ is
again an It\^o process that satisfies the SDE
\begin{align}
dy_t &= \frac{\partial f(x_t,t)}{\partial t} \, dt +
\frac{\partial f(x_t,t)}{\partial x_t} \, dx_t + \frac{1}{2}
\frac{\partial^2 f(x_t,t)}{\partial x_t^2} \, (dx_t)^2. \nonumber
\end{align}
Mnemonically, we need to expand to first order in time, but to
second order in $x_t$. The quantity $(dx_t)^2$ is found using the
rules
\begin{align}
(dt)^2 = dt \, dB_t = dB_t \, dt = 0, \quad (dB_t)^2 = dt.
\nonumber
\end{align}
Thus $(dx_t)^2 = v^2(x_t,t) \, dt$, and the SDE for $y_t$ becomes
the {\bf It\^o formula}:
\begin{align}
dy_t &= \Bigl( \frac{\partial f(x_t,t)}{\partial t} + \frac{1}{2}
v^2(x_t,t) \frac{\partial^2 f(x_t,t)}{\partial x_t^2} + u(x_t,t)
\frac{\partial f(x_t,t)}{\partial x_t} \Bigr) \, dt \nonumber\\ &
\quad + v(x_t,t) \frac{\partial f(x_t,t)}{\partial x_t} \, dB_t.
\label{Ito}
\end{align}

Often one encounters ``time-homogeneous'' processes when both
$u(x_t)$ and $v(x_t)$ do not depend explicitly on time. In this
case the process $x_t$ is called a diffusion with drift $u(x_t)$
and diffusion coefficient $v^2(x_t)$. For a function $y_t =
f(x_t)$ of a diffusion the previous formula slightly simplifies:
\begin{align}
dy_t & = \Bigl( \frac{1}{2} v^2(x_t) \frac{d^2 f(x_t)}{d x_t^2} +
u(x_t) \frac{d f(x_t)}{d x_t} \Bigr) \, dt + v(x_t) \frac{d
f(x_t)}{d x_t} \, dB_t. \label{Ito-homogeneous}
\end{align}
The coefficients here do not depend explicitly on $t$, which means
that in this case the process $y_t$ is also a diffusion. The
differential operator ${\hat A}$ appearing in the first term here
is called the {\bf generator} of the diffusion $x_t$:
\begin{align}
{\hat A}f(x) = \frac{1}{2} v^2(x) \frac{d^2 f(x)}{d x^2} + u(x)
\frac{d f(x)}{dx}. \nonumber
\end{align}

we can integrate Eq. (\ref{Ito-homogeneous}):
\begin{align}
\label{Y-homogeneous} f(x_t) & = f(x_0) + \int_0^t \!\! {\hat A}
f(x_s) \, ds + \int_0^t \!\! v(x_s) \frac{d f(x_s)}{d x_s} \,
dB_s.
\end{align}

\subsection{Stopping times and Dynkin formula}
\label{subsec:Dynkin}

It is often interesting to study functions of stochastic processes
at various random times. Such a random time $\tau$ is called a
{\bf stopping time} if at any moment $t$ we can decide whether
$\tau < t$ or not, or, in other words, whether $\tau$ has happened
before time $t$. In our basic example we can consider, for
example, the escape time, that is, the first time $\tau$ when the
diffusing particle leaves the interval $(a,b)$:
\begin{align}
\label{exit time} \tau_{(a,b)} = \inf[t: x_t \notin (a,b)].
\end{align}
In this example in each realization of our random process we are
able to say whether the particle is in the interval at time $t$ or
outside it. Similarly, we can consider the first hitting time of a
closed set $A \subset \mathbb{R}$:
\begin{align}
\tau_A = \inf[t: x_t \in A]. \nonumber
\end{align}

Now we evaluate Eq. (\ref{Y-homogeneous}) at some stopping time
$\tau$:
\begin{align}
f(x_\tau) & = f(x_0) + \int_0^\tau \!\! {\hat A} f(x_s) \, ds +
\int_0^\tau \!\! v(x_s) \frac{d f(x_s)}{d x_s} \, dB_s. \nonumber
\end{align}
It can be shown that an analog of the martingale property
(\ref{E-Ito-integral}) holds for It\^o integrals with limits that
are stopping times (this is related to the so-called strong Markov
property of the BM, which states that even for a stopping time
$\tau$ the increment $B_{t+\tau} - B_\tau$ is a standard BM
independent from $B_t$ for $t \in [0,\tau]$). Then, taking the
expectation values on both sides of the last equation we get the
so-called {\bf Dynkin formula}:
\begin{align}
\label{Dynkin formula} {\mathbf E}^x[f(x_\tau)] & = f(x) +
{\mathbf E}^x\Bigl[\int_0^\tau \!\! {\hat A} f(x_s) \, ds\Bigr].
\end{align}
we assumed here that $x_0 = x$. Also, to really prove this
formula, one needs to assume that ${\mathbf E}^x[\tau] < \infty$.

The Dynkin formula is extremely useful when we need to find
various escape probabilities. Let me consider one example in
detail. Suppose, we have a diffusion $x_t$ started at $x_0 = x \in
(a,b)$. Then at the exit time $\tau = \tau_{(a,b)}$ (see Eq.
(\ref{exit time})) the particle can only escape the interval
$(a,b)$ either at the point $a$ or at the point $b$. Then we can
ask the question: ``What is the probability that the escape
happens through the point $a$?'' Formally, we need to find one of
the quantities
\begin{align}
P_a = {\mathbf P}[x_\tau = a], && P_b = {\mathbf P}[x_\tau = b],
\nonumber
\end{align}
where ${\mathbf P}[X]$ denotes the probability of the event $X$.
It is obvious that these two probabilities add to one:
\begin{align}
\label{sum=1} P_a + P_b = 1.
\end{align}
We will find another equation relating $p_a$ and $p_b$ using the
Dynkin formula.

To do this, we consider the expectation value
\begin{align}
{\mathbf E}^x[f(x_\tau)] = P_a f(a) + P_b f(b). \nonumber
\end{align}
For any function $f(x_t)$ the LHS of this equation is given by the
Dynkin formula. But if we find a function that satisfies the
equation
\begin{align}
{\hat A} f(x) = 0, \nonumber
\end{align}
then the It\^o formula (\ref{Ito-homogeneous}) implies that
$f(x_t)$ is a martingale (no drift term in the equation), and the
Eq. (\ref{Dynkin formula}) simplifies to ${\mathbf E}^x[f(x_\tau)]
= f(x)$, and for such a function we get
\begin{align}
P_a f(a) + P_b f(b) = f(x). \nonumber
\end{align}
Combining this with Eq. (\ref{sum=1}), we finally find
\begin{align}
P_a = \frac{f(x) - f(b)}{f(a) - f(b)}, && P_b = \frac{f(a) -
f(x)}{f(a) - f(b)}. \label{escape-probabilities}
\end{align}
Since ${\hat A}$ is a linear differential operator, the function
$f(x)$ can usually be found explicitly. Often it is expressed in
terms of the hypergeometric function.

Let us note that to use the formulas (\ref{escape-probabilities}),
we need {\it any non-constant} zero mode of $\hat A$. There is a
continuum of such solutions parametrized by two constants of
integration ($\hat A$ is a second order differential operator),
but both the additive and the multiplicative constants cancel when
a zero mode is substituted into Eq. (\ref{escape-probabilities}).

\subsection{Backward and forward Kolmogorov equations}

Let us denote
\begin{align}
b(x,t) = {\mathbf E}^x[f(x_t)]. \nonumber
\end{align}
Then taking ${\mathbf E}^x$ of both sides in Eq.
(\ref{Y-homogeneous}) and differentiating with respect to $t$, we
get
\begin{align}
\frac{\partial b}{\partial t} &= {\mathbf E}^x[{\hat A} f(x_t)].
\nonumber
\end{align}
It turns out that the right hand side here can be expressed in
terms of $b(x,t)$ also. Roughly speaking (this is not very
trivial), the expectation value and the operator $\hat A$ can be
interchanged (after this $\hat A$ acts on the variable $x$),
giving the so called {\bf backward Kolmogorov equation}:
\begin{align}
\frac{\partial b}{\partial t} &= {\hat A} b, & b(x,0) &= f(x).
\label{Kolmogorov-back}
\end{align}
Note that this is different from the more familiar Fokker-Planck
equation. In fact, the Fokker-Plank equation (called the {forward
Kolmogorov equation} in mathematics) involves the operator ${\hat
A}^*$ that is adjoint to ${\hat A}$:
\begin{align}
{\hat A}^* f(x) = \frac{1}{2} \frac{d^2}{d x^2} (v^2(x) f(x)) -
\frac{d}{dx} (u(x) f(x)). \nonumber
\end{align}

The forward Kolmogorov equation involving ${\hat A}^*$ appears as
follows. The process $x_t$ has the transition measure density
$p_t(y,x)$, which means that the expectation values of functions
of $x_t$ can be found like this:
\begin{align}
{\mathbf E}^x[f(x_t)] = \int \!\! f(y) p_t(y,x) \, dy. \nonumber
\end{align}
This is equivalent to $p_t(y,x) = {\mathbf E}^x[\delta(x_t - y)]$,
which is a familiar definition of the probability density for the
process $x_t$. The density $p_t(y,x)$ is also the kernel or the
Green's function of the diffusion $x_t$. It is this function that
satisfies the forward Kolmogorov equation with respect to the
final coordinate $y$:
\begin{align}
\frac{\partial}{\partial t}p_t(y,x) = {\hat A}_y^* p_t(y,x).
\nonumber
\end{align}
Because the operator ${\hat A}^*$ has all the derivatives on the
left, the total probability is conserved: $\int \!\! p_t(y,x) \,
dy = {\mathbf E}^x[1] = 1$.

\subsection{Feynman-Kac formula}
\label{subsec:FK}

A simple generalization of the backward Kolmogorov equation
(\ref{Kolmogorov-back}) is the so-called Feynman-Kac (FK) formula.
It concerns the expectation value
\begin{align}
c(x,t) &= {\mathbf E}^x \Big[\exp \Big(-\int_0^t \!\! V(x_s) ds
\Big) f(x_t) \Big], \nonumber
\end{align}
where $V(x)$ is a continuous function such that the integral in
the exponent converges as $t \to \infty$, $f(x)$ is as before, and
$x_t$ is a time-homogeneous It\^o process (a diffusion). The FK
formula is the following partial differential equation for
$c(x,t)$:
\begin{align}
\frac{\partial c}{\partial t} &= {\hat A}c - V c, & c(x,0) & =
f(x). \label{FK-formula}
\end{align}
This formula is obtained (schematically) as follows. We define
\begin{align}
D_t(x) &= \int_0^t \!\! V(x_s) ds, & C(x_t, t) &= e^{-D_t(x)}
f(x_t). \label{martingale}
\end{align}
The process $C(x_t, t)$ explicitly depends on $t$ through its
first factor, and the It\^o equation for it is obtained from the
formula (\ref{SDE}):
\begin{align}
dC(x_t, t) &=  \big[{\hat A} - V(x_t)\big] C(x_t, t) dt + v(x_t)
\frac{\partial C(x_t, t)}{\partial x_t} \, dB_t. \label{Ito-C}
\end{align}
Upon averaging the last term vanishes, as usual, and we get
\begin{align}
\frac{\partial c}{\partial t} &= {\mathbf E}^x \big[\big({\hat A}
- V(x_t)\big) C(x_t, t) \big]. \nonumber
\end{align}
Similar to the case of the backward Kolmogorov equation, the right
hand side can be expressed in terms of $c(x,t)$, which results in
Eq. (\ref{FK-formula}).

There is a variant of the FK formula that we can call a stationary
FK formula. Namely, we can choose the function $f(x)$ in Eq.
(\ref{martingale}) to satisfy the stationary version of Eq.
(\ref{FK-formula}):
\begin{align}
\big[{\hat A} - V(x)\big] f(x) = 0. \label{stationary FK formula}
\end{align}
Then the process $C(x_t,t)$ defined in Eq. (\ref{martingale}) with
$f(x)$ being a solution of (\ref{stationary FK formula}) is a
martingale, since the drift term in the It\^o formula
(\ref{Ito-C}) vanishes! The expectation value $c(x,t)$ is then
really a function of $x$ only, and is equal to $f(x)$ for all
times.

Now if we know that the process $x_t$ is transient, that is,
$\lim_{t \to \infty} x_t = \infty$, we normalize $f(x)$ such that
$f(\infty) = 1$, and denote $D(x) = D_\infty(x) < \infty$ we get
\begin{align}
f(x) &= \lim_{t \to \infty} {\mathbf E}^x[C(x_t, t)] = {\mathbf
E}^x\big[e^{-D(x)}\big]. \label{stationary FK formula 1}
\end{align}
It should be clear now that in this situation we can compute the
characteristic function $\chi(k,x) = {\mathbf E}^x\big[e^{i k
D(x)}\big]$ of the random variable $D$. In addition, if the
variable $D(x)$ is known to be non-negative, the same approach
gives the Laplace transform $L(s,x)$ of its probability
distribution function $p(D,x)$:
\begin{align}
L(s,x) &= {\mathbf E}^x\big[e^{- s D(x)}\big] = \int_0^\infty e^{-
s D(x)} p(D,x) dD. \label{Laplace transform}
\end{align}
Notice that all the quantities $D(x)$, $\chi(k,x)$, $L(s,x)$, and
$p(D,x)$ implicitly depend on $x$, the initial value of the random
process $x_t$.

\subsection{Conditional probabilities and expectation values}
\label{subsec:conditioning}

Sometimes in the study of random variables and stochastic
processes it is interesting or necessary to restrict the
statistical ensemble of realizations to a sub-ensemble satisfying
a certain condition. This condition may depend on the outcome of a
certain random event. For example, for a diffusion on the real
line we may consider only trajectories that always stay on the
positive semi-axis, or the ones that happen to be on the positive
semi-axis at a certain time. Such a restriction of an ensemble is
called conditioning.

Within a restricted or conditioned ensemble we can ask for
probabilities of various events or expectation values of random
quantities. These are called conditional probabilities and
expectation values. In words we can say: ``What is the probability
of an event $A$ given that an event $B$ happened?'' Such
probability is denoted by ${\mathbf P}[A \mathop{|} B]$. It is
well known in probability theory that conditional probabilities
are easily calculated by the formula
\begin{align}
{\mathbf P}[A \mathop{|} B] &= \frac{{\mathbf P}[A \text{ and }
B]}{{\mathbf P}[B]}, \label{conditional probability}
\end{align}
where ${\mathbf P}[A \text{ and } B]$ is the unconditioned
probability that the events $A$ and $B$ both happen, and ${\mathbf
P}[B]$ is the unconditioned probability that the event $B$
happens. Notice that the Eq. (\ref{conditional probability}) only
makes sense if the event $B$ has non-zero probability ${\mathbf
P}[B] > 0$.

Similarly, given that an event $B$ occurs, we may want to find the
expectation value of a random variable $X$, denoted ${\mathbf E}[X
\mathop{|} B]$. Conditional expectation values have many known
properties, but there is no general explicit formula for them
similar to Eq. (\ref{conditional probability}).

\section{Basic properties of SLE}
\label{sec:basic properties}

In this Section, based mainly on Refs. \cite{Rohde-Schramm, LSW-1,
LSW-restriction}, we consider the basic properties of SLE. Some of
them have already been mentioned, and they will be here illustrated
by plausible arguments. These arguments already require some
calculations typical for SLE. The main idea of most simple
calculations with SLE is to look at various random events in the
physical plane, then see what happens at the same time in the
mathematical plane. Then we choose a simple real function of the SLE
process and study the values this function assumes during the
interesting events.

\subsection{Scaling}

The scaling property of the Brownian motion (\ref{Brownian
scaling}) immediately implies the scaling for the SLE processes
$g_t(z)$ and the growing SLE hulls $K_t$. Namely, we have the
following stochastic equivalence:
\begin{align}
g_t(z) &= \frac{1}{a} g_{a^2t}(az), & w_t(z) &= \frac{1}{a}
w_{a^2t}(az), &  \textrm{in law}. \label{SLE scaling}
\end{align}
By this we mean that the random quantities on both sides of these
equations have the same probability distribution. Eq. (\ref{SLE
scaling}) is easily derived by observing that the SDE for the
right hand side contains $\frac{\sqrt{\kappa}}{a} B_{a^2t}$ as the
driving function. The scaling for the SLE processes (\ref{SLE
scaling}) immediately implies a similar scaling for the SLE hulls:
\begin{align}
K_t &= \frac{1}{a} K_{a^2t} \qquad \textrm{in law}. \label{SLE
hulls scaling}
\end{align}

\subsection{Phases on SLE}
\label{subsec:SLE phases}

The phases of SLE were already described above in Section
\ref{sec:Chordal SLE}. Here we provide a crude derivation of the
phases and phase transitions between them.

\subsubsection{Transition at $\kappa = 4$.}
\label{subsubsec:kappa=4}

First we discuss the transition at $\kappa = 4$. To this end we
will fix a point $x \in {\mathbb R}$ on the real axis in the
physical plane and consider the motion of its image $x_t = w_t(x)$
up to the time when it hits $0$ (which may never happen):
\begin{align}
d x_t &= \frac{2}{x_t} dt - \sqrt{\kappa} dB_t, & x_0 &= x.
\nonumber
\end{align}

In the mathematical plane, we fix points $a$ and $b$ on the real
axis so that
\begin{align}
0 < a < x < b < \infty. \nonumber
\end{align}
Let $\tau$ be the exit time from the interval $[a,b]$. Being
continuous, the process $x_t$ can exit $[a,b]$ either through $a$
(with probability $P_a$), or through $b$ (with probability $1 -
P_a$). As described in Sec. \ref{subsec:Dynkin}, the probability
$P_a$ can be found if we know a non-constant zero mode $f(x)$ of
the generator of diffusion $x_t$, see Eq.
(\ref{escape-probabilities}).

Now if we take the limits $a  \to 0,\, b \to \infty,$ it becomes
the probability for $x_t$ to hit $0$ in a finite time, which is
the probability for the point $x$ on the physical plane to belong
to the hull:
\begin{align}
P = \lim_{b \to \infty} \lim_{a \to  0} \frac{f(x) - f(b)}{f(a) -
f(b)}. \label{p}
\end{align}
In general, the order of limits matters here. If it is reversed,
\begin{align}
\tilde{P} = \lim_{a \to 0}\lim_{b \to \infty}\frac{f(x) -
f(b)}{f(a) - f(b)} \label{ptilde}
\end{align}
is the probability for $x_t$ to come arbitrarily close to 0, that
is, for the hull to come arbitrarily close to the boundary. Note
that in order to determine these probabilities, only the behavior
of a zero mode of $\hat A$ at zero and at infinity is necessary.

The generator for the process $x_t$ is
\begin{align}
{\hat A} = \frac{\kappa}{2}\frac{d^2}{dx^2} +
\frac{2}{x}\frac{d}{dx}. 
\nonumber
\end{align}
A non-constant zero mode of this operator is
\begin{align}
f(x) = \left\{
\begin{array}{ll}
|x|^{1 - \frac{4}{\kappa}} & \textrm{for $\kappa\neq 4$}, \\
\log |x| & \textrm{for $\kappa = 4$}.
\end{array}
\right. \nonumber
\end{align}
Substituting this into Eqs. (\ref{p}, \ref{ptilde})  we find that
the answer is independent of $x$, thus the probability for the
hull to touch the boundary
\begin{align} P = \left\{
\begin{array}{ll}
0 & \textrm{for $\kappa \leqslant 4$}, \\
1 & \textrm{for $\kappa > 4$}.
\end{array}
\right. \nonumber
\end{align}
For $\kappa=4$ the order of limits is important and we find that
\begin{align} \tilde{P} = \left\{
\begin{array}{ll}
0 & \textrm{for $\kappa < 4$}, \\
1 & \textrm{for $\kappa \geqslant 4$}.
\end{array}
\right. \nonumber
\end{align}
These formulas clearly exhibit a sort of ``phase transition'' at
$\kappa = 4$.

\subsubsection{Transition at $\kappa = 8$.}

This transition is more subtle. To study it, we fix two points $0
< x < y < \infty$ on the real axis in the physical plane, and
compare the times $\tau_x$ and $\tau_y$ when they enter the
growing SLE hull. For $\kappa > 4$ both these times are finite.

It happens that for $\kappa < 8$ there is a finite probability
that the points $x$ and $y$ are swallowed simultaneously:
${\mathbf P}[\tau_x = \tau_y] > 0$. On the other hand, for $\kappa
\geqslant 8$, with probability one, $\tau_x < \tau_y$. In this
case the points on the real axis are added to the trace
sequentially. The same is true for points in $\mathbb H$.

To make these statements plausible (without giving a real proof),
let us consider $x_t = w_t(x)$, $y_t = w_t(y)$, and $q_t = \log
\frac{y_t}{x_t}$. By continuity it is clear that $0 \leqslant x_t
\leqslant y_t \leqslant \infty$ for all times, so $0 \leqslant q_t
\leqslant \infty$.

If $x$ joins the hull before $y$ (that is, $\tau_x < \tau_y$),
then $q_{\tau_x} = \infty$, and the probability ${\mathbf
P}[q_{\tau_x} = \infty] > 0$ (as well as ${\mathbf P}[q_{\tau_x} =
0] > 0$). If the points $x$ and $y$ join the hull simultaneously,
then $q_t$ stays finite (bounded) for all times up to $\tau_x =
\tau_y$, and ${\mathbf P}[q_{\tau_x} = \infty] = 0$. So we need to
consider the motion of $q_t$.

Using It\^o formula for $q_t$ we get
\begin{align}
d q_t &= \frac{1}{y_t} dy_t - \frac{1}{x_t} dx_t -
\frac{1}{2y_t^2} (dy_t)^2 + \frac{1}{2x_t^2} (dx_t)^2. \nonumber
\end{align}
Here we need to substitute, as usual,
\begin{align}
d x_t &= \frac{2}{x_t} dt - \sqrt{\kappa} dB_t, & d y_t &=
\frac{2}{y_t} dt - \sqrt{\kappa} dB_t. \nonumber
\end{align}
This gives the following SDE:
\begin{align}
d q_t &= \Big(\frac{\kappa}{2} -2 \Big) \Big(\frac{1}{2x_t^2} -
\frac{1}{2y_t^2}\Big) dt + \sqrt{\kappa} \Big( \frac{1}{x_t} -
\frac{1}{y_t}\Big) dB_t. \nonumber
\end{align}
Notice that this equation is not of standard It\^o type, since the
coefficients of the right hand side depend separately on $x_t$ and
$y_t$, but not on $q_t$. This is easily remedied by a trick that
is called ``a random time change''.

This time change amounts to consider a new time variable
\begin{align}
{\tilde t} &= \int_0^t \frac{ds}{x_s^2}, \nonumber
\end{align}
which is a monotonous function of $t$ (since we integrate a
positive quantity $x_s^{-2}$). In differential form the time
change is
\begin{align}
d {\tilde t} &= dt/x_t^2. \nonumber
\end{align}
We also need to consider the stochastic process
\begin{align}
{\tilde B}_t &= \int_0^t \frac{dB_s}{x_s}, & d{\tilde B}_t &=
\frac{dB_t}{x_t}. \nonumber
\end{align}
Notice that
\begin{align}
\big(d{\tilde B}_t\big)^2 &= (dB_t)^2/x_t^2 = dt/x_t^2 = d {\tilde
t}. \nonumber
\end{align}
Therefore, the process $B_{\tilde t} = {\tilde B}_t$ is the
standard Brownian motion with respect to the new time $\tilde
t\,$!

In terms of the new variables the SDE for $q_{\tilde t}$ takes the
standard It\^o form:
\begin{align}
d q_{\tilde t} &= \Big(\frac{\kappa}{2} -2 \Big) \big(1 -
e^{-2q}\big) d{\tilde t} + \sqrt{\kappa} \big(1 - e^{-q}\big)
dB_{\tilde t}. \nonumber
\end{align}
The generator of diffusion for this process is
\begin{align}
{\hat A} &= \frac{\kappa}{2} \big(1 - e^{-q}\big)^2
\frac{d^2}{dq^2} + \Big(\frac{\kappa}{2} -2 \Big) \big(1 -
e^{-2q}\big) \frac{d}{dq}, \nonumber
\end{align}
and we need to find a zero mode of this operator to study the
probability ${\mathbf P}[q_t = \infty]$.

The equation ${\hat A} f = 0$ can be easily solved by rewriting it
as (prime denotes the derivative with respect to $q$)
\begin{align}
\frac{f''}{f'} = (\log f')' = \Big(\frac{4}{\kappa} - 1 \Big)
\coth \frac{q}{2}. \nonumber
\end{align}
When we integrate this equation, we can drop the integration
constants, which are inessential, as was explained in the end of
Section \ref{subsec:Dynkin}:
\begin{align}
\log f' &= \Big(\frac{8}{\kappa} - 2 \Big) \log \Big(\sinh
\frac{q}{2}\Big), & f' &= \Big(\sinh
\frac{q}{2}\Big)^{\frac{8}{\kappa} - 2}. \nonumber
\end{align}
Since we now only consider $\kappa > 4$, the function $f'(q)$
exponentially decays as $q \to \infty$, and we can choose
(ignoring a multiplicative constant)
\begin{align}
f(q) = \int_q^\infty \Big(\sinh \frac{s}{2}\Big)^{\frac{8}{\kappa}
- 2} ds. \nonumber
\end{align}
When $q \to 0$, this integral converges at the lower limit when
$\kappa < 8$, and in this case we get ${\mathbf P}[q_{\tau_x} = 0]
> 0$. On the other hand, when $\kappa \geqslant 8$, the function
$f(q)$ diverges as $q \to 0$, and ${\mathbf P}[q_{\tau_x} = 0] =
0$, implying that ${\mathbf P}[q_{\tau_x} = \infty] = 1$.

\subsection{Locality}
\label{subsec:locality}

Many properties of SLE can be discovered by studying how SLE gets
perturbed by distortions of the boundary of the domain where it
evolves. Such distortions can be described by conformal maps. This
setting is similar to the definition of SLE in an arbitrary
simply-connected domain $D$ in Section \ref{sec:Chordal SLE}, but
there are important differences.

Specifically, let us consider a usual SLE$_\kappa$ evolving in the
upper half plane. Consider a hull $A$ located a finite distance
away from the origin. Then the SLE trace may hit the hull and have
a non-zero overlap with its interior. Note that this would not
happen for SLE defined in the domain ${\mathbb H} \setminus A$ as
in Section \ref{sec:Chordal SLE}. Let the hitting time of the hull
$A$ be $\tau_A$.

\begin{figure}[t]
\centering
\includegraphics[width=0.8\textwidth]{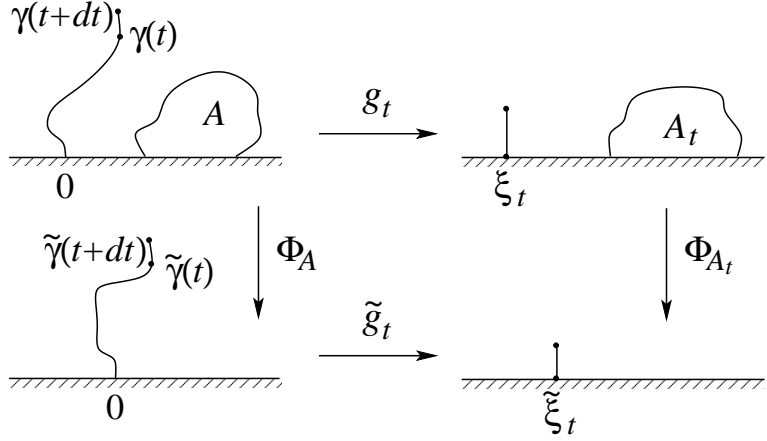}
\caption{Various maps in the definitions of locality and
restriction. Similar figure illustrating a commutative diagram of
conformal maps first appeared in Ref. \cite{LSW-restriction}.}
\label{fig:locality}
\end{figure}
Next we consider the image of the SLE under the map $\Phi_A$
removing the hull $A$ from the upper half plane and normalized as
in Eq. (\ref{mapPhi}):
\begin{align}
\Phi_A(0) &= 0, & \Phi_A(\infty) &= \infty, & \Phi_A'(\infty) &=
1, \label{mapPhiA}
\end{align}
The SLE hulls $K_t$ get mapped to hulls ${\tilde K}_t =
\Phi_A(K_t)$, and these can be removed from $\mathbb H$ by the
family of maps ${\tilde g}_t$. Though this procedure works for any
value of $\kappa$, for simplicity we illustrate it in Fig.
\ref{fig:locality} for $\kappa \leqslant 4$, in which case SLE
hulls are the traces. The maps ${\tilde g}_t$ are normalized as
\begin{align}
{\tilde g}_t(z) &= z + \frac{a_t}{z} + o(z^{-1}), & z \to \infty,
\nonumber
\end{align}
where $a_t$ is the capacity of ${\tilde K}_t$, and evolve
according to the Loewner equation
\begin{align}
\partial_t {\tilde g}_t(z) &=
\frac{\partial_t a_t}{{\tilde g}_t(z) - {\tilde \xi}_t}.
\label{Loewner-g-tilde}
\end{align}

We want now to find the properties of the driving function ${\tilde
\xi}_t$ and see if there is a time change that would make it a
Brownian motion $\sqrt{\kappa} B_{\tilde t}$. If this is the case,
then ${\tilde g}_t(z)$ is the standard SLE$_\kappa$, at least for $t
\leqslant \tau_A$. This implies that the original SLE $g_t$ is the
same as the SLE defined in the domain ${\mathbb H} \setminus A$
according to Section \ref{sec:Chordal SLE}, see Fig.
\ref{fig:SLE-region-D} and Eq. (\ref{mapF}). Loosely speaking, we
can say that in this situation the SLE $g_t$ does not feel the
presence of the hull $A$ until $K_t$ hits $A$. This justifies the
name {\bf locality} for this property. As we will see next, this
happens only for $\kappa = 6$, which was rigorously established in
Ref. \cite{LSW-1} (Ref. \cite{LSW-restriction} contains a simpler
proof that we follow here), and is in perfect agreement with the
statement that SLE$_6$ describes the scaling limit of critical
percolation interfaces. Even on the lattice such interfaces are
determined locally, since every site is black or white independently
of the others.

To find the capacity $a_t$ and the driving function ${\tilde
\xi}_t$ we notice that the hull $K_t \cup A$ can be removed by a
different sequence of maps. Namely, we first remove the hull $K_t$
of the original SLE by $g_t$. Doing this, we deform the hull $A$
into a hull $A_t$. Next we remove the hull $A_t$ by the map
$\Phi_t = \Phi_{A_t}$ normalized the same way as $\Phi_A$, see
Fig. \ref{fig:locality} where an infinitesimal step from $t$ to $t
+ dt$ is also shown. Then we have a commutative diagram of maps,
meaning that
\begin{align}
{\tilde g}_t(\Phi_A(z)) &= \Phi_t(g_t(z)). \label{commutative
diagram}
\end{align}
In particular, for the image of the tip of the trace $\gamma(t)$
we have
\begin{align}
{\tilde \xi}_t &= {\tilde g}_t(\Phi_A(\gamma(t))) =
\Phi_t(g_t(\gamma(t))) = \Phi_t(\xi_t), \label{tilde xi}
\end{align}
where $\xi_t = \sqrt{\kappa} B_t$ is the usual SLE driving
function.

From the derivation of Loewner equation in Section
\ref{sec:Loewner} we already know that the capacity of the
infinitesimal vertical segment $g_t(\gamma[t,t+dt])$ is $2dt$, see
Eq. (\ref{capacity}). When this segment (together with the hull
$A_t$) is mapped to $\mathbb H$ by $\Phi_t$, it simply gets
rescaled by $\Phi_t'(\xi_t)$. (Note that the map $\Phi_t$ is
regular away from the hull $A_t$, and in particular, at the point
$\xi_t$. Also, since the map preserves a portion of the real axis
near $\xi_t$, its derivative $\Phi_t'(\xi_t) > 0$). Therefore, the
scaling property of the capacity, Eq. (\ref{scaling}), implies
\begin{align}
\partial_t a_t &= 2 \Phi_t'(\xi_t)^2.
\label{capacity a t}
\end{align}

Taking time derivative of Eq. (\ref{commutative diagram}) and
using Eqs. (\ref{Loewner-g-tilde}, \ref{capacity a t}) we get
\begin{align}
\frac{2 \Phi_t'(\xi_t)^2}{{\tilde g}_t(\Phi_A(z)) - {\tilde
\xi}_t} =
\partial_t \Phi_t(g_t(z)) + \Phi_t'(g_t(z)) \frac{2}{g_t(z) -
\xi_t}. \nonumber
\end{align}
Denoting $w = g_t(z)$ and using Eq. (\ref{tilde xi}) we simplify
this to
\begin{align}
\partial_t \Phi_t(w) &=
\frac{2 \Phi_t'(\xi_t)^2}{\Phi_t(w) - \Phi_t(\xi_t)} - \frac{2
\Phi_t'(w)}{w - \xi_t}. \nonumber
\end{align}
The right hand side of this equation is non-singular in the limit
$w \to \xi_t$. To see this we expand it in powers of $(w - \xi_t)$
(using Mathematica):
\begin{align}
\partial_t \Phi_t(w) &= -3 \Phi_t''(\xi_t) +
\Big(\frac{1}{2} \frac{\Phi_t''(\xi_t)^2}{\Phi_t'(\xi_t)} -
\frac{4}{3}\Phi_t'''(\xi_t)\Big) (w - \xi_t) + O\left((w -
\xi_t)^2\right). \label{expansion}
\end{align}
The first term of the expansion gives $\partial_t
\Phi_t(w)\big|_{w = \xi_t} = -3 \Phi_t''(\xi_t)$.

Finally, using It\^o formula for ${\tilde \xi}_t = \Phi_t(\xi_t)$,
we have
\begin{align}
d {\tilde \xi}_t &= \partial_t \Phi_t(w)\big|_{w = \xi_t} dt +
\Phi_t'(\xi_t) d\xi_t + \frac{1}{2} \Phi_t''(\xi_t)
(d\xi_t)^2 \nonumber \\
&= \Phi_t'(\xi_t) d\xi_t + \Big(\frac{\kappa}{2} - 3 \Big)
\Phi_t''(\xi_t) dt. \nonumber
\end{align}
For $\kappa = 6$ and only for this value the drift term in the
above equation vanishes, and $d{\tilde \xi}_t$ becomes
$\sqrt{\kappa} dB_{\tilde t}$ after the random time change
$d{\tilde t} = \Phi_t'(\xi_t)^2 dt$, proving locality for SLE$_6$.

\subsection{Restriction}
\label{subsec:restriction}

In this Section we consider the same setup as in the previous one,
but only for $\kappa \leqslant 4$. In this case SLE traces are
simple curves, and there is a non-zero probability ${\mathbf
P}[\gamma \cap A = \varnothing]$ that a trace $\gamma[0,\infty)$
does not intersect the hull $A$. This is also the probability that
the hitting time $\tau_A = \infty$. The collection of these
probabilities for all hulls $A$ located a finite distance away
from the origin completely characterizes the distribution of the
curve $\gamma$, since it determines the likelihoods for the curve
to go through various places. As we will show in this Section,
these probabilities can be found for SLE$_{8/3}$ in term of the
map $\Phi_A$ normalized again as in Eq. (\ref{mapPhiA}).

Given that an SLE$_\kappa$ trace does not intersect $A$, we can map
the whole trace together with ${\mathbb H} \setminus A$ to $\mathbb
H$ by the map $\Phi_A$ and ask what the probability distribution of
the image $\Phi_A(\gamma)$ is. If this distribution happens to be
the same SLE$_\kappa$, we say that SLE$_\kappa$ satisfies the {\bf
restriction} property. Following Ref. \cite{LSW-restriction}, where
conformal restriction was introduced and rigorously studied, we will
show below that this happens only for $\kappa = 8/3$.

We start by studying the rescaling factor $\Phi_t'(\xi_t)$ as a
function of time. The second term in the expansion
(\ref{expansion}) leads to
\begin{align}
\partial_t \Phi_t'(w)|_{w = \xi_t} &= \frac{1}{2}
\frac{\Phi_t''(\xi_t)^2}{\Phi_t'(\xi_t)} -
\frac{4}{3}\Phi_t'''(\xi_t), \nonumber
\end{align}
and then It\^o formula gives
\begin{align}
d\Phi_t'(\xi_t) &= \partial_t \Phi_t'(w)|_{w = \xi_t} dt +
\Phi_t''(\xi_t) d\xi_t + \frac{1}{2} \Phi_t'''(\xi_t)
(d\xi_t)^2 \nonumber \\
&= \Phi_t''(\xi_t) d\xi_t + \Big(\frac{1}{2}
\frac{\Phi_t''(\xi_t)^2}{\Phi_t'(\xi_t)} + \Big(\frac{\kappa}{2} -
\frac{4}{3}\Big)\Phi_t'''(\xi_t)\Big) dt. \nonumber
\end{align}
The drift term in this equation cannot be removed by any choice of
$\kappa$. However, if we apply It\^o formula again to $M_t^{(h)} =
\Phi_t'(\xi_t)^h$, we get
\begin{align}
\frac{dM_t^{(h)}}{h M_t^{(h)}} &=
\frac{d\Phi_t'(\xi_t)}{\Phi_t'(\xi_t)} + \frac{h - 1}{2}
\frac{[d\Phi_t'(\xi_t)]^2}{\Phi_t'(\xi_t)^2}
\nonumber \\
&= \frac{\Phi_t''(\xi_t)}{\Phi_t'(\xi_t)} d\xi_t + \Big( \frac{(h
- 1)\kappa + 1}{2} \frac{\Phi_t''(\xi_t)^2}{\Phi_t'(\xi_t)^2} +
\Big(\frac{\kappa}{2} -
\frac{4}{3}\Big)\frac{\Phi_t'''(\xi_t)}{\Phi_t'(\xi_t)} \Big) dt.
\label{M t alpha}
\end{align}
If we now choose $\kappa = 8/3$ and $h = 5/8$, the drift term in
the last equation vanishes, which implies that $M_t^{(5/8)}$ is a
martingale. Then, on the one hand, the expectation value of
$M_t^{(5/8)}$ is given by the value of this process at $t=0$:
\begin{align}
{\mathbf E}[M_t^{(5/8)}] & = M_0^{(5/8)} = \Phi_0'(\xi_0)^{5/8} =
\Phi_A'(0)^{5/8}. \nonumber
\end{align}
On the other hand, let us consider the expectation value at the
stopping time $\tau$ when the trace $\gamma$ hits for the first
time either the hull $A$ or the semicircular arc $C_R$ of radius
$R$ centered at the origin and completely enclosing $A$. The
following argument is very similar to the application of Dynkin
formula to diffusions on an interval in Section
\ref{subsec:Dynkin}.

At time $\tau$ we have two options: either $\gamma(\tau)$ hits the
hull $A$ (in which case $\tau = \tau_A$) or it hits the arc $C_R$
(in which case $\tau_A > \tau$). Then
\begin{align}
{\mathbf E}[M_\tau^{(5/8)}] &= {\mathbf P}[\tau = \tau_A] \,
\Phi_{\tau_A}'(\xi_{\tau_A})^{(5/8)} + {\mathbf P}[\tau_A > \tau]
\, \Phi_{\tau}'(\xi_{\tau})^{(5/8)}. \nonumber
\end{align}
In the first case the point $\xi_{\tau_A}$ hits a ``side'' of the
hull $A_{\tau_A}$ (see Fig. \ref{fig:locality}) where the
derivative $\Phi_{\tau_A}'(\xi_{\tau_A})$ vanishes. Thus, only the
second option contributes to the expectation value ${\mathbf
E}[M_\tau^{(5/8)}]$. Then if we take the radius $R$ to be very
large, the point $\xi_{\tau}$ becomes very far from the hull
$A_\tau$, and the derivative $\Phi_\tau'(\xi_\tau)$ tends to 1.
Since the expectation value of a martingale does not depend on
time, by taking the limit $R \to \infty$ we obtain
\begin{align}
{\mathbf P}[\gamma \cap A = \varnothing] &= {\mathbf P}[\tau_A =
\infty] = \Phi_A'(0)^{5/8}. \label{restriction}
\end{align}
(To make this argument rigorous we would need to show that
$M_t^{(5/8)}$ is what is called a local martingale.)

The beautiful Eq. (\ref{restriction}) can now be used to show that
SLE$_{8/3}$ satisfies restriction property. To do this, let us
consider two different hulls $A$ and $B$ (as in Fig.
\ref{fig:composition}), both a finite distance from the origin,
and calculate the conditional probability that the image
$\Phi_A(\gamma)$ of an SLE trace does not intersect the hull $B$,
given that the original trace $\gamma$ does not intersect the hull
$A$. This is done using Eq. (\ref{conditional probability}) for
conditional probabilities:
\begin{align}
{\mathbf P}[\Phi_A(\gamma) \cap B = \varnothing \mathop{|} \gamma
\cap A = \varnothing] &= \frac{{\mathbf P}[\gamma \cap (A \cup
\Phi_A^{-1}(B))= \varnothing]}{{\mathbf P}[\gamma \cap A =
\varnothing]}. \nonumber
\end{align}
According to Eq. (\ref{restriction}) the denominator is equal to
$\Phi_A'(0)^{5/8}$. The hull $A \cup \Phi_A^{-1}(B)$ that appears
in the numerator in the last equation is removed by the
composition $\Phi_B \circ \Phi_A$ (see Fig.
\ref{fig:composition}), so the probability in the numerator is
equal to $(\Phi_B \circ \Phi_A)'(0)^{5/8} =
\Phi_B'(\Phi_A(0))^{5/8} \Phi_A'(0)^{5/8} = \Phi_B'(0)^{5/8}
\Phi_A'(0)^{5/8}$, where we used the product rule for the
derivative and the normalization $\Phi_A(0) = 0$. Combining all
this, we get
\begin{align}
{\mathbf P}[\Phi_A(\gamma) \cap B = \varnothing \mathop{|} \gamma
\cap A = \varnothing] &= \Big(\frac{\Phi_B'(0)
\Phi_A'(0)}{\Phi_A'(0)}\Big)^{5/8} = \Phi_B'(0)^{5/8}. \nonumber
\end{align}
Thus, the image under $\Phi_A$ of the subset of the SLE$_{8/3}$
traces that avoid the hull $A$ has the same distribution as
SLE$_{8/3}$, which, by definition implies restriction property for
SLE$_{8/3}$.

The notion of restriction can be applied to probability measures
on sets of random hulls in the upper half plane that are more
general than simple curves. These hulls $K$ must be connected,
unbounded, and such that ${\overline K} \cap {\mathbb R} = 0$ and
${\mathbb C} \setminus {\overline K}$ is connected. As was proven
in Ref. \cite{LSW-restriction}, there is a one-parameter family of
conformally-invariant measures on such hulls that have the
restriction property. For all these measures the probabilities of
avoiding a fixed hull $A$, as before, are given by
\begin{align}
{\mathbf P}[K \cap A = \varnothing] &=  \Phi_A'(0)^h,
\label{restriction-alpha}
\end{align}
where the restriction exponents $h \geqslant 5/8$ characterizes a
particular probability measure. Restriction measures with
exponents $h > 5/8$ are not realized on simple curves. An
important example is given by the so called Brownian excursions
(two-dimensional Brownian motions conditioned to always stay in
the upper half plane) with filled-in holes which satisfy
restriction property with $h = 1$.

For any value of $\kappa$ in the interval $[0,4]$ other than
$8/3$, SLE$_\kappa$ does not have restriction property, since no
value of $h$ makes $M_t^{(h)}$ a martingale, see Eq. (\ref{M t
alpha}), and the relation (\ref{restriction-alpha}) is not
satisfied. However, the amount by which SLE$_\kappa$ fails to
satisfy restriction property can be quantified. Namely, let $S f$
denote the Schwarzian derivative of the function $f$:
\begin{align}
S f(z) & = \frac{f'''(z)}{f'(z)} - \frac{3}{2} \Big(
\frac{f''(z)}{f'(z)} \Big)^2. \nonumber
\end{align}
Then
\begin{align}
\Phi_t'(\xi_t)^h \exp \Big( - \frac{c_\kappa}{6} \int_0^t S
\Phi_s(\xi_s) ds \Big) \nonumber
\end{align}
is a martingale (this is an easy consequence of It\^o formula and
Eq. (\ref{M t alpha})), if we choose
\begin{align}
h &= \frac{6 - \kappa}{2\kappa}, & c_\kappa &= \frac{(8 -
3\kappa)(\kappa - 6)}{2\kappa}. \label{alpha c kappa}
\end{align}
Notice the appearance of the central charge $c_\kappa$ of the CFT
corresponding to SLE$_\kappa$. This is quite natural, since
distortions of the boundary of the domain in which a CFT is defined
cause changes in its partition function,  that depend on the central
charge. The exponent $h$ in Eq. (\ref{alpha c kappa}) also has a
meaning in CFT: it is the dimension $h_{1,2}$ of a primary operator
that creates a critical curve when inserted on the boundary, see Eq.
(\ref{h12}) in Section \ref{sec:Coulomb gas}. Restriction property
has been used to relate SLE with CFT in Refs. \cite{FW-1, FW-2, FK,
Friedrich, Bauer-Bernard-partition-function}.

Finally, let me mention that restriction measures with exponents
$h > 5/8$ can be constructed by adding certain Brownian
``bubbles'' (subsets of 2D Brownian motions that are closed loops)
to an SLE curve with $\kappa \leqslant 4$, see Ref.
\cite{LSW-restriction}.

\section{Calculations with SLE}
\label{sec:SLE calculations}

In this section I will give detailed examples of calculations of
various probabilities and geometric characteristics of critical
curve using SLE.

\subsection{Left passage probability}
\label{subsec:Left passage}

This section is adapted from Ref. \cite{Schramm-percolation}.

Let us fix a point $z = x + i y \in {\mathbb H}$ in the upper half
of the physical plane. We may ask whether the trace $\gamma$
passes to the right or to the left of this point. Formally, this
is defined using the winding numbers, as follows. We can close the
curve $\gamma[0,t]$ by drawing the arc with the radius
$|\gamma(t)|$ from the tip of the trace to the point $|\gamma(t)|$
on the positive real axis, and then draw the straight segment in
$\mathbb R$ to 0. Then the trace $\gamma$ passes to the left of
$z$ if the winding number of the closed curve defined above is 1
for all large times $t$. Since $\gamma$ is transient a.s., there
is some random time $\tau$ such that the winding number is
constant for $t \in (\tau, \infty)$. This constant is either 0 or
1, since $\gamma$ does not cross itself. The random time $\tau_z$
for a given $z$ is either $\infty$ (if $\kappa \leqslant 4$), or
the swallowing time of $z$. Then the trace $\gamma$ satisfies
\begin{align}
&{\mathbf P}[\gamma \text{ passes to the left of } z] 
= \frac{1}{2} + \frac{\Gamma \bigl(\frac{4}{\kappa}
\bigr)}{\sqrt{\pi} \Gamma \bigl(\frac{4}{\kappa} - \frac{1}{2}
\bigr)} \frac{x}{y} \, {}_2 F_1 \Bigl(\frac{1}{2},
\frac{4}{\kappa}; \frac{3}{2}; - \frac{x^2}{y^2} \Bigr). \nonumber
\end{align}

The idea of the proof of this statement, as well as similar
statements about crossing probability, is to see what happens in
the mathematical plane, and relate events in the mathematical
plane with some real-valued random functions of the SLE process.
For this purpose we consider the image of $z$ under the shifted
function $w_t = w_t(z)$, and define:
\begin{align}
u_t &= \mathop{\rm Re} w_t, & v_t &=  \mathop{\rm Im} w_t, & q_t
&= \frac{u_t}{v_t}. \nonumber
\end{align}

Then it is almost obvious, that, $\gamma$ is to the left of $z$ if
and only if $\lim_{t \nearrow \tau_z} q_t = \infty$, and $\gamma$
is to the right of $z$ if and only if $\lim_{t\nearrow \tau_z} q_t
= -\infty$. A heuristic argument for $\kappa \leqslant 4$ is as
follows. In this case the point $z$ is not swallowed ($\tau_z =
\infty$). If the trace $\gamma$ passes to the left of $z$, then a
particle which starts an un-biased isotropic two-dimensional
diffusion from $z$ will hit ${\mathbb R} \cup \gamma(0,\infty)$
either in $[0,\infty)$ or from the right side of $\gamma$ with
probability one. By definition (see Section \ref{sec:harmonic
measure} for details), this probability is the harmonic measure of
$\gamma(0,\infty) \cup [0,\infty)$ from $z$. It is conformally
invariant, which means that the harmonic measure in the
mathematical UHP of $[\xi_t, \infty)$ from $g_t(z)$ tends to 1 as
$t \to \infty$. And this, in turn, means that $\lim_{t \to \infty}
q_t = \infty$.

In the case $4 < \kappa < 8$ the point $z$ is swallowed at some
time $\tau_z$ which is finite with probability 1. At this time the
curve $\gamma$ closes a loop around $z$. Then the issue is whether
the loop is clockwise or counter-clockwise. In the first case for
times $t$ close to $\tau_z$ the harmonic measure of $\gamma(0,t)
\cup {\mathbb R}$ is mostly concentrated on $[0,\infty)$ and the
right side of the curve $\gamma(0,t)$, which implies that $\lim_{t
\nearrow \tau_z} q_t = \infty$. For a counter-clockwise loop the
same reasoning gives $\lim_{t \nearrow \tau_z} q_t = -\infty$.

Next we will use the It\^o formula to write the stochastic
equation for $q_t$, and then the Dynkin formula to find the
necessary probability ${\mathbf P}[\lim_{t \to \infty} q_t =
\infty]$. So, first we have equations for $u_t$ and $v_t$ which
are just real and imaginary parts of Eq. (\ref{shiftedLE}):
\begin{align}
d u_t &= \frac{2u_t}{u_t^2 + v_t^2}dt - \sqrt{\kappa}dB_t, 
& d v_t &= - \frac{2v_t}{u_t^2 + v_t^2}dt. \label{Re f, Im f}
\end{align}
The two-dimensional variant of the It\^o formula now gives (no
explicit time dependence)
\begin{align}
d q_t &= \frac{1}{v_t}du_t - \frac{u_t}{v_t^2}dv_t 
= \frac{4q_t}{u_t^2 + v_t^2}dt - \frac{\sqrt{\kappa}}{v_t}dB_t.
\nonumber
\end{align}
This does not look like a standard It\^o equation, so we redefine
the time variable and the noise. First, the new time ${\tilde t}$
is introduced as
\begin{align}
{\tilde t}(t) &= \int_0^t \frac{dt}{v_t^2}, & d{\tilde t} &=
\frac{dt}{v_t^2}. \label{time change}
\end{align}
This is indeed a time change, since ${\tilde t}(t)$ is a
monotonously increasing function. Also, notice that as $t \to
\tau_z$, $v_t \to 0$ sufficiently fast so that ${\tilde t}(\tau_z)
= \infty$. In the new time the equation for $q_{\tilde t}$ becomes
\begin{align}
d q_{\tilde t} &= \frac{4q_{\tilde t}}{q_{\tilde t}^2 + 1}d{\tilde
t} - \frac{\sqrt{\kappa}}{v_t}dB_t. \nonumber
\end{align}

Next we set $d {\tilde B}_t = dB_t/v_t$ and note that $(d {\tilde
B}_t)^2 = (dB_t)^2/v_t^2 = dt/v_t^2 = d{\tilde t}$. This means
that $B_{\tilde t} = {\tilde B}_t$ is the standard Brownian motion
with respect to the new time variable ${\tilde t}$. The equation
for $q_{\tilde t}$ now has the standard It\^o form:
\begin{align}
d q_{\tilde t} &= \frac{4q_{\tilde t}}{q_{\tilde t}^2 + 1}d{\tilde
t} -\sqrt{\kappa}dB_{\tilde t}. \label{SDE q}
\end{align}

Next we find the diffusion generator for this equation:
\begin{align}
{\hat A}f(q) = \frac{\kappa}{2} \frac{d^2 f(q)}{d q^2} +
\frac{4q}{q^2 + 1} \frac{d f(q)}{dq}. \nonumber
\end{align}
If we study this diffusion on an interval $(a,b)$, where $a < q <
b$, then the probability $p_b$ that $q_{\tilde t}$ escapes this
interval through the right hand point rather then through the left
is given by the second of the formulas
(\ref{escape-probabilities}) where $f(q)$ should satisfy the
equation
\begin{align}
\label{eq. for f} \frac{\kappa}{2} f''(q) + \frac{4q}{q^2 + 1}
f'(q) = 0.
\end{align}
This equation has a constant solution, but this solution is not
what we need, obviously. The other solution is found by
straightforward separation of variables (and some specific choice
of the constants of integration):
\begin{align}
f(q) &= \int_0^q \frac{dr}{(r^2 + 1)^{4/\kappa}}. \label{f(q)}
\end{align}
We expand the integrand in powers of $r$ and integrate term by
term:
\begin{align}
f(q) &= \int_0^q \!\! dr \,\, \sum_{m=0}^\infty (4/\kappa)_m
\frac{(-r^2)^m}{m!} = q \sum_{m=0}^\infty
\frac{(4/\kappa)_m}{2m+1} \frac{(-q^2)^m}{m!}. \nonumber
\end{align}
Here $(a)_m$ denotes the Pochhammer symbol $(a)_m = \Gamma(a +
m)/\Gamma(a)$. Using this notation we can write
\begin{align}
\frac{1}{m + a} &= \frac{\Gamma(m+a)}{\Gamma(m+1+a)} =
\frac{\Gamma(a)}{\Gamma(1+a)} \frac{(a)_m}{(1+a)_m} = \frac{1}{a}
\frac{(a)_m}{(1+a)_m}, \label{fraction-Pochhammer}
\end{align}
and, in particular, $1/(2m+1) = (1/2)_m/(3/2)_m$. This gives
\begin{align}
f(q) &= q \sum_{m=0}^\infty \frac{(1/2)_m (4/\kappa)_m}{(3/2)_m}
\frac{(-q^2)^m}{m!} = q \, {}_2F_1\Bigl(\frac{1}{2},
\frac{4}{\kappa}; \frac{3}{2}; -q^2 \Bigr). \nonumber
\end{align}

Using asymptotics of the hypergeometric function we see that the
solution $f(q)$ has finite limits
\begin{align}
\lim_{q \to \pm \infty} f(q) &= \pm \frac{\sqrt{\pi}}{2}
\frac{\Gamma \bigl(\frac{4}{\kappa} - \frac{1}{2} \bigr)}{\Gamma
\bigl(\frac{4}{\kappa} \bigr)}. \nonumber
\end{align}
This shows that the considered diffusion of $q_{\tilde t}$ is
transient, meaning that with a finite probability $\lim_{{\tilde
t} \to \infty} q_{\tilde t} = \infty$. Thus, we can finally take
limits $a \to -\infty, b \to \infty$ and get the result
\begin{align}
& {\mathbf P}[\gamma \text{ passes to the left of } z] = {\mathbf
P}[\lim_{t
\nearrow \tau_z} q_t = \infty] \nonumber \\
& \quad = \frac{f(x/y) - f(-\infty)}{f(\infty) - f(-\infty)} =
\frac{1}{2} + \frac{\Gamma \bigl(\frac{4}{\kappa}
\bigr)}{\sqrt{\pi} \Gamma \bigl(\frac{4}{\kappa} - \frac{1}{2}
\bigr)} \frac{x}{y} \, {}_2 F_1 \Bigl(\frac{1}{2},
\frac{4}{\kappa}; \frac{3}{2}; - \frac{x^2}{y^2} \Bigr). \nonumber
\end{align}
When $\kappa = 2, 8/3, 4$ and $8$, the last formula simplifies to
\begin{align}
1 + \frac{x y}{\pi |z|^2} - \frac{\arg z}{\pi}, && \frac{1}{2} +
\frac{x}{2 |z|}, && 1 - \frac{\arg z}{\pi}, && \frac{1}{2},
\nonumber
\end{align}
respectively.

The value $1/2$ obtained for $\kappa = 8$ is somewhat misleading.
The point is that if $\kappa \geqslant 8$, then the curve $\gamma$
densely fills the upper half plane, as was mentioned in Section
\ref{subsec:SLE phases}, and goes {\it through} every point, not
to the left or right of it. This is reflected in the fact that for
$\kappa \geqslant 8$ the function $f(q)$ in Eq. (\ref{f(q)})
diverges as $q \to \pm \infty$. This divergence means that to
determine the fate of the process $q_{\tilde t}$ as ${\tilde t}
\to \infty$, we need to start with a finite interval $(a,b)$ ($a <
x/y < b$) and take the limits $a \to -\infty$ and $b \to \infty$
separately. In both cases we find that ${\mathbf P}[\lim_{t
\nearrow \tau_z} q_t = \pm \infty] = 0$, meaning that $q_{\tilde
t}$ always stays bounded. See a related discussion in Section
\ref{subsec:dimensions}.

\subsection{Cardy's formula for crossing probability}

The problem is first posed in a rectangle ABCD. We need to find
the probability that there is a percolation cluster connecting the
left side AB and the right side CD of the rectangle, where we
impose the fixed boundary condition ($p=1$). Note that from the
point of view suggested by SLE, we need to consider not the
cluster itself, but one of its ``boundaries'', upper or lower. Let
us consider the lower boundary, which in the continuous limit is
described by SLE$_6$. Then we see that if there is a spanning
cluster, then the boundary will necessarily start at the point B,
and will reach the side CD without touching the upper side AD. In
the opposite case, when there is no spanning cluster, the boundary
will touch AD before touching CD.

In fact, this reformulation of the problem can be generalized to
any $\kappa > 4$, and we will assume this has been done.

Next we conformally map the rectangle to the upper half plane
using the Schwarz-Christoffel formula. The direct mapping
$\Phi(z)$ (from rectangle to the UHP) is given by an elliptic
function, and the inverse mapping---by an elliptic integral.
Postponing the details until the end of this section, let us
assume for now that the images of the vertices of the rectangle
are
\begin{align}
\Phi(A) &= a < 0, & \Phi(B) &= 0, & \Phi(C) &= c > 0, & \Phi(D) &=
\infty. \label{vertices-images}
\end{align}

Since the crossing probability is conformally invariant (as a
property of SLE), we are now interested in the following question.
Since $\kappa > 4$, both points $a$ and $c$ will be swallowed at
some finite random times $\tau_a$ and $\tau_c$. The crossing
probability then is ${\mathbf P}[\tau_c < \tau_a]$, that is, the
probability that the point $c$ is swallowed before the point $a$.

As should be obvious by now, we need to study the motion of the
images of the points $a,b$ under the Loewner map. In this case it
is easier to use the original map (before the shift), so we define
\begin{align}
a_t &= g_t(a), & c_t &=g_t(c), & r_t &= \frac{\xi_t - a_t}{c_t -
a_t}. \nonumber
\end{align}
The variable $r_t$ is normalized to lie between 0 and 1, and we
are essentially interested in the probability ${\mathbf P}[c_\tau
= \xi(\tau)] = {\mathbf P}[r_\tau = 1]$, where $\tau$ is the
escape time from $[0,1]$ for $r_t$.

The calculations are straightforward:
\begin{align}
d(\xi - a_t) &= d\xi - \frac{2}{a_t - \xi} dt, & d(c_t - a_t) &=
\Big(\frac{2}{c_t - \xi} - \frac{2}{a_t - \xi} \Big) dt. \nonumber
\end{align}
Then
\begin{align}
d r_t &= \frac{d(\xi - a_t)}{c_t - a_t} - \frac{\xi - a_t}{(c_t -
a_t)^2} d(c_t - a_t) \nonumber \\
&= \Big(\frac{1}{r_t} - \frac{1}{1 - r_t}\Big) \frac{2 dt}{(c_t -
a_t)^2} + \frac{\sqrt{\kappa}}{c_t - a_t} dB_t. \nonumber
\end{align}
Again, this SDE is not of the It\^o type, and we perform a time
change:
\begin{align}
d{\tilde t} &= dt/(c_t - a_t)^2, & d{\tilde B}_t &= dB_t/(c_t -
a_t). \nonumber
\end{align}
Then the process ${\tilde r}_t = r_{\tilde t}$ satisfies the It\^o
equation
\begin{align}
d r_{\tilde t} &= 2 \Big(\frac{1}{r_{\tilde t}} - \frac{1}{1 -
r_{\tilde t}}\Big) d{\tilde t} + \sqrt{\kappa} dB_{\tilde t}.
\nonumber
\end{align}

The generator of diffusion for this process is
\begin{align}
{\hat A} = \frac{\kappa}{2} \frac{d^2}{dr^2} + 2 \Big(\frac{1}{r}
- \frac{1}{1 - r}\Big) \frac{d}{dr}, \nonumber
\end{align}
and its zero mode $f(r)$ is found by simple integrations as
before:
\begin{align}
f(r) &= \int_0^r \frac{ds}{\big(s(1-s)\big)^{4/\kappa}}. \nonumber
\end{align}
Since $\kappa > 4$, the last integral converges both at the lower
and the upper limits, when $r \to 1$. As in the previous section,
using Eq. (\ref{fraction-Pochhammer}) this integral can be
expressed in terms of the Gauss hypergeometric function:
\begin{align}
f(r) &= \int_0^r \!\! ds \,\, s^{-4/\kappa} \sum_{m=0}^\infty
(4/\kappa)_m \frac{s^m}{m!} = r^{1-4/\kappa} \sum_{m=0}^\infty
\frac{(4/\kappa)_m}{m+1-\frac{4}{\kappa}}
\frac{r^m}{m!} \nonumber \\
& = \frac{1}{1 - \frac{4}{\kappa}} r^{1-4/\kappa}
\sum_{m=0}^\infty \frac{(4/\kappa)_m (1 - 4/\kappa)_m}{(2 -
4/\kappa)_m} \frac{r^m}{m!} \nonumber \\ & = \frac{1}{1 -
\frac{4}{\kappa}} r^{1-4/\kappa}\, {}_2F_1\Big(\frac{4}{\kappa}, 1
- \frac{4}{\kappa}; 2 - \frac{4}{\kappa}; r \Big). \nonumber
\end{align}
At the ends of the interval for diffusion of $r_t$ this function
takes the values $f(0) = 0$ and $f(1) = \Gamma^2(1 -
4/\kappa)/\Gamma(2 - 8/\kappa)$.
Substituting this into Eq. (\ref{escape-probabilities}) with $a=0,
b =1$, we get the final result
\begin{align}
{\mathbf P}[\text{crossing}]
&= \frac{\Gamma\big(2 - \frac{8}{\kappa} \big)}{\Gamma\big(2 -
\frac{4}{\kappa}\big) \Gamma\big(1 - \frac{4}{\kappa}\big)} r^{1 -
4/\kappa} {}_2 F_1\Big(\frac{4}{\kappa}, 1 - \frac{4}{\kappa}; 2 -
\frac{4}{\kappa}; r \Big). \label{Cardy's formula}
\end{align}
As usual, here $r$ means the initial value of the process $r_t$,
that is, $r = -a/(c-a)$. For $\kappa = 6$ this reduces to Cardy's
formula for crossing probability for percolation \cite{Cardy3}.

Now we can discuss how to map a given rectangle to the UHP.
Suppose the horizonal and vertical sides of the rectangle have
lengths $L$ and $L'$. It is obvious that the crossing probability
is invariant under rescaling. Then we need to find the (unique)
number $0 < k < 1$ (the so called elliptic modulus) from the
equation
\begin{align}
\frac{L'}{L} &= \frac{K'(k)}{2K(k)}, \nonumber
\end{align}
where $K(k)$ is the complete elliptic integral of the first kind,
and $K'(k) = K\big(\sqrt{1-k^2}\big)$ (in the following we
simplify these to $K, K'$). Next we rescale the rectangle and
place its vertices as follows:
\begin{align}
A &= -K + i K', & B &= -K, & C &= K, & D &= K + i K'. \nonumber
\end{align}
It is easy to see then that the function
\begin{align}
\Phi(z) &= k \frac{1 + \mathop{\rm sn}(z,k)}{1 - k \mathop{\rm
sn}(z,k)} \nonumber
\end{align}
maps the interior of our rectangle to the UHP, and its vertices to
\begin{align}
\Phi(A) &= -\frac{1-k}{2}, & \Phi(B) &= 0, & \Phi(C) &=
\frac{2k}{1-k}, & \Phi(D) &= \infty. \nonumber
\end{align}
Comparing this with Eq. (\ref{vertices-images}) we obtain $r =
\Big( \dfrac{1-k}{1+k}\Big)^2$, which we need to substitute to Eq.
(\ref{Cardy's formula}) to get the crossing probability for the
rectangle.

\subsection{Fractal dimensions of SLE curves}
\label{subsec:dimensions}

SLE curves are fractal objects. Their fractal dimension can be
estimated by the box counting dimension. Namely, we can ask how
the number of small disks of radius $\epsilon$ required to cover
an SLE$_\kappa$ curve scales with $\epsilon$:
\begin{align}
N_\epsilon \sim \epsilon^{-d_f(\kappa)}, \nonumber
\end{align}
where $d_f(\kappa)$ is the box counting fractal dimension.
Strictly speaking, this definition is applicable only for finite
curves, but it can be applied for any segment of a chordal SLE
curve, since all the segments should be statistically similar.

The fractal dimension $d_f(\kappa)$ is related to multifractal
exponents of the harmonic measure, and can be obtained from them,
as explained in Section \ref{sec:harmonic measure}. In this
section we use a probabilistic approach.

The dimension $d_f(\kappa)$ can be estimated in the spirit of
Monte-Carlo methods by throwing disks of radius $\epsilon$ randomly
onto the domain containing the critical curve, and then counting the
fraction of the disks which intersect the curve. Alternatively, we
can look for the probability that an SLE curve intersects a {\it
given} disk. It is clear that this probability should scale as
$\epsilon^{2-d_f(\kappa)}$ (2 here is the dimensionality of the
physical plane), and it is this scaling that can be relatively
easily obtained from SLE, with the result (rigorously established in
Refs. \cite{Beffara-1, Beffara-2}, see also an earlier discussion in
Ref. \cite{Rohde-Schramm})
\begin{align}
d_f(\kappa) &= \min \Big(1 + \frac{\kappa}{8}, 2 \Big).
\label{d_f}
\end{align}

To derive this scaling we need to introduce some notation and
properties of conformal maps. First, let $D$ be a domain in the
complex plain, $\partial D$ its boundary, and $z$ a point inside
$D$. Denote by $\mathop{\rm dist}(z, \partial D)$ the Euclidean
distance between $z$ and the domain boundary.

If the domain $D$ is mapped conformally to a domain ${\tilde D}$
by a function ${\tilde z} = f(z)$, then the distance between close
points $z$ and $z + dz$ gets multiplied by a rescaling factor:
$|d{\tilde z}| = |f'(z)| |dz|$. The same is roughly speaking true
for finite distances. More precisely, if $d = \mathop{\rm dist}(z,
\partial D)$ and ${\tilde d} = \mathop{\rm dist}({\tilde z},
\partial {\tilde D})$, then a corollary to the famous Koebe 1/4
theorem states that
\begin{align}
\frac{{\tilde d}}{4d} &\leqslant |f'(z)| \leqslant \frac{4{\tilde
d}}{d} & {\rm or} && \frac{{\tilde d}}{4|f'(z)|}  \leqslant d
\leqslant \frac{4{\tilde d}}{|f'(z)|}. \label{Koebe}
\end{align}
Let us denote these bounds by $d \asymp {\tilde d}/|f'(z)|$ and
say that both quantities are comparable.

Now we apply this to the Loewner map $w_t(z)$ to estimate the
limit of the distance $d_t(z) = \mathop{\rm dist}(z, \gamma(0,t)
\cup {\mathbb R})$ between a point $z$ and an SLE curve in the
physical plane, as the time goes up to the swallowing time
$\tau_z$ (which may be intinite). We use the same notation as in
section \ref{subsec:Left passage}, and write $w_t(z) = w_t = u_t +
i v_t$ for the image of the point $z$. In the mathematical plane
the distance from the image to the boundary is simply $\mathop{\rm
Im} w_t(z) = v_t$. If we introduce the process
\begin{align}
D_t(z) &= \log \frac{|w_t'(z)|}{\mathop{\rm Im} w_t(z)}, \nonumber
\end{align}
Eq. (\ref{Koebe}) gives $d_t(z) \asymp e^{-D_t(z)}$. Let us find
the SDE for $D_t(z)$. First, the $z$-derivative of the basic SLE
equation (\ref{shiftedLE}) gives
\begin{align}
\partial_t \log w_t'(z) &= - \frac{2}{w_t^2(z)}. \nonumber
\end{align}
The real part of this equation is
\begin{align}
\partial_t \log |w_t'(z)| &= - \frac{2 \mathop{\rm Re}
[w_t^2(z)]^*}{|w_t(z)|^4} = 2 \frac{v_t^2 - u_t^2}{(v_t^2 +
u_t^2)^2}. \label{log-derivative-bulk}
\end{align}
Combining this with the Eq. (\ref{Re f, Im f}) for $v_t$, we get
\begin{align}
\partial_t D_t(z) = \frac{4v_t^2}{(v_t^2 + u_t^2)^2} \geqslant 0. \nonumber
\end{align}
Thus, $D_t(z)$ increases with $t$, and to estimate $d(z) =
\mathop{\rm dist}(z, \gamma(0,\infty) \cup {\mathbb R}) \asymp
e^{-D(z)}$ we need to look at
\begin{align}
D(z) &= \lim_{t \nearrow \tau_z} D_t(z) = \int_0^{\tau_z}
\frac{4v_t^2}{(v_t^2 + u_t^2)^2} dt. \nonumber
\end{align}
As in Section \ref{subsec:Left passage} we change time according
to (\ref{time change}) and get
\begin{align}
D(x/y) &= 4 \int_0^\infty \frac{d{\tilde t}}{(q_{\tilde t}^2 +
1)^2}, \label{D(z)}
\end{align}
where the process $q_{\tilde t} = u_{\tilde t}/v_{\tilde t}$
satisfies the SDE (\ref{SDE q}) and has the initial value $x/y$.
As we discussed in the end of Section \ref{subsec:Left passage},
if $\kappa \geqslant 8$ the process $q_{\tilde t}$ stays bounded
as ${\tilde t} \to \infty$. Then the integral in Eq. (\ref{D(z)})
diverges, and $D(x/y) = \infty$. This immediately gives that $d(z)
= 0$ and
\begin{align}
d_f(\kappa \geqslant 8) = 2, \label{d_f-kappa>8}
\end{align}
consistent with the curve $\gamma$ densely filling the upper half
plane.

Now consider the case $0 \leqslant \kappa < 8$. Since $d(z) \asymp
e^{-D(x/y)}$, the probability ${\mathbf P}[\Delta(z) \leqslant
\epsilon]$ that the SLE curve intersects the disc of radius
$\epsilon$ centered at the point $z$ is comparable to (scales in
the same way with $\epsilon$ as) the probability ${\mathbf
P}[D(x/y) \geqslant - \log \epsilon]$. The latter probability can
be estimated if we find the asymptotics of the probability
distribution function $p(D,x/y)$ for $D(x/y)$.

We expect that the scaling of ${\mathbf P}[d(z) \leqslant
\epsilon]$ with $\epsilon$ should not depend on the actual
position of $z$. In fact, the SLE scaling property (\ref{SLE
scaling}) implies that $d(x + iy)$ has the same distribution as $y
d\big(\frac{x}{y} +i\big)$, and thus we are free to choose the
point $z$ anywhere. To simplify the formulas below, we will now
take the point $z$ to be $x + i$. Then the process $q_{\tilde t}$
starts at $q_0 = x$ and is transient, that is, goes to $\infty$ or
$-\infty$. In both cases the integral (\ref{D(z)}) is convergent
and non-negative, and we can use the stationary FK formulas
(\ref{stationary FK formula}, \ref{stationary FK formula 1}) from
Section \ref{subsec:FK} to find $p(D,x)$ though its Laplace
transform $L(s,x)$. Namely, $L(s,x)$ should satisfy
\begin{align}
\frac{\kappa}{2} \frac{d^2 L}{dx^2} + \frac{4x}{x^2+1} \frac{d
L}{dx} - \frac{4s}{(x^2+1)^2} L &= 0. \nonumber
\end{align}
The change of variables $y = x^2/(x^2 + 1)$ leads to the
hypergeometric equation
\begin{align}
y(1-y) \frac{d^2 L}{dy^2} + \Big[\frac{1}{2} +
\Big(\frac{4}{\kappa} - 2\Big) y \Big] \frac{d L}{dy} -
\frac{2s}{\kappa} L &= 0. \nonumber
\end{align}
The solution of this equation normalized as $L(s,x=\infty) = 1$ is
\begin{align}
L(s,x) &= \frac{\Gamma\big(\frac{1}{2} - a_+ \big)
\Gamma\big(\frac{1}{2} - a_- \big)}{\Gamma\big(\frac{1}{2}\big)
\Gamma\big(\frac{4}{\kappa} - \frac{1}{2} \big)} {}_2 F_1 \Big(
a_+, a_-; \frac{1}{2}; \frac{x^2}{x^2 + 1} \Big), \label{L} \\
a_\pm(s) &= \frac{1}{2} - \frac{2}{\kappa} \pm
\sqrt{\Big(\frac{1}{2} - \frac{2}{\kappa} \Big)^2 -
\frac{2s}{\kappa}}. \nonumber
\end{align}
The inverse Laplace transform  give the probability density for
$D$:
\begin{align}
p(D,x) &= \frac{1}{2\pi i} \int_{c-i\infty}^{c+i\infty} e^{sD(x)}
L(s,x) ds. \nonumber
\end{align}
The integration contour should lie to the right of all the
singularities of $L(s,x)$ in the $s$-plane. If we deform the
contour by moving it to the left, it will encircle the poles of
$L(s,x)$, and for large $D(x)$ the leading behavior of $p(D,x)$
will be determined by the pole with the largest real part.

Let us now find the singularities of $L(s,x)$ given by Eq.
(\ref{L}). Since the hypergeometric function ${}_2 F_1 (a,b;c;x)$
is an entire function of its parameters $a, b, c$, the only
singularities of $L(s,x)$ are in the prefactor in Eq. (\ref{L}).
Gamma functions have poles when their arguments are non-positive
integers: $1/2 - a_\pm(s) = -n$, $n \geqslant 0$, which gives the
poles at real positions
\begin{align}
s_n &= -1 + \frac{\kappa}{8} - 2n - \frac{\kappa}{2}n^2. \nonumber
\end{align}
The largest pole is at $s_0 = - 1 + \kappa/8$, which gives for
large $D$
\begin{align}
p(D,x) \propto e^{-(1 - \kappa/8)D}. \nonumber
\end{align}
Finally, we have the estimate
\begin{align}
{\mathbf P}[d(x+i) \leqslant \epsilon] \asymp {\mathbf P}[D(x)
\geqslant -\log \epsilon] = \int_{- \log \epsilon}^\infty p(D,x)
dD \propto \epsilon^{1 - \kappa/8}, \nonumber
\end{align}
which gives
\begin{align}
d_f(\kappa < 8) = 1 + \frac{\kappa}{8}. \label{d_f-kappa<8}
\end{align}
Together with (\ref{d_f-kappa>8}) this establishes Eq.
(\ref{d_f}).

\subsection{Derivative expectation}
\label{subsec:derivative}

The absolute value of the derivative of the SLE map $|w_t'(z)|$
and its moments are useful quantities. As for any conformal map
$|w_t'(z)|$ is the local measure of rescaling introduced by the
map. For a critical curve described by SLE the moments ${\mathbf
E}[|w_t'(z)|^h]$ where $h \in \mathbb R$, are also related to the
spectrum of multifractal exponents of the harmonic measure, as
explained in Section \ref{sec:harmonic measure}. In this context
the derivative should be estimated at a certain distance from an
SLE curve. Alternatively, of all the SLE curves we should choose
only the ones that pass closer than a certain small distance
$\epsilon$ from a given point $z$. This is an example of
conditioning introduced in Section \ref{subsec:conditioning}.

Finding the conditional expectation value ${\mathbf
E}[|w_t'(z)|^h; d(z) \leqslant \epsilon]$ at point $z$ in the bulk
is a difficult problem that has not been solved so far (see,
however, how similar quantities are calculated in Refs.
\cite{Lawler-book, Rohde-Schramm}). However, each SLE curve starts
at the origin in the physical plane, and no conditioning is
required to estimate the derivative of the SLE map at real point
$x$ on the boundary.

Indeed, in this case Eq. (\ref{log-derivative-bulk}) simplifies
(since $v_t = 0$) and gives in the notation of Section
\ref{subsubsec:kappa=4}
\begin{align}
\partial_t \log |w_t'(x)| &= - \frac{2}{x_t^2}, & |w_t'(x)|^h &=
\exp \Big(-2h\int_0^t \frac{ds}{x_s^2} \Big).
\label{log-derivative-boundary}
\end{align}
According to the Feynman-Kac formula (\ref{FK-formula}) the
expectation value $c(x,t) = {\mathbf E}[|w_t'(x)|^h]$ satisfies
\begin{align}
\partial_t c(x,t) &= \frac{\kappa}{2}\partial_x^2 c(x,t) +
\frac{2}{x}\partial_x c(x,t) - \frac{2h}{x^2} c(x,t), & c(x,0) & =
1. \nonumber
\end{align}

From the SLE scaling law (\ref{scaling}) we know that $x$ and $t$
must appear in the combination $x/{\sqrt t}$. With some hindsight
we denote $y = x^2/2\kappa t$ and $c(x,t) = f(y)$. Then the
equation and the boundary value for $f(y)$ become
\begin{align}
y^2 f''(y) + y \Big(\frac{2}{\kappa} + \frac{1}{2} + y \Big) f'(y)
- \frac{h}{\kappa}f(y) &= 0, & \lim_{y \to \infty} f(y) = 1.
\nonumber
\end{align}
In the limit $y\to 0$ (long times) we can neglect $y$ in the
brackets in front of $f'$, and the equation simplifies to an Euler
equation with the solution $y^{\Delta(h)/2}$, where
\begin{align}
\Delta(h) &= \frac{\kappa - 4 + \sqrt{(\kappa - 4)^2 + 16 \kappa
h}}{2\kappa} \label{Delta(h)}
\end{align}
is a solution of the indicial equation that vanishes as $h \to 0$.

In the context of the problem of multifractal exponents of
harmonic measure the scaling
\begin{align}
{\mathbf E}[|w_t'(x)|^h] \sim \Big(\frac{|x|}{\sqrt{2\kappa t}}
\Big)^{\Delta(h)} \nonumber
\end{align}
is all we need. But we can also solve the problem completely.
Namely, if we write $f(y) = e^{-y} y^{\Delta(h)/2} \psi(y)$, then
$\psi(y)$ satisfies
\begin{align}
& y \psi''(y) + (c - y) \psi'(y) - a \psi(y) = 0, \qquad \psi(y
\to \infty) \to e^y y^{-\Delta(h)/2}, \nonumber \\
& a = \frac{2}{\kappa} + \frac{1}{2} + \frac{\Delta(h)}{2}, \qquad
c = \frac{2}{\kappa} + \frac{1}{2} + \Delta(h), \nonumber
\end{align}
which is the standard form of the differential equation for the
confluent hypergeometric function. The solution with the required
asymptotic behavior is $[\Gamma(a)/\Gamma(c)] \Phi(a,c;y)$, and we
finally get
\begin{align}
{\mathbf E}[|w_t'(x)|^h] &= \frac{\Gamma\big( a \big)}{\Gamma\big(
c \big)} \Big(\frac{|x|}{\sqrt{2\kappa t}} \Big)^{\Delta(h)}
e^{-x^2/2\kappa t} \Phi\Big( a, c; \frac{x^2}{2\kappa t} \Big).
\nonumber
\end{align}

\section{Critical curves and bosonic fields (Coulomb gas)}
\label{sec:Coulomb gas}

In the rest of this review I will provide a connection between SLE
and a more traditional approach to critical 2D systems, namely,
conformal field theory (CFT). In this Section we will se how
critical curves can be described within a CFT of a scalar field.
Closely related discussions have appeared before in Refs.
\cite{Cardy-SLEkr, Bauer-Friedrich-toolbox, MRR-boundary-Coulomb}.

\subsection{From loop models to bosonic fields}
\label{subsec:loop models}

The relation between critical curves and operators of a boundary CFT
is most transparent in their representation by a Gaussian boson
field $\varphi(z,\bar z)$ \cite{cft,nienhuis,Schulze,kawai}. This
representation is commonly known as the Coulomb gas method.
Specifically, let us consider the O$(n)$ model on a honeycomb
lattice. In the hope of describing the critical point by a local
field theory, we need to have a description in terms of local
weights on the lattice.

To reproduce the partition function (\ref{O(n) partition}) we
randomly assign orientations to loops and then sum over all
possible arrangements. The sum of weights for two orientations of
every loop should give $n$. This is achieved by giving the local
weight $e^{\pm i e_0 \pi/6}$ to each lattice site where an
oriented loop makes right (left) turn. The weight of an oriented
closed loop is the product of all local site weights, and is equal
to $e^{\pm i e_0 \pi}$ since for a closed loop the difference
between the numbers of right and left turns is $\pm 6$. The sum
over the orientations reproduces the correct weight $n$ for an
un-oriented loop if we choose
\begin{align}
n = 2 \cos \pi e_0. \nonumber
\end{align}
The range of $-2 \leqslant n \leqslant 2$ where the loops of
O$(n)$ are critical can be covered once by $e_0 \in [0, 1]$.
However, as we will see, to describe both the dilute and the dense
phases we need to allow for a wider range $e_0 \in [-1, 1]$, with
positive $e_0$ for the dense phase and negative $e_0$ for the
dilute phase.

For each configuration of oriented loops we can define a real
height variable $H$ that resides on the dual lattice and takes
discrete values conventionally chosen to be multiples of $\pi$. To
define $H$ we start at some reference point where we set $H=0$,
and then every time we cross an oriented loop, we change $H$ by
$\pm \pi$ depending on whether we cross the loop from its left to
its right side or vice versa. Since the orientation of the loops
was introduced randomly, the height function has to be
compactified with radius ${\mathcal R} = 1$:
\begin{align}
\label{compac} H \simeq H + 2\pi,
\end{align}
which means that the heights $H$ and $H + 2\pi$ correspond to the
same configuration of un-oriented loops.

At criticality, the coarse-grained height function becomes a
continuous scalar field (boson), believed to be described by the
Gaussian action $(g/4\pi)\int_Dd^2x\, (\nabla H)^2$, where the
fluctuation strength parameter $g$ is not yet determined. This can
be done either by comparison with exact solutions of a related
six-vertex model, or by an elegant argument due to Kondev and Henley
\cite{kondev, Kondev-Henley} (which, unfortunately, only works in
the dense phase). Here's the argument.

If the system is defined on a domain with boundaries, some loops
may not be counted with the correct statistical weight. For
example, the difference between the numbers of left and right
turns for a loop that wraps around a cylinder is 0 rather than 6.
Therefore, without modifications all such loops will be counted
with a wrong weight 2 in the partition function . This is fixed by
adding to the action a boundary term $(ie_0/2\pi)\int_{\partial
D}dl\, K H $, where $K$ is the geodesic curvature of the boundary.
Each loop wrapped around the cylinder introduces an additional
height difference $\Delta H = \pm\pi$ between the ends thus
acquiring the correct weight.

A similar situation occurs if the critical system lives on a
surface with curvature, which microscopically can be viewed as
existence of defects on the honeycomb lattice (pentagons and
heptagons correspond to positive and negative curvature,
correspondingly). The correct weight for a loop that surrounds a
region of non-zero curvature is obtained only if we include in the
action the so-called background charge term
$(ie_0/8\pi)\int_Dd^2x\, R H$, where $R$ is the scalar curvature.

Yet another necessary term in the action is the {\it locking
potential}  of the form $\lambda \int d^2x\,V(H)$ which would
force the discrete values $H = k\pi$ in the limit $\lambda \to
\infty$. It must be, therefore, a $\pi$-periodic function of $H$,
the most general form of it being $V=\sum_{k \in {\mathbb Z}, k
\neq 0} v_k e^{2ikH}$. Each term of $V$ is a vertex operator whose
dimension is \cite{nienhuis}
\begin{align}
x_k=\frac{2}{g}k(k - e_0). \nonumber
\end{align}
Most of these terms are irrelevant at the Gaussian fixed point,
and we can ignore them. The most relevant term has $k=1$ if $0 <
e_0 < 1$, and it has to be strictly marginal ($x_k = 2$) in order
to retain the conformal invariance of the action. This gives the
relations
\begin{align}
e_0 &= 1 - g, & n = -2\cos \pi g. \label{n-g-relation}
\end{align}
In this case $0 < g < 1$, which is known to describe the dense
phase of the O$(n)$ model. In the dilute phase the second relation
(\ref{n-g-relation}) still holds \cite{nienhuis}, but with $1
\leqslant g \leqslant 2$. This range is not possible to obtain
from the previous argument since for $-1 < e_0 < 0$ we would need
to pick $k = -1$ term as the most relevant, and it would still
give us $g = 1 + e_0 = 1 - |e_0| < 1$. With some amount of
hindsight we will assume {\it both} relations (\ref{n-g-relation})
to be valid for the whole range $g \in (0,2]$ encompassing both
the dense and the dilute phase. The point $g=1$ separating the
phases is somewhat special: there we need to keep both $k=1$ and
$k= -1$ terms in the locking potential since they have the same
dimension.

The failure of Kondev's argument in the dilute phase has a very
significant geometric meaning. Namely, upon the coarse-graining
the O$(n)$ loops become level lines of the bosonic field. However
this identification can only be made for the dense phase, where
the loops come close to themselves and each other on the lattice,
translating to them becoming non-simple curves (with double
points) in the continuum limit, resembling the traces of SLE with
$\kappa > 4$. The relation between critical lines and the bosonic
field is quite different in the dilute phase, and this difference
is related to quite a few subtleties in the treatment of both the
dilute and the dense phases of a {\it bounded} system in the
Coulomb gas formalism. For details see our paper
\cite{RBGW-long-paper}.

We now introduce the parametrization
\begin{align}
g &= \frac{4}{\kappa}, & 2 \leqslant \kappa < \infty,
\label{g-kappa}
\end{align}
where $\kappa$ can be identified with the SLE parameter by
comparing calculations of some quantity within the two approaches.
A typical example is the distribution of winding angles of
critical curves on a cylinder, which is known through both the
Coulomb gas method and SLE. Another good example is the
multifractal exponents related to derivative expectations, which
we compute in the Coulomb gas formalism in Section
\ref{sec:harmonic measure}. Notice that $\kappa < 4$ and $\kappa
> 4$ describe the dilute and the dense phases, correspondingly,
while $\kappa = 4$ gives the point $g = 1$ separating the two
phases. All this is quite consistent with the SLE phases
determined in Section \ref{subsec:SLE phases}.

In the CFT literature it is customary to rescale the field
$\varphi = \sqrt{2g} H$ and make the coupling constant fixed
$g_{\text{new}} = 1/2$, at the expense of varying the
compactification radius of $\varphi$:
\begin{align}
{\mathcal R} = \sqrt{8/\kappa}. \label{comp radius}
\end{align}
This is the normalization that we adopt from now on. For the
rescaled field the action with all the terms becomes
\begin{align}
S[\varphi] &= \frac{1}{8\pi} \int_D d^2 x \,\big[(\nabla\varphi)^2
+ i 2\sqrt{2}\alpha_0 R \varphi \big] + i
\frac{\sqrt2\alpha_0}{2\pi} \int_{\partial D} dl \, K \varphi
\nonumber \\
& \quad + \int_D d^2 x \, e^{i\sqrt{2} \alpha_ + \varphi},
\label{bigaction}
\end{align}
where we use the notation
\begin{align}
2\alpha_0 &= \frac{\sqrt\kappa}{2} - \frac{2}{\sqrt\kappa}, &
\alpha_\pm &= \alpha_0 \pm \sqrt{\alpha_0^2 + 1}, \nonumber \\
\alpha_+ &= \sqrt\kappa/2, & \alpha_- &= - 2/\sqrt\kappa.
\label{alphas}
\end{align}
Notice that $\alpha_0$ is proportional to $e_0$, and can be both
positive and negative, its sign being different in the two phases
of the loop model.

\subsection{Coulomb gas CFT in the bulk}

Consider now our bosonic theory on the infinite plane (the Riemann
sphere), dropping for now the boundary term in Eq.
(\ref{bigaction}). The action $S[\varphi]$ does not describe a
free field beacuse of the presence of the locking potential. In
practice, however, this potential is always treated
perturbatively, and any correlation function is expanded as
\begin{align}
\langle X \rangle_S &= \sum_{n=0}^\infty \frac{1}{n!}\int d^2 x_1
\ldots \int d^2 x_n \langle e^{i\sqrt2\alpha_+ \varphi(x_1)}
\ldots e^{i\sqrt2\alpha_+ \varphi(x_n)} X \rangle, \label{X
correlator}
\end{align}
where $\langle \ldots \rangle$ stands for correlators in the free
theory with action
\begin{align}
S_0 &= \frac{1}{8\pi} \int_D d^2x\,\big[(\nabla\varphi)^2 + i
2\sqrt{2}\alpha_0 R \varphi \big]. \label{free action}
\end{align}
The neutrality condition discussed below makes sure that for a
given operator $X$, at most one term survives in the sum in Eq.
(\ref{X correlator}). The free action $S_0$ is known to describe a
CFT with the central charge
\begin{align}
c_\kappa &= 1 - 24 \alpha_0^2 = 1 - 3 \frac{(\kappa - 4)^2}{2
\kappa}, \label{central charge 2}
\end{align}
which is the same as Eq. (\ref{central charge 1}). The holomorphic
part of the stress-energy tensor corresponding to the central
charge (\ref{central charge 2}) is
\begin{align}
T &= -\frac{1}{2} \nol (\partial\varphi)^2 \nor +
i\sqrt2\alpha_0\partial^2\varphi, \label{stress}
\end{align}
where $\partial = \partial/\partial z$, and semicolons stand for
normal ordering.

Notice that by the appropriate choice of metric, the curvature $R$
may be made to vanish everywhere in the finite region of the
plane. The curvature is then concentrated at infinity, and its
effect is represented by insertion of a certain vertex operator
($V_{-2\alpha_0,-2\alpha_0}$ in the notation of Eq. (\ref{vertex
operators}) below) in the correlation functions, thereby changing
the neutrality condition, see discussion below. This prescription,
due to Dotsenko and Fateev \cite{DF}, allows to calculate
correlators of primary fields (vertex operators) using simple free
boson with $c=1$ described by the action (\ref{free action}) but
with $\alpha_0 = 0$:
\begin{align}
S_0 &= \frac{1}{8\pi} \int_D d^2x\, (\nabla\varphi)^2. \label{free
action c=1}
\end{align}

In the complex coordinates $z = x + iy, {\bar z} = x - iy$ the
field $\varphi$ separates into the holomorphic and antiholomorphic
parts:
\begin{align}
\varphi(z,\bar z) &= \phi(z) + \bar\phi(\bar z), \nonumber
\end{align}
and the basic correlators of these fields follow from (\ref{free
action c=1}):
\begin{align}
\langle \phi(z)\phi(z') \rangle &= -\log(z - z'), & \langle
\bar\phi(\bar z) \bar\phi(\bar z') \rangle &= -\log(\bar z - \bar
z'), & \langle \phi(z) \bar\phi(\bar z) \rangle = 0.
\label{propagators}
\end{align}

Primary fields in the theory (\ref{free action}) are electric
(vertex) and magnetic (vortex) operators, and their combinations
also called vertex operators for simplicity (they all are
implicitly assumed to be normal ordered):
\begin{align}
V_{e,0}(z,\bar z) &= e^{i\sqrt2e\varphi(z,\bar z)}, \qquad
V_{0,m}(z,\bar z) = e^{-\sqrt2 m \widetilde\varphi(z,\bar z)},
\nonumber\\
V_{e,m}(z,\bar z) &= e^{i\sqrt2 e \varphi(z,\bar z)}e^{-\sqrt2 m
\widetilde\varphi(z,\bar z)}, \nonumber
\end{align}
where we introduced the Cauchy-Riemann dual
\begin{align}
\widetilde\varphi(z,\bar z) = -i \phi(z) + i \bar\phi(\bar z)
\nonumber
\end{align}
of the field $\varphi$, as well as electric and magnetic charges
$e$ and $m$. A general vertex operator can also be written as a
product of holomorphic and antiholomorphic components:
\begin{align}
V_\alpha(z) & = e^{i\sqrt 2 \alpha \phi(z)}, \qquad {\bar
V}_{\bar\alpha}(\bar z) = e^{i \sqrt2 \bar\alpha \bar\phi(\bar
z)},
\nonumber\\
V_{\alpha,\bar\alpha}(z,\bar z) &= V_\alpha(z) {\bar
V}_{\bar\alpha}(\bar z) = e^{i\sqrt 2 \alpha \phi(z)} e^{i \sqrt2
\bar\alpha \bar\phi(\bar z)}, \label{vertex operators}
\end{align}
where the holomorphic and antiholomorphic charges are:
\begin{align}
\alpha &= e + m, & \bar\alpha = e - m. \nonumber
\end{align}

The holomorphic and antiholomorphic dimensions of the vertex
operators follow from the anomalous stress-energy tensor
(\ref{stress}):
\begin{align}
h(\alpha) &= \alpha(\alpha - 2\alpha_0) =
h(e,m) = (e + m)(e + m - 2\alpha_0), \nonumber \\
{\bar h}(\bar \alpha) &= \bar\alpha (\bar\alpha - 2\alpha_0) =
{\bar h}(e,m) = (e - m)(e - m - 2\alpha_0). \nonumber
\end{align}
From this we see that a vertex operator is spinless (meaning that
$h = \bar h$) if either $\bar\alpha = \alpha$ or $\bar\alpha =
2\alpha_0 - \alpha$. In the first case the operator is purely
electric ($m=0$), and in the second case it can have an arbitrary
magnetic charge, but the electric charge should be $e = \alpha_0$.
We then introduce the notation
\begin{align}
V^{(\alpha)}(z, \bar z) &= V_\alpha(z) {\bar V}_{2\alpha_0 -
\alpha}(\bar z). \label{spinless bulk operator}
\end{align}

Notice also a certain duality: the dimensions of the operators
$V_\alpha$ and $V_{2\alpha_0 - \alpha}$ are the same. This is
consistent with the correlator
\begin{align}
\langle V_\alpha(z) V_{2\alpha_0 - \alpha}(z') \rangle = (z -
z')^{-2h_\alpha}. \nonumber
\end{align}
We see that the sum of the charges of the operators within the
correlator is $2\alpha_0$, which is the negative of the background
charge $-2\alpha_0$ placed at infinity. This is true in general:
in the theory with a background charge correlators of vertex
operators do not vanish only if the following neutrality condition
is satisfied --- the total sum of charges should equal to
$2\alpha_0$, in which case the chiral correlator is given by
\begin{align}
\langle V_{\alpha_1}(z_1) V_{\alpha_2}(z_2) \ldots
V_{\alpha_n}(z_n) \rangle &= \prod_{i < j} (z_i - z_j)^{2\alpha_i
\alpha_j}, & \sum_i \alpha_i = 2\alpha_0. \label{chiral
correlator}
\end{align}
Similarly, the correlator of vertex operators
$V_{\alpha,\alpha}(z,\bar z)$ is
\begin{align}
\big\langle \prod_i V_{\alpha_i,\alpha_i}(z_i, \bar z_i)
\big\rangle &= \prod_{i < j} |z_i - z_j|^{4\alpha_i \alpha_j}, &
\sum_i \alpha_i = 2\alpha_0. \label{non-chiral correlator}
\end{align}

While the global behavior of correlators of vertex operators is
affected by the background charge, their local properties are
completely encoded in the short-distance operator product
expansions (OPE)
\begin{align}
V_{\alpha_1}(z_1) V_{\alpha_2}(z_2) = (z_1 - z_2)^{2 \alpha_1
\alpha_2} V_{\alpha_1 + \alpha_2}(z_2) + \ldots, \label{OPE}
\end{align}
where the dots stand for subleading terms.

Finally, let us mention that in CFT literature it is customary to
label holomorphic charges and weights by two numbers $r,s$
according to
\begin{align}
\alpha_{r,s} &= \frac{1}{2}(1-r)\alpha_+ +
\frac{1}{2}(1-s)\alpha_-, \nonumber \\
h_{r,s} &= \alpha_{r,s}(\alpha_{r,s} - 2\alpha_0) =
\frac{1}{4}\Bigl[\Bigl(r \frac{\sqrt\kappa}{2} - s
\frac{2}{\sqrt\kappa} \Bigr)^2 - \Bigl(\frac{\sqrt\kappa}{2} -
\frac{2}{\sqrt\kappa} \Bigr)^2\Bigr] \nonumber \\
&= \frac{(r \kappa - 4 s)^2 - (\kappa - 4)^2}{16 \kappa}.
\label{Kac weights}
\end{align}
We stress that we use it as a shorthand notation and do not impose
any restrictions on $r$ or $s$. The holomorphic primary field of
weight $h_{r,s}$ is denoted by $\psi_{r,s}(z)$. In the notation
(\ref{vertex operators}), this is the field $V_{\alpha_{r,s}}(z)$.
The corresponding spinless bulk operator $V^{(\alpha_{r,s})}(z,
\bar z)$ will be denoted as $\psi_{r,s}(z,\bar z)$.

\subsection{Coulomb gas CFT in the upper half plane}

We now consider modifications to the Coulomb gas description that
result when the boson lives in a bounded region. Due to conformal
invariance we may choose the simplest possible case: the upper half
plane $\mathbb H$, and most formulas will be written for this case.
For basics on boundary CFT see Refs. \cite{cft, cardyoperators-1,
cardyoperators-2}.

First we note that since the boson field $\varphi$ is defined by
the orientation of the loops, it is a pseudoscalar, meaning that
it changes sign under parity transformation, or reflection in a
boundary ($z \to \bar z$ for the upper half plane). This implies
that we must impose the Dirichlet boundary condition
\begin{align}
\partial_l\varphi |_{\partial D} = 0,
\label{D}
\end{align}
where the derivative is taken along the boundary (for $\mathbb H$
the boundary is the real axis $z = \bar z$). Each boundary is then
a level line of $\varphi$, in correspondence with lattice loops.

The Dirichlet boundary condition glues together the holomorphic
and antiholomorphic sectors of the theory in a way that is easiest
to describe in terms of the image charges, see, for example,
Chapter 11 in Ref. \cite{cft}. The dependence of correlators of
primary fields on the antiholomorphic coordinates $\bar z_i$ in
the upper half plane can be regarded as the dependence on
holomorphic coordinates $z_i^*$ in the lower half plane, after the
parity transformation is performed. In our case the parity
transformation for the chiral boson is simply
\begin{align}
\bar\phi(\bar z) \to -\phi(z^*), \nonumber
\end{align}
which gives the following prescription for bulk vertex operators
in the upper half plane:
\begin{align}
V_{e,0}(z,\bar z) &= e^{i \sqrt{2} e \varphi(z,\bar z)} = e^{i
\sqrt{2} e [\phi(z) + \bar\phi(\bar z)]} \to
e^{i \sqrt{2} e \phi(z)} e^{-i \sqrt{2} e \phi(z^*)}, \nonumber \\
V_{0,m}(z,\bar z) & = e^{- \sqrt{2} m \widetilde\varphi(z,\bar z)}
= e^{i \sqrt{2} m [\phi(z) - \bar\phi(\bar z)]} \to
e^{i \sqrt{2} m \phi(z)}  e^{i \sqrt{2} m \phi(z^*)}, \nonumber \\
V_{e,m}(z,\bar z) &\to e^{i \sqrt{2} (m + e) \phi(z)}
e^{i \sqrt{2} (m - e) \phi(z^*)}, \nonumber \\
V_{\alpha,\bar\alpha}(z,\bar z) &= e^{i\sqrt 2 \alpha \phi(z)}
e^{i \sqrt2 \bar\alpha \bar\phi(\bar z)} \to e^{i \sqrt{2} \alpha
\phi(z)} e^{-i \sqrt{2} \bar\alpha \phi(z^*)}. \label{parity}
\end{align}
The right hand sides of these equations should be viewed as
products of two holomorphic operators.

This situation can be summarized by saying that under reflection
the electric charges change sign, while the magnetic ones do not.
This implies, in particular, that on the boundary (the real axis)
only magnetic operators survive, electric ones being rendered
trivial by the Dirichlet boundary condition. Indeed, as $z$ and
$z^*$ both approach a point $x$ on the real axis, the bulk
operator $V_{e,m}(z, \bar z)$ (with any electric charge $e$)
reduces to
\begin{align}
V^{(2m)}(x) &= e^{i 2\sqrt{2} m\phi(x)}. \label{V2m}
\end{align}
Such boundary operator is characterized only by one weight which
is the same as the holomorphic weight $h(0,2m)$ of a bulk
operator. This situation can also be described by saying that the
fusion on the boundary of a holomorphic operator with its image
produces a magnetic boundary operator.

\subsection{Creation of critical curves}
\label{subsec:curve creation}

As we have seen, in a microscopic description a critical curve
starting at a boundary is created by a change in boundary
conditions. In the effective language of CFT such change is
implemented by insertion of a certain operator at a point on the
boundary.

Microscopic definition of height function $H$ implies that if we
have $n$ critical curve with the same orientation that start on a
boundary at some point, the boundary values of the field $\varphi$
on the two sides of the point should differ by $\pm n\pi {\mathcal
R}$. The curves should be oriented in the same way to prevent them
from reconnecting with each other. For a system on the UHP, $n$
such curves starting at the origin are then created by the
following boundary condition:
\begin{align}
\varphi(x) = \left\{
\begin{array}{ll} n\pi\mathcal{R},
& x\leqslant 0, \\ 0, & x>0. \\
\end{array} \right.
\label{bc}
\end{align}
The operator creating such a jump in the value of $\varphi$ at
$x=0$ is a certain magnetic boundary operator. To find out what it
is, we first consider magnetic operators in the bulk.

\begin{figure}[t]
\centering
\includegraphics[width=3in]{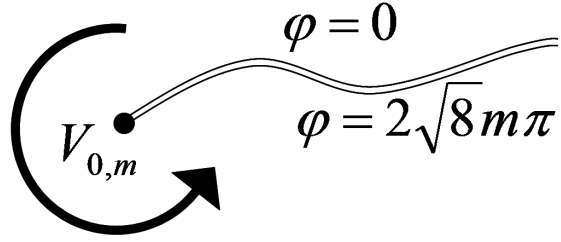}
\caption{A bulk magnetic operator $V_{0,m}$ creates a vortex
configuration of the field $\varphi$. The field changes by
$2\sqrt{8} \pi m$ when going around $V_{0,m}$.}
\label{fig:bulkmagnet}
\end{figure}

Note that the insertion of a magnetic operator in the bulk creates
a vortex in the field $\varphi$. Indeed, the bulk OPE (\ref{OPE})
gives
\begin{align}
V_{e,0}(z, \bar z) V_{0,m}(z', \bar z') &= \Big( \frac{z -
z'}{\bar z - \bar z'} \Big)^{2em} V_{e+m}(z') {\bar V}_{e-m}(\bar
z') + \ldots \nonumber \\
&= e^{4i e m \arg(z - z')} V_{e+m}(z') {\bar V}_{e-m}(\bar z') +
\ldots. \nonumber
\end{align}
This means that when $z$ goes around $z'$, the field $\varphi$
changes by $4\sqrt{2} \pi m$, and hence, a discontinuity line with
this jump arises, see Fig. \ref{fig:bulkmagnet}.
If this vortex corresponds to a star of $n$ critical curves joined
at the point $z'$, the change in $\varphi$ should be equal to
$n\pi {\mathcal R}$, and then the discontinuity is not physical
due to the compactification of $\varphi$. This gives the magnetic
charge of the bulk curve creating operator:
\begin{align}
m = \frac{\sqrt{2}}{8} n {\mathcal R} = \frac{n}{2 \sqrt{\kappa}}
= - \frac{n}{4} \alpha_-, \label{magnetic charge}
\end{align}
where we used the value (\ref{comp radius}) of the
compactification radius and the definition (\ref{alphas}) of
$\alpha_-$. In order to be spinless (otherwise the operator would
transform under rotations, giving non-trivial dependence on the
winding number of curves), the bulk curve creating operator should
also have electric charge $\alpha_0$.

We have, therefore, found the holomorphic charge of the bulk curve
creation operator to be
\begin{align}
\alpha &= \alpha_0 - \frac{n}{4} \alpha_- = \alpha_{0,n/2}
\nonumber
\end{align}
in the notation of Eq. (\ref{Kac weights}). The operator itself is
then $\psi_{0,n/2}(z,\bar z)$, and its holomorphic weight is
\begin{align}
h_{0,n/2} = \frac{4n^2 - (\kappa - 4)^2}{16 \kappa}. \nonumber
\end{align}
In particular, a single critical curve going through a point $z$
is created by the operator $\psi_{0,1}(z,\bar z)$ with the
holomorphic weight $h_{0,1} = (8 - \kappa)/16$. Notice that this
weight is related to the fractal dimension of the critical curve
by $d_f = 2 - 2h_{0,1}$, see Ref. \cite{BB-zigzag} for details.

\begin{figure}[t]
\centering
\includegraphics[width=4in]{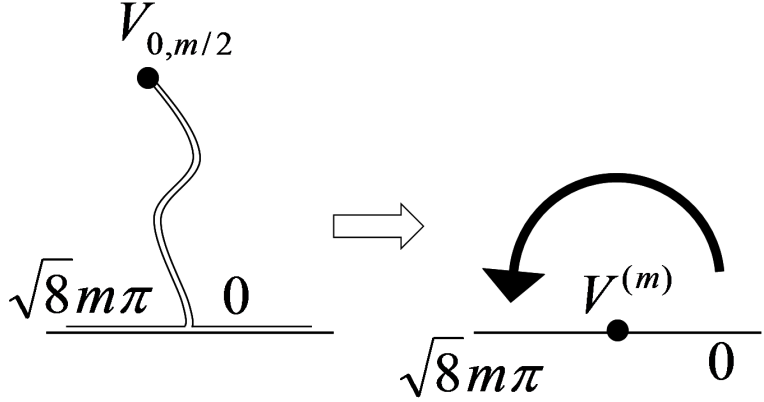}
\caption{A boundary magnetic operator is obtained as the boundary
limit of the bulk magnetic operator $V_{0,m/2}$.} \label{magnet}
\end{figure}

Now we go back to the boundary. According to Eq. (\ref{parity}),
as $z$ approaches a point $x$ on the real axis, the bulk operator
$V_{e,m}(z, \bar z)$ (with any electric charge $e$) reduces to
$e^{i 2\sqrt{2} m\phi(x)}$. In this process the two sides of the
discontinuity line become parts of the real axis separated by $x$
(see Fig. \ref{magnet}).
Thus, when we go from one side of $x$ to the other along a
semicircle, the field changes by the same amount as in making a
full circle around a bulk magnetic operator. Then to create the
boundary condition (\ref{bc}), and, correspondingly, $n$ critical
curves starting at the origin, we need to insert there the
magnetic operator
\begin{align}
V^{(m)}(x) = e^{i\sqrt{2} m\phi(x)}, \nonumber
\end{align}
with the magnetic charge determined by the condition $\sqrt{8} \pi
m = n \pi {\mathcal R}$, which gives
\begin{align}
m = \frac{1}{\sqrt{8}} n {\mathcal R} = \frac{n}{\sqrt{\kappa}} =
-\frac{n}{2} \alpha_-. \label{magnetic charge boundary}
\end{align}
In the notation of Eq. (\ref{Kac weights}) this is
$\alpha_{1,n+1}$, so the boundary curve creating operator is
$\psi_{1,n+1}(x)$ with dimension
\begin{align}
h_{1,n+1} = \frac{2n^2 + n(4 - \kappa)}{2\kappa}. \nonumber
\end{align}
In particular, a single curve is created by the insertion of
$\psi_{1,2}(x)$ with the dimension
\begin{align}
h_{1,2} = \frac{6 - \kappa}{2\kappa}, \label{h12}
\end{align}
which appeared before in the discussion of restriction property of
SLE, see Eq. (\ref{alpha c kappa}).

\section{Harmonic measure of critical curves}
\label{sec:harmonic measure}

Harmonic measure is a useful quantity describing geometry of
complicated plane domains \cite{harmonic measure-1, harmonic
measure-2}. In the following sections I define it and the related
spectrum of multifractal exponents, and then show how to compute
these exponents for harmonic measure on critical curves using CFT.

\subsection{Definitions of harmonic measure}
\label{subsec:definition harmonic measure}

Let $D$ be a domain (an open connected subset of the complex
plane), $\partial D$ its boundary, and $z \in D$.

Harmonic measure in $D$ from $z$, denoted as $\omega_D(z,\Gamma)$
where $\Gamma \in
\partial D$, is a probability measure on $\partial D$, which can
be defined as the probability that the standard two-dimensional
Brownian motion $B_t$ that starts at $z$ hits $\partial D$ in a
given portion $\Gamma \subset \partial D$ of the boundary:
\begin{align}
\omega_D(z,\Gamma) &= {\mathbf P}^z[B_{\tau_D} \in \Gamma].
\nonumber
\end{align}
Here $\tau_D$ is the escape time from $D$, that is, the first time
when the Brownian motion $B_t$ hits the boundary $\partial D$.

Harmonic measure $\omega_D(z,\Gamma)$ can also be characterized as
the unique harmonic function (solution of the Laplace's equation)
$u(z)$ in $D$ with the Dirichlet boundary conditions
\begin{align}
u(\zeta) &= 1, \quad \zeta \in \Gamma, &&
u(\zeta) = 0, \quad \zeta \notin \Gamma. \nonumber
\end{align}

Here we will only be interested in harmonic measure from infinity
of the domain $D$ exterior to a closed curve $\gamma$. In this
case we will denote it simply by $\omega(\Gamma)$. Harmonic
measure $\omega(\Gamma)$ has an electrostatic interpretation.
Imagine that $D$ is a charged metallic cluster with the total
charge one. The charge is concentrated on the boundary $\partial
D$. Then $\omega(\Gamma)$ is the charge located on the portion
$\Gamma$ of $\partial D$.

Harmonic measure is conformally-invariant: if $f: D \to D'$ is a
conformal map that is continuous and one-to-one on $D \cup
\partial D$, then
\begin{align}
\omega_D(z,\Gamma) = \omega_{D'}(f(z),f(\Gamma)). \nonumber
\end{align}

\subsection{Moments of harmonic measure and multifractal spectrum}

\begin{figure}[]
\centering
\includegraphics[width=0.6\textwidth]{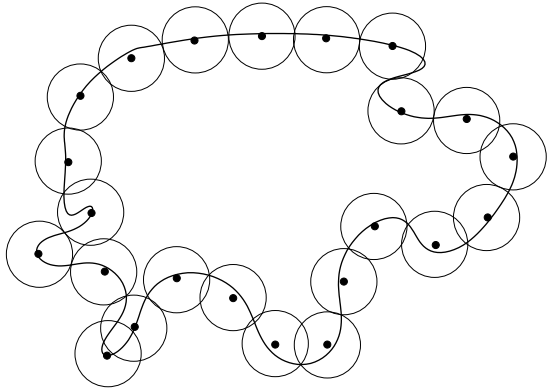}
\caption{A curve covered by discs.}
\end{figure}

Consider a closed curve $\gamma$. One can cover the curve $\gamma$
with discs $B(z_i, r)$ of radius $r$ centered at some points $z_i
\in \gamma$ ($z_i$ form a discrete subset of $\gamma$). Let
\begin{align}
p(z_i, r) = \omega(\gamma \cap B(z_i, r)) \nonumber
\end{align}
be the harmonic measure (from infinity) of the portion of the
curve covered by the disc $B(z_i, r)$. Then we consider the
moments
\begin{align}
M_h = \sum_{i=1}^N p(z_i, r)^h, \label{moments}
\end{align}
where $h$ is a real power, and $N$ is the number of discs needed
to cover $\gamma$. As the radius $r$ gets smaller and the number
of discs $N$ gets larger, these moments scale as
\begin{align}
M_h \sim \Bigl(\frac{r}{L}\Bigr)^{\tau(h)}, \quad
\frac{r}{L}\rightarrow 0. \label{scaling_of_moments}
\end{align}
The size $L$ (diameter) of the curve is used to make the right
hand side of this equation dimensionless.

The function $\tau(h)$ is called the {\bf multifractal spectrum}
of the curve $\gamma$. This function encodes a lot of information
of the curve $\gamma$. It also has some simple properties. First
of all, since all $0 < p(z_i, r) \leqslant 1$, the moments $M_h$
are well defined for any real $h$, and the function $\tau(h)$ is
non-decreasing: $\tau(h) \leqslant \tau(h')$ for any $h < h'$.
Secondly, if $h=1$, the sum in (\ref{moments}) is equal to the
total charge of the cluster, and therefore does not scale with
$r$, producing the normalization condition
\begin{align}
\tau(1) = 0. \nonumber
\end{align}
Third, if we set $h=0$, $M_0$ is simply the number $N$ of discs of
radius $r$ necessary to cover the curve $\gamma$, so that by
definition the fractal (Hausdorff) dimension of $\gamma$ is
\begin{align}
d_f = -\tau(0). \nonumber
\end{align}
If the curve $\gamma$ were smooth, we would have a simple relation
$\tau(h) = h-1$. For a fractal curve one defines the anomalous
exponents $\delta(h)$ by
\begin{align}
\tau(h) &= h - 1 + \delta(h). \nonumber
\end{align}
Also, the generalized multifractal dimensions of a fractal curve
$\gamma$ are defined as $D(h) = \tau(h)/(h-1)$ (so that $D(0) =
d_f$). A non-trivial theorem due to N. Makarov \cite{Makarov-1,
Makarov-2} states that
\begin{align}
D(1) = \tau'(1) = 1. \nonumber
\end{align}

If the curve $\gamma$ is a member of an ensemble of curves
generated in some random way, the moments $M_h$ become random
variables, and we can study their distribution functions. If the
distribution functions are narrow, then the mean moments
$\overline{M_h}$, where the overline denotes the ensemble
averaging, characterize them well. This is usually the case for
$|h|$ that is not too large. In this case the mean and the typical
moments scale in the same way, which is a statement about
self-averaging of the moments. This is what we assume here.

In this situation it is also natural to assume some sort of
ergodicity, meaning that the summation over the points $z_i$ in
Eq. (\ref{moments}) for a typical curve is equivalent to the
ensemble average. Hence we can write
\begin{align}
M_h = N \overline{p(z_0, r)^h} \sim
\Bigl(\frac{r}{L}\Bigr)^{\tau(0)} \overline{p(z_0, r)^h},
\nonumber
\end{align}
where now the harmonic measure $p(z_0, r)$ is evaluated at any
point $z_0 \in \gamma$. We define the local multifractal exponent
$\widetilde\tau(h)$ at a point $z_0$ by
\begin{align}
\overline{p(z_0, r)^h} &\sim
\Bigl(\frac{r}{L}\Bigr)^{\widetilde\tau(h)}.
\label{local exponent}
\end{align}
Similar to $\tau(h)$, for a smooth curve we have
$\widetilde\tau(h) = h$, so in general we define the local
anomalous exponents $\Delta^{(2)}_{\text{bulk}}(h)$ by
\begin{align}
\widetilde\tau(h) &= h + \Delta^{(2)}_{\text{bulk}}(h). \nonumber
\end{align}
The reason for the superscript $(2)$ and the subscript ``bulk''
will become clear in the next subsection.

It is obvious from the definitions that $\widetilde\tau(0) =
\Delta^{(2)}_{\text{bulk}}(0) = 0$, and we deduce simple
ergodicity relations
\begin{align}
\label{ergodicity} \widetilde\tau(h) &= \tau(h) - \tau(0),
&& \tau(h) = \widetilde\tau(h) - \widetilde\tau(1), \nonumber \\
\Delta^{(2)}_{\text{bulk}}(h) &= \delta(h) - \delta(0), &&
\delta(h) = \Delta^{(2)}_{\text{bulk}}(h) -
\Delta^{(2)}_{\text{bulk}}(1), \nonumber \\
d_f &= 1 + \Delta^{(2)}_{\text{bulk}}(1).
\end{align}

\subsection{Critical curves and uniformizing maps}

So far we have considered arbitrary closed curves. An example of
such a curve is the exterior perimeter $\gamma$ of a critical
cluster. One can imagine that the cluster is made of a conducting
material and carries the total unit electric charge. The harmonic
measure of any part of $\gamma$ is then equal to the electric
charge of this part. Since exterior perimeters are always dilute
curves in the sense of Section \ref{subsec:loop models} (see also
discussion of duality around Eq. (\ref{duality})), in the
remaining part of this paper we will always assume $\kappa
\leqslant 4$.

The critical clusters and their boundaries appear as members of
statistical ensembles, which is the situation suitable for local
multifractal analysis of the previous section. We then will pick a
point of interest $z_0$ on the curve $\gamma$ and consider a disc
of a small radius $r \ll L$ centered at $z_0$. It surrounds a
small part of $\gamma$, and we will study the mean moments of the
harmonic measure $p(z_0,r)$ and their scaling as in Eq.
(\ref{local exponent}).

There are a few generalizations of the simple closed critical
curve considered above. First of all, the curve $\gamma$ need not
be closed or stay away from system boundaries. If $\gamma$ touches
a boundary, we can supplement it with the image $\bar\gamma$
(reflected in the boundary) and take the union $\gamma \cup
\bar\gamma$ to be the charged conducting object. The electrostatic
definition of $p(z_0,r)$ can be naturally extended to the cases
when $z_0$ is the endpoint of $n$ critical curves on the boundary
or in the bulk. If $n$ is even, the latter case can be also seen
as $n/2$ critical curves passing through $z_0$. In particular,
$n=2$ corresponds to the situation of a single curve in the bulk,
considered above.

When $z_0$ is the endpoint of $n$ critical curves on the boundary
or the bulk we define the corresponding scaling exponents similar
to Eq. (\ref{local exponent}):
\begin{align}
\overline{p(z_0, r)^h} &\sim r^{h + \Delta^{(n)}(h)}, &
\overline{p(z_0, r)^h} &\sim r^{h +
\Delta^{(n)}_{\mathrm{bulk}}(h)}. \label{exponent definition}
\end{align}
In the case of a single curve we will drop the superscript, so,
for example, $\Delta(h)\equiv \Delta^{(1)}(h)$ is the same
exponent as obtained in Eq. (\ref{Delta(h)}).

These exponents were first obtained by means of quantum gravity in
\cite{Duplantier-PRL,Duplantier}. For a critical system with
parameter $\kappa$ the results read
\begin{align}
\Delta(h) &= \frac{\kappa-4 + \sqrt{(\kappa-4)^2 + 16\kappa h}}
{2\kappa} = \frac{\sqrt{1-c+24h} - \sqrt{1-c}} {\sqrt{25-c} -
\sqrt{1-c}}, \label{Delta(h)-again}
\\
\Delta^{(n)}(h) &= n \Delta(h), \label{simple boundary exponent}
\\
\Delta^{(n)}_{\mathrm{bulk}}(h) & = - \frac{h}{2} +
\Big(\frac{1}{16} + \frac{n-1}{4\kappa}\Big) \big(\kappa - 4 +
\sqrt{(\kappa - 4)^2 + 16\kappa h}\big). \label{simple bulk
exponent}
\end{align}
Remarkably, $\Delta(h)$ is the gravitationally dressed dimension
$h$, as given by the KPZ formula of 2D quantum gravity
\cite{Knizhnik:1988ak, DiFrancesco-QG-review}. Starting in the
next section, we will show how to obtain these exponents in the
framework of Coulomb gas CFT, where they are also written in a
more transparent way.

\begin{figure}[t]
\centering
\includegraphics[width=0.8\textwidth]{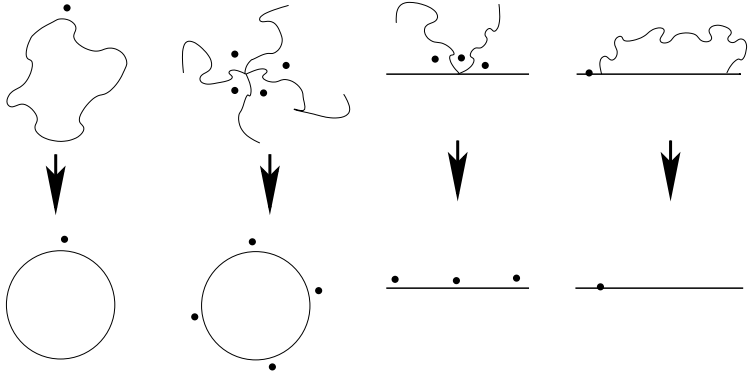}
\caption{The uniformizing conformal maps for various cases
considered. The dots denote that points where the electric field
is measured.} \label{mappings}
\end{figure}

The basic property that allows ro calculate the multifractal
exponents using CFT is the conformal invariance of the harmonic
measure. Let us consider a conformal map $w(z)$ of the exterior of
$\gamma$ to a standard domain. Usually we will choose the upper
half plane but sometimes the exterior of a unit circle is more
convenient. We normalize the map so that the point of interest
$z_0$ is mapped onto itself, choose it to be the origin $z_0=0$
and demand that at infinity $w(z) = z + o(z).$ Examples of $w(z)$
for several cases are shown in Fig. \ref{mappings}.

The scaling of $w(z)$ near the origin is directly related to that
of the harmonic measure. Indeed, since $p(0,r)$ is the charge
inside the disc of radius $r$, by Gauss theorem it is equal to the
flux of the electric field through the boundary of this disc.
This, in turn, should scale as the circumference of this disc
times a typical absolute value of the electric field at the
distance $r$ from the origin, i.e. $|w'(r)|$.  This leads to
scaling relation
\begin{align}
p(0,r) \sim r|w'(r)|, \nonumber
\end{align}
which allows to rewrite the definitions (\ref{exponent
definition}) as
\begin{align}
\overline{|w'(r)|^h} &\sim r^{\Delta^{(n)}(h)}, &
\overline{|w'(r)|^h} &\sim r^{\Delta^{(n)}_{\mathrm{bulk}}(h)}.
\label{real definition}
\end{align}

The relation of the scaling of the harmonic measure and the
derivative of a uniformizing map allows for further
generalizations. Namely, we can measure the electric field in more
than one point. Close to the origin $n$ curves divide the plain
into $n$ sectors in the bulk and $n + 1$ on the boundary. Then we
can study objects like
\begin{align}
& \overline{|w'(z_1)|^{h_1} \ldots
|w'(z_{n+1})|^{h_{n+1}}} & \textrm{(boundary)}, \nonumber \\
& \overline{|w'(z_1)|^{h_1}\ldots |w'(z_n)|^{h_n}} &
\textrm{(bulk)}, \label{higher boundary case}
\end{align}
where no two $z_i$'s lie in the same sector. The case when the
electric field is not measured in some sectors is done by setting
$h_i=0$ in them. We will see how to express these quantities as
CFT correlation functions. In the case when $z_i$ are all at the
distance $r$ from the origin ($z_i = r e^{i\theta_i}$,
$\theta_i=\mathrm{const}$), these averages scale as
$r^{\Delta^{(n)}(h_1,\ldots h_{n+1})}$ and
$r^{\Delta^{(n)}_{\mathrm{bulk}}(h_1,\ldots h_n)}$ with the {\it
higher multifractal exponents} \cite{Duplantier}
\begin{align}
\Delta^{(n)}(h_1,\ldots h_{n+1}) & = \sum_{i=1}^{n+1}
\Delta^{(n)}(h_i) + \frac{\kappa}{2} \sum_{i<j}^{n+1} \Delta(h_i)
\Delta(h_j),
\label{higher boundary exponent} \\
\Delta^{(n)}_{\mathrm{bulk}}(h_1,\ldots h_n) & = \sum_{i=1}^n
\Delta_{\mathrm{bulk}}^{(n)}(h_i) + \frac{\kappa}{4} \sum_{i<j}^n
\Delta(h_i) \Delta(h_j). \label{higher bulk exponent}
\end{align}

\subsection{Derivative expectations and CFT in fluctuating geometry}

Here we begin the Coulomb gas derivation of results
(\ref{Delta(h)-again} -- \ref{simple bulk exponent}, \ref{higher
boundary exponent}, \ref{higher bulk exponent}). It is easiest to
start with a point where a single curve $\gamma$ connects with the
system boundary. We assume that the critical system occupies the
upper half plane, so that the real axis is the boundary.

The partition function $Z(0,L)$, restricted to configurations that
contain a curve $\gamma$ connecting the points 0 and $L$, is given
by the correlator of two boundary curve creating operators, in
this case $\psi_{1,2}$ (see section (\ref{subsec:curve
creation})):
\begin{align}
\frac{Z(0,L)}{Z} = \langle
\psi_{1,2}(0)\psi_{1,2}(L)\rangle_\mathbb{H}, \nonumber
\end{align}
where $Z$ is the unrestricted partition function. This correlation
function can be computed in two steps. In the first step we pick a
particular realization of the curve $\gamma$. Within each
realization, it is the boundary separating two independent
systems---the interior and the exterior of $\gamma$. In both these
systems we can sum over microscopic degrees of freedom to obtain
the partition functions $Z^\mathrm{int}_\gamma$ and
$Z^\mathrm{ext}_\gamma$, respectively. These are stochastic
objects that depend on the fluctuating geometry of $\gamma$. In
the second step we average over the ensemble of curves of
$\gamma$. We thus obtain
\begin{align}
Z(0,L) = \overline{Z^\mathrm{int}_\gamma Z^\mathrm{ext}_\gamma}.
\nonumber
\end{align}

Next, we insert an additional boundary primary operator $O_h(r)$
of dimension $h$ close to 0, and another one $O_h(\infty)$ at
infinity. The first one serves as a ``probe'' of harmonic measure,
and the second is necessary to ensure the charge neutrality. We
thus consider the correlation function
\begin{align}
\label{thecorr} \langle\psi_{1,2}(0)O_h(r)\psi_{1,2}(L)
O_h(\infty) \rangle_{\mathbb{H}}.
\end{align}
Since we are only interested in the $r$-dependence of the
correlation function, we can fuse together the distant primary
fields: $\psi_{1,2}(L) O_h(\infty) \to \Psi(\infty)$. We therefore
consider the $r$-dependence of a 3-point function
\begin{align}
\label{the3corr} \langle\psi_{1,2}(0)O_h(r)\Psi(\infty)
\rangle_\mathbb{H},
\end{align}
and show that it yields the statistics of the harmonic measure.

Decomposing the upper half plane into the exterior and the
interior of $\gamma$ as before, we can rewrite (\ref{thecorr}) as
the average over the fluctuating geometry of $\gamma$:
\begin{align}
\label{7} \overline{\langle O_h(r) O_h(\infty)
\rangle^\mathrm{ext}_\gamma Z^\mathrm{int}_\gamma
Z^\mathrm{ext}_\gamma}.
\end{align}
Here the domain of the definition of the  correlation function of
primary fields is the exterior of $\gamma$. This correlation
function is statistically independent from the other two factors
in the numerator of (\ref{7}) in the limit $r\ll |L|$, and we are
left with the correlation function $\langle
O_h(r)O_h(\infty)\rangle^\mathrm{ext}_\gamma$ of two primary
fields of boundary CFT, further averaged over all configurations
of the boundary $\gamma$. This average is proportional to the
3-point correlation function (\ref{the3corr}).

Now we  apply  the conformal transformation $w(z)$ which maps the
exterior of $\gamma$ onto the upper half plane. Being a primary
operator of weight $h$,  $O_h(r)$ transforms as $O_h \to |w'(r)|^h
O_h(w(r))$, while $O_h(\infty)$ does not change because of the
normalization of $w(z)$ at infinity. The transformation relates
the correlation function in the exterior of $\gamma$ to a
correlation function in the upper half plane:
\begin{align}
\langle O_h(r)O_h(\infty)\rangle_\gamma^\mathrm{ext}=|w'(r)|^h
\langle O_h(w(r))O_h(\infty)\rangle_\mathbb{H }. \label{2pt
correlator}
\end{align}
The latter does not depend on $r$ and can be neglected.

Summing up, we obtain a scaling relation between the moments of
the harmonic measure near the boundary and correlation functions
of primary boundary fields \cite{Bauer-Bernard-1}:
\begin{align}
\label{goodrelation} \overline{|w'(r)|^h} &\propto \langle O_h(r)
\psi_{1,2}(0) \Psi(\infty)\rangle_\mathbb{H}, & r &\ll |L|.
\end{align}
The primary field $\Psi(\infty)$ should be chosen in such a way as
to render the correlation function non-zero. The choice is made
unique by picking the conformal block which satisfies simple
physical condition $\Delta(0)=0$. The $r$-dependence of the
correlation function (\ref{goodrelation}) is found from the OPE of
the fields $O_h(r)$ and $\psi_{1,2}(0)$:
\begin{align}
O_h(r)\psi_{1,2}(0)=\sum_{k=1}^\infty r^{\Delta_k}\Phi^{(k)}(0).
\label{OPE-1}
\end{align}
The exponent $\Delta(h)$ is then identified as the lowest power
$\Delta_k$ such that $\langle\Phi^{(k)}(0)\Psi(\infty)\rangle\neq
0$.

Several remarks are in order. As presented, this argument produces
the scaling exponent $\Delta(h)$ for a single curve on the
boundary. But it can be easily modified for studying the scaling
behavior in all other cases. The case of $n$ curves starting from
a point on the boundary is obtained by simple replacement of the
curve creating operators: $\psi_{1,2} \to \psi_{1,n+1}$ (see
Section (\ref{subsec:curve creation})).

Also, the argument can be repeated for the case of $\gamma$
connected to the real axis only at one point (as in SLE). In this
case no separation in two systems is necessary. Finally, nothing
compels us measure the electric field on the real axis. We could
take instead a bulk primary field $O_{h', h'}(z, \bar z),$ where
$|z|=r$. The weight $h'$ should be chosen such that when the
holomorphic part $O_{h'}(z)$ is fused with its image
$O_{h'}(z^*)$, the boundary field $O_h(\frac{z + z^*}{2})$ with
dimension $h$ is obtained, similar to Eq. (\ref{V2m}). This fusion
will be used below.

\subsection{Calculation of boundary multifractal exponents}

In practice we view both $O_h$ and $\psi_{1,2}$ in Eq.
(\ref{OPE-1}) as boundary vertex operators $V^{(\alpha_h)}$ and
$V^{(\alpha_{1,2})}$, see Eq. (\ref{V2m}), with charges
\begin{align}
\alpha_h &= \alpha_0 - \sqrt{\alpha_0^2 + h}, & \alpha_{1,2} &=
-\frac{\alpha_-}{2} = \frac{1}{\sqrt\kappa}. \nonumber
\end{align}
The leading term in the OPE of these two operators corresponds to
simple addition of charges, see Eq. (\ref{OPE}). Hence, the
scaling relation (\ref{goodrelation}) immediately gives the result
(\ref{Delta(h)-again}) written in a compact and suggestive form:
\begin{align}
\label{eb} \Delta(h) = 2 \alpha_{1,2} \alpha_h.
\end{align}
It is interesting that written in this form, the KPZ formula for
gravitationally dressed dimensions amounts to OPE of vertex
operators in a simple Coulomb gas CFT, without any quantum
gravity.

An immediate generalization to the statistics of harmonic measure
of ${n}$ curves reaching the system boundary at the same point is
obtained by replacing $\psi_{1,2} \to \psi_{1, n + 1}$. Since
$\alpha_{1, n + 1} = -n\alpha_-/2 = n \alpha_{1,2}$, this
immediately leads to
\begin{align}
\Delta^{(n)}(h) = 2 \alpha_{1, n + 1} \alpha_h = n \Delta(h),
\nonumber
\end{align}
which is the same as Eq. (\ref{simple boundary exponent}).

To calculate the higher boundary multifractal exponents, we
consider $n$ non-intersecting critical curves growing from the
origin on the boundary (the real axis). It will be convenient to
assume that the curves end somewhere in the bulk thus forming a
boundary star (e.g. the third picture in Fig. \ref{mappings}). Let
$w(z)$ be the conformal map of the exterior of the star to the
upper half plane with the usual normalization $w(z) = z + o(z)$ at
$z \to \infty$.

We want to find the scaling of the average
\begin{align}
\overline{|w'(z_1)|^{h_1} \ldots |w'(z_{n+1})|^{h_{n+1}}},
\nonumber
\end{align}
where $z_i$ are all close to the origin, no two of them lying in
the same sector. The latter condition will be automatically
satisfied in the subsequent calculation due to the following: if
in a particular realization two points $z_i$ and $z_j$ happen to
be in the same sector, then $w(z_i) - w(z_j) \to 0$ as $z_i - z_j
\to 0,$ but if they lie in different sectors, $w(z_i) - w(z_j)$
remains large in the same limit.

Since $n$ curves starting from the origin on the boundary are
produced by the operator $\psi_{1,n+1}(0)$, we now consider a
boundary CFT correlation function with several ``probes'' of the
harmonic measure:
\begin{align}
C = \big\langle \prod_{i=1}^{n+1} O_{h_i',h_i'}(z_i, \bar z_i)
\psi_{1,n+1}(0)\Psi(\infty)\big\rangle_{\mathbb{H}}. \label{C-1}
\end{align}
The primary field at infinity represents the fusion of all fields
far from the origin and should be chosen by the charge neutrality
condition. As before, this correlation function is equal to the
statistical average of a certain correlator in the fluctuating
domain, and we further apply the uniformizing map $w(z)$ to
transform this domain into the UHP:
\begin{align}
C = \overline{\prod_i |w'(z_i)|^{2h_i'} \big\langle \prod_i
O_{h_i',h_i'}\big(w(z_i), \bar w(\bar z_i)\big)
\Psi(\infty)\big\rangle_{\mathbb{H}}}. \label{C-2}
\end{align}
Unlike Eq. (\ref{2pt correlator}), the correlator under the
average cannot be neglected since it does depend on the short
scale $r$, as we shall see soon.

The correlator $C$ can now be evaluated in two ways. As before, we
view the primaries $O_{h_i',h_i'}(z_i, \bar z_i)$ as vertex
operators with charges
\begin{align}
\alpha_i' = \alpha_0 - \sqrt{\alpha_0^2 + h_i'}. \label{alpha
prime}
\end{align}
Then, using the prescription (\ref{parity}), we can rewrite $C$ in
Eq. (\ref{C-1}) as a full plane chiral correlator, which then is
evaluated using Eq. (\ref{chiral correlator}):
\begin{align}
C &= \big\langle \prod_i O_{h_i'}(z_i) O_{h_i'}(z_i^*)
\psi_{1,n+1}(0)\Psi(\infty)\big\rangle \nonumber \\
&\propto \prod_i |z_i|^{2\alpha_{1,n+1} \alpha_i'} \prod_{i<j}
|z_i - z_j|^{4 \alpha_i' \alpha_j'} \prod_{i,j} (z_i - z_j^*)^{2
\alpha_i' \alpha_j'}. \nonumber
\end{align}
When all $z_i$ are at the same distance $r$ from the origin, the
last expression scales as
\begin{align}
C \propto r^{4 \alpha_{1,n+1} \sum_i \alpha_i' + 8 \sum_{i<j}
\alpha_i' \alpha_j' + 2 \sum_i \alpha_i'{}^2}. \label{C-scaling-1}
\end{align}

On the other hand, we can evaluate the correlator that appears
inside the average in Eq. (\ref{C-2}) in the same way:
\begin{align}
& \big\langle \prod_i O_{h_i',h_i'}\big(w(z_i), \bar w(\bar
z_i)\big) \Psi(\infty)\big\rangle_{\mathbb{H}} = \big\langle
\prod_i O_{h_i'}(w(z_i)) O_{h_i'}(w^*(z_i))\Psi(\infty)
\big\rangle \nonumber \\
&\propto \prod_i (w(z_i)-w^*(z_i))^{2\alpha_i'{}^2} \prod_{i<j}
|w(z_i) - w(z_j)|^{4 \alpha_i' \alpha_j'} \prod_{i \neq j}
\big(w(z_i) - w(z_j^*)\big)^{2 \alpha_i' \alpha_j'}. \nonumber
\end{align}
We specifically separated the diagonal ($i=j$) terms, since only
they contribute to the necessary short-distance behavior. All the
other term insure that the realizations of the curves in which any
two points $z_i$ end up in the same sector are suppressed (since
the distances $w(z_i) - w(z_j)$ are then small), and we can
consider only the case when all $w(z_i)$ are far apart. Then the
relevant short-distance dependence of Eq. (\ref{C-2}) is
\begin{align}
C \propto \overline{\prod_i |w'(z_i)|^{2h_i'}
(w(z_i)-w^*(z_i))^{2\alpha_i'{}^2}}. \nonumber
\end{align}
Insofar as the scaling with $r$ is concerned, we further
approximate $w(z_i)-w^*(z_i) \sim |z_i| |w'(z_i)| \sim r
|w'(z_i)|$. This gives
\begin{align}
C \propto r^{2\sum_i \alpha_i'{}^2} \overline{\prod_i
|w'(z_i)|^{2h_i' + 2\alpha_i'{}^2}}. \label{C-scaling-2}
\end{align}
The exponents in the last factor
\begin{align}
2h_i' + 2\alpha_i'{}^2 &= 2 \alpha_i' (\alpha_i' - 2\alpha_0) +
2\alpha_i'{}^2 = 2 \alpha_i' (2 \alpha_i' - 2\alpha_0) =
h_{\alpha_i} = h_i \nonumber
\end{align}
are the dimensions of the boundary operators with charges
\begin{align}
\alpha_i = 2 \alpha_i' = \alpha_0 - \sqrt{\alpha_0^2 + h_i},
\nonumber
\end{align}
appearing in the OPE of two chiral operators with charges
$\alpha_i'$.

Finally, comparing Eqs. (\ref{C-scaling-1}) and
(\ref{C-scaling-2}), we get the result
\begin{align}
\overline{|w'(z_1)|^{h_1} \ldots |w'(z_{n+1})|^{h_{n+1}}} \propto
r^{\Delta^{(n)}(h_1,\ldots h_{n+1})}, \nonumber
\end{align}
with the higher multifractal exponent
\begin{align}
\Delta^{(n)}(h_1,...h_{n+1}) &=
2\alpha_{1,n+1}\sum_{i=1}^{n+1}\alpha_i +
2\sum_{i<j}^{n+1}\alpha_i \alpha_j, \nonumber
\end{align}
which is the formula (\ref{higher boundary exponent}).

\subsection{Calculation of bulk multifractal exponents}

Calculation of bulk multifractal behavior is done in much the same
way as on the boundary, so we go straight to the general case of
higher bulk exponents.

Let the critical system, occupying the whole complex plane, be
restricted to having $n$ critical curves growing from a single
point, in which we place the origin $z=0$. We will assume that
$z=0$ is the only common point of these curves, since the local
results around this points are unaffected by the curves' behavior
at large distances. We define the conformal map $w(z)$ of the
exterior of the ``star" to the exterior of a unit circle with the
the normalization $w(z)=z+o(z)$ at $z\rightarrow\infty$.

Close to the origin the curves divide the plane into $n$ sectors.
We consider a quantity
\begin{align}
\overline{|w'(z_1)|^{h_1}\ldots |w'(z_n)|^{h_n}}, \nonumber
\end{align}
where $z_i$ are points close to the origin, no two of them lying
in one sector. As before, if two points $z_i$ and $z_j$ happen to
be in the same sector, $w(z_i)-w(z_j)\rightarrow 0$ when
$z_i-z_j\rightarrow 0,$ but if they lie in different sectors,
$w(z_i)-w(z_j)$ remains large.

Since $n$ curves starting from the origin in the bulk are produced
by the operator $\psi_{0,n/2}(0),$ we introduce a CFT correlation
function
\begin{align}
C_{\text{bulk}} = \big\langle \prod_{i=1}^{n} O_{h_i',h_i'}(z_i,
\bar z_i) \psi_{0,n/2}(0)\Psi(\infty)\big\rangle, \label{C-bulk-1}
\end{align}
where, as before, $h'_i$ is the weight of a primary field such
that the result of its fusion with its image has the weight $h_i$:
\begin{align}
\alpha_i = 2 \alpha_i'. \nonumber
\end{align}
Proceeding exactly as in the boundary case, we first rewrite the
correlator $C_{\text{bulk}}$ as the ensemble average of another
correlator in the exterior of the star of critical curves. Then we
map that exterior to the exterior of a unit disk ${\mathbb
C}\setminus D$ (see the second picture on Fig. \ref{mappings}):
\begin{align}
C_{\text{bulk}} = \overline{\prod_i |w'(z_i)|^{2h_i'} \big\langle
\prod_i O_{h_i',h_i'}\big(w(z_i), \bar w(\bar z_i)\big)
\Psi(\infty)\big\rangle_{{\mathbb C}\setminus D}}.
\label{C-bulk-2}
\end{align}

Next, we evaluate $C_{\text{bulk}}$ as defined in Eq.
(\ref{C-bulk-1}), using Eq. (\ref{non-chiral correlator}):
\begin{align}
C_{\text{bulk}} = \prod_i |z_i|^{4\alpha_{0,n/2} \alpha_i'}
\prod_{i<j} |z_i - z_j|^{4 \alpha_i' \alpha_j'} \propto
r^{4\alpha_{0,n/2} \sum_i \alpha_i' + 4 \sum_{i<j} \alpha_i'
\alpha_j'}. \nonumber
\end{align}
Alternative evaluation starting from Eq. (\ref{C-bulk-2}) gives
the same result as Eq. (\ref{C-scaling-2}). Combining the two
results for $C_{\text{bulk}}$, we obtain
\begin{align}
\overline{|w'(z_1)|^{h_1}\ldots |w'(z_n)|^{h_n}} \propto
r^{\Delta_{\mathrm{bulk}}^{(n)}(h_1,\ldots h_n)}, \nonumber
\end{align}
with the higher bulk exponent
\begin{align}
\Delta_{\mathrm{bulk}}^{(n)}(h_1,\ldots h_n) = \sum_{i=1}^n
\Delta_{\mathrm{bulk}}^{(n)}(h_i)+\sum_{i<j}^
{n}\alpha_{h_i}\alpha_{h_j}, \nonumber
\end{align}
where
\begin{align}
\Delta_{\mathrm{bulk}}^{(n)}(h) = 2\alpha_{0,n/2} \alpha_h -
\frac{1}{2} \alpha_h^2 = (2\alpha_{0,n/2}-\alpha_0)\alpha_h -
\frac{h}{2} \nonumber
\end{align}
is the scaling exponent of a single $\overline{|w'(z)|^h}$ in the
presence of $n$ critical curves in the bulk. These are the results
quoted in Eqs. (\ref{simple bulk exponent}, \ref{higher bulk
exponent}).

\section{Omitted topics: guide to the literature}
\label{sec:omitted topics}

The current literature on SLE and related subject is already quite
large. In this paper I had to omit many interesting and important
topics. In this section I simply list the topics and give
appropriate references.

For the overall logic of this paper the biggest omission is the
discussion of the relation of SLE and CFT through the identification
of CFT correlators and SLE martingales. This identification was
established and developed by Bauer and Bernard (see review
\cite{Bauer-Bernard} and references there). We have further
developed this correspondence \cite{RBGW-long-paper}, showing that
one can recover all familiar objects of the Coulomb gas CFT, such as
bosonic field, its current, vertex operators, and the stress-energy
tensor, by focusing on SLE martingales. Alternative and independent
versions of SLE-CFT correspondence and generalizations were given by
Friedrich, Kalkkinen and Kontsevich, see Refs. \cite{FK, Friedrich,
Kontsevich}.

Chordal SLE considered in this paper has been generalized in many
ways. First of all, one can define SLE in other simply-connected
geometries than that of the UHP. The corresponding processes are
known as radial SLE \cite{Lawler-book, radial SLE}, whole plane
SLE \cite{Lawler-book}, dipolar SLE \cite{BB-zigzag, dipolar SLE}.
All these variants happen to be closely related
\cite{Schramm-Wilson}.

Secondly, one can consider SLE in multiply-connected domains
including arbitrary Riemann surfaces \cite{FK,Friedrich,
multiply-connected SLE-1, multiply-connected SLE-2,
multiply-connected SLE-3, multiply-connected SLE-4,
multiply-connected SLE-5}, though there is some amount of
arbitrariness involved in the definition, since in this setting the
conformal type of the domain changes during the evolution (one moves
in the moduli space) \cite{Lawler-EMP}.

Third, there is a way to modify the dynamics of the growth of a
random curve by including certain moving points (``spectators'') on
the boundary of the domain, that influence the evolution by
supplying a drift term in the (analog of) Loewner equation. This
generalization is known as SLE$_{\kappa,\rho}$, where $\rho$ stands
for a vector of parameters describing the coupling of the
``spectators'' \cite{LSW-restriction, Cardy-SLEkr,
Bauer-Friedrich-toolbox, MRR-boundary-Coulomb, SLEkr-1, SLEkr-2,
SLEkr-3, SLEkr-4}.

The fourth generalization allows for multiple curves to grow
simultaneously. This is called multiple SLE \cite{multiple SLE-1,
multiple SLE-2, multiple SLE-3, multiple SLE-4, multiple SLE-5,
multiple SLE-6}.

We can also generalize SLE by dropping the demand that the forcing
stochastic function be continuous, but keeping the requirement of
stationary and statistically-independent increments. This leads to
a much broader class of forcing processes including, in
particular, the so called L\'evy processes. This generalization
might be a useful description of tree-like stochastic growth
\cite{ROKG-SLE-Levy, Levy-SLE}.

Still another generalization is to combine the evolution of
conformal maps with some stochastic process in a Lie algebra or some
other algebraic structure. This leads to generalized SLE processes
describing CFT with additional symmetries, such as Wess-Zumino
models \cite{WZW-SLE-1, WZW-SLE-2}.

Recently, a few papers appeared that used SLE as a tool to probe
conformal invariance in systems that are not described by
traditional statistical mechanics models. A remarkable example is
Ref. \cite{SLE-turbulence} which numerically demonstrated that
zero vorticity lines in highly developed 2D turbulence are SLE$_6$
with high accuracy. Similar conclusions were presented for domain
walls in spin glasses \cite{SLE-spin-glass} and nodal domains of
some chaotic maps \cite{SLE-chaos}.

Finally, I would like to mention that there is generalization of
Loewner equation that describes evolving 2D domains which may grow
with a specified rate at every point on the boundary. These are
called Loewner chains (see Refs. \cite{Bauer-Bernard,
Bauer-Bernard-big-review} for a review of this enormous field in
relation to SLE) and can describe various non conformally-invariant
growth processes such as Laplacian growth, diffusion-limited
aggregation, dielectric breakdown, etc.

\section*{Acknowledgements}

First I want to thank my collaborators on the projects related to
this article: E. Bettelheim, L. Kadanoff, P. Oikonomou, I.
Rushkin, and P. Wiegmann.

Other people generously shared their ideas and knowledge of SLE
and related things with me, and I thank them all: M. Bauer, M. K.
Berkenbusch, D. Bernard, J. Cardy, B. Duplaniter, W. Kager, B.
Nienhuis, O. Schramm, and S. Sheffield.

Additional thanks go to the following people who kindly permitted
me to use figures from their papers and reviews: M. Bauer, D.
Bernard, L. Kadanoff, W. Kager, B. Nienhuis, and W. Werner.

This work was supported by the NSF MRSEC Program under
DMR-0213745, the NSF Career award DMR-0448820, the Sloan Research
Fellowship from Alfred P. Sloan Foundation, and the Research
Innovation Award from Research Corporation.


\newpage

\begin{thebibliography}{99}

\bibitem{Schramm}
O. Schramm, Scaling limits of loop-erased random walks and uniform
spanning trees, Israel J. Math. {\bf 118}, 221 (2000); arXiv:
math.PR/9904022.

\bibitem{Duplantier-PRL}
B. Duplantier, Conformally invariant fractals and potential
theory, Phys. Rev. Lett. {\bf 84}, 1363 (2000).

\bibitem{Duplantier}
B. Duplantier, Conformal fractal geometry and boundary quantum
gravity, in {\it  Fractal geometry and applications: a jubilee of
Benoît Mandelbrot, Part 2},  365, Proc. Sympos. Pure Math., 72,
Part 2, AMS, 2004; arXiv: math-ph/0303034.

\bibitem{Knizhnik:1988ak}
V.~G.~Knizhnik, A.~M.~Polyakov and A.~B.~Zamolodchikov, Fractal
structure of 2d-quantum gravity, Mod.\ Phys.\ Lett.\ A {\bf 3},
819 (1988).

\bibitem{Bauer-Bernard}
M. Bauer and D. Bernard, Loewner chains, in {\it String theory: from
gauge interactions to cosmology}, 41, NATO Sci. Ser. II Math. Phys.
Chem., {\bf  208}, Springer, Dordrecht, (2006); arXiv:
cond-mat/0412372.

\bibitem{Bauer-Bernard-big-review}
M. Bauer and D. Bernard, 2D growth processes: SLE and Loewner
chains, arXiv: math-ph/0602049.

\bibitem{Cardy1}
J. Cardy, Conformal invariance in percolation, self-avoiding walks
and related problems, arXiv: cond-mat/0209638.

\bibitem{Cardy2}
J. Cardy, SLE for theoretical physicists, Ann. Phys. {\bf 318}, 81
(2005); arXiv: cond-mat/0503313.

\bibitem{Gruzberg-Kadanoff}
I. A. Gruzberg and L. P. Kadanoff, The Loewner equation: maps and
shapes, J. Stat. Phys. {\bf 114}, 1183 (2004); arXiv:
cond-mat/0309292.

\bibitem{Kager-Nienhuis}
W. Kager, B. Nienhuis, A guide to stochastic loewner evolution and
its applications, J. Stat. Phys. {\bf 115}, 1149 (2004); arXiv:
math-ph/0312056.

\bibitem{Lawler-book}
G. F. Lawler, {\it Conformally invariant processes in the plane}.
Mathematical Surveys and Monographs, 114. American Mathematical
Society, Providence, RI, 2005.

\bibitem{Lawler1}
G. F. Lawler, Conformally invariant processes in the plane,
available online at URL
\verb|http://www.math.cornell.edu/~lawler/papers.html|

\bibitem{Lawler2}
G. F. Lawler, An introduction to the stochastic Loewner evolution,
in {\it Random walks and geometry}, 261, Walter de Gruyter GmbH \&
Co. KG, Berlin 2004; available online at URL
\verb|http://www.math.cornell.edu/~lawler/papers.html|

\bibitem{Schramm-review}
O. Schramm, Scaling limits of random processes and the outer
boundary of planar Brownian motion, Current developments in
mathematics, 2000, 233, Int. Press, Somerville, MA, 2001.

\bibitem{Schramm-overview}
O. Schramm, Conformally invariant scaling limits (an overview and
a collection of problems), to appear in the ICM 2006 Madrid
Proceedings, arXiv: math.PR/0602151.

\bibitem{Sheffield-review}
S. Sheffield, Gaussian free fields for mathematicians, arXiv:
math.PR/0312099.

\bibitem{Werner1}
W. Werner, Random planar curves and Schramm-Loewner evolutions, in
{\it Lectures on probability theory and statistics}. Lecture Notes
in Mathematics, {\bf 1840}. Springer-Verlag, Berlin, 2004; arXiv:
math.PR/0303354.

\bibitem{Werner2}
W. Werner, Conformal restriction and related questions, arXiv:
math.PR/0307353.

\bibitem{cft}
P. Di Francesco, P. Mathieu, D. Senechal, {\it Conformal field
theory}, Springer, 1999.

\bibitem{BRGW-harmonic-measure-PRL}
E. Bettelheim, I. Rushkin, I. A. Gruzberg, and P. Wiegmann, On
harmonic measure of critical curves, Phys. Rev. Lett. {\bf 95},
170602 (2005); arXiv: hep-th/0507115.

\bibitem{RBGW-long-paper}
I. Rushkin, E. Bettelheim, I. A. Gruzberg, and P. Wiegmann,
Critical curves in conformally invariant statistical systems, in
preparation.

\bibitem{nienhuis}
B. Nienhuis, Coulomb gas formulation of 2D phase transitions, in
{\it Phase Transitions and Critical Phenomena}, vol. 11, edited by
C. Domb, Academic Press, 1987.

\bibitem{DF}
Vl. S. Dotsenko, V. A. Fateev,  Conformal algebra and multipoint
correlation functions in 2D statistical models, Nucl. Phys.
{\textbf{B240}}, 312 (1984).

\bibitem{loop models-1}
H. Saleur, Lattice models and conformal field theories, Phys. Rep.
{\bf 184}, 177 (1989).

\bibitem{loop models-2}
B. Duplantier, two-dimensional fractal geometry, critical phenomena
and conformal invariance, Phys. Rep. {\bf 184}, 177 (1989).

\bibitem{loop models-3}
C. Vanderzande, {\it Lattice models of polymers}, Cambridge
University Press, 1998.

\bibitem{Polyakov:1970xd}
A.~M.~Polyakov, Conformal symmetry of critical fluctuations, JETP
Lett.\  {\bf 12}, 381 (1970) [Pisma Zh.\ Eksp.\ Teor.\ Fiz.\ {\bf
12}, 538 (1970)].


\bibitem{Belavin:1984vu}
A.~A.~Belavin, A.~M.~Polyakov and A.~B.~Zamolodchikov, Infinite
conformal symmetry in two-dimensional quantum field theory, Nucl.\
Phys.\ B {\bf 241}, 333 (1984).

\bibitem{Ahlfors1}
L. V. Ahlfors, {\it Complex analysis}, McGraw-Hill, 1979.

\bibitem{Loewner}
P. Duren. {\it Univalent functions}, Springer-Verlag, 1983.

\bibitem{Gong}
Sheng Gong, {\it The Bieberbach conjecture}, Providence, American
Mathematical Society, 1999.

\bibitem{Loewner-exact-solutions}
W. Kager, B. Nienhuis, and L. P. Kadanoff, Exact solutions for
Loewner evolutions, J. Stat. Phys. {\bf 115}, 805 (2004); arXiv:
math-ph/0309006.

\bibitem{Bauer-Bernard-1}
M. Bauer, D. Bernard, Conformal field theories of stochastic
Loewner evolutions, Commun. Math. Phys. {\bf 239}, 493 (2003);
arXiv: hep-th/0210015.

\bibitem{ROKG-SLE-Levy}
I. Rushkin, P. Oikonomou,  L. P. Kadanoff, and I. A. Gruzberg,
Stochastic Loewner evolution driven by L\'evy processes, J. Stat.
Mech., P01001 (2006); arXiv: cond-mat/0509187.

\bibitem{Rohde-Schramm}
S. Rohde and O. Schramm, Basic properties of SLE, Ann. of Math.
(2) {\bf 161}, 883 (2005); arXiv: math.PR/0106036.

\bibitem{Beffara-1}
V. Beffara, Hausdorff dimensions for $SLE_6$, Ann. Probab. {\bf 32},
2606--2629 (2004); arXiv: math.PR/0204208.

\bibitem{Beffara-2}
V. Beffara, The dimension of the SLE curves, arXiv: math.PR/0211322.

\bibitem{LSW-LERW}
G.~F.~Lawler, O.~Schramm, W.~Werner, Conformal invariance of
planar loop-erased random walks and uniform spanning trees, Ann.
Probab. {\bf 32}, 939 (2004); arXiv: math.PR/0112234.

\bibitem{LSW-SAW}
G. F. Lawler, O. Schramm, W. Werner, On the scaling limit of
planar self-avoiding walk, in {\it Fractal geometry and
applications: a jubilee of Benoît Mandelbrot, Part 2}, 339, Proc.
Sympos. Pure Math., 72, Part 2, Amer. Math. Soc., Providence, RI,
2004; arXiv: math.PR/0204277.

\bibitem{Kenyon}
R. Kenyon, Dominos and the Gaussian free field, Ann. Probab. {\bf
29}, 1128 (2001); arXiv: math-ph/0002027.

\bibitem{Schramm-Sheffield-harmonic-explorer}
O. Schramm and S. Sheffield, The harmonic explorer and its
convergence to SLE$_4$, Ann. Probab. {\bf 33}, 2127 (2005); arXiv:
math.PR/0310210.

\bibitem{Schramm-Sheffield-level-lines}
O. Schramm and S. Sheffield, Contour lines of the two-dimensional
discrete Gaussian free field, arXiv: math.PR/0605337.

\bibitem{Smirnov}
S. Smirnov, Critical percolation in the plane: conformal
invariance, Cardy's formula, scaling limits, C.~R.~Acad.~Sci.
Paris S\'er. I Math. {\bf 333}, 239 (2001).

\bibitem{Smirnov-Werner}
S. Smirnov, W. Werner, Critical exponents for two-dimensional
percolation, Math. Res. Lett. {\bf 8}, 729, 2001; arXiv:
math.PR/0109120.

\bibitem{Oksendal}
B. \O ksendal, {\it Stochastic differential equations},
Springer-Verlag, Berlin, 2003.

\bibitem{Klebaner}
F. C. Klebaner, {\it Introduction to stochastic calculus with
applications}, Imperial College Press, London, 1998.

\bibitem{LSW-1}
G. F. Lawler, O. Schramm, W. Werner, Values of Brownian intersection
exponents I: Half-plane exponents, Acta Math. {\bf 187}, 237 (2001);
arXiv: math.PR/9911084.

\bibitem{LSW-restriction}
G. F. Lawler, O. Schramm, W. Werner, Conformal restriction: the
chordal case, J. Amer. Math. Soc. {\bf 16}, 917 (2003); arXiv:
math.PR/0209343.

\bibitem{FW-1}
R. Friedrich, W. Werner, Conformal fields, restriction properties,
degenerate representations and SLE.  C. R. Math. Acad. Sci. Paris
{\bf 335}, 947 (2002); arXiv: math.PR/0209382.

\bibitem{FW-2}
R. Friedrich, W. Werner, Conformal restriction, highest-weight
representations and SLE, Comm. Math. Phys. {\bf 243}, 105--122
(2003); arXiv: math.PR/0301018.

\bibitem{FK}
R. Friedrich, J. Kalkkinen, On conformal field theory and stochastic
Loewner evolution, Nucl. Phys. B {\bf 687}, 279 (2004); arXiv:
hep-th/0308020.

\bibitem{Friedrich}
R. Friedrich, On connections of conformal field theory and
stochastic Loewner evolution, arXiv: math-ph/0410029.

\bibitem{Bauer-Bernard-partition-function}
M. Bauer, D. Bernard, Conformal transformations and the SLE
partition function martingale, Ann. Henri Poincar\'e {\bf 5}, 289
(2004); arXiv: math-ph/0305061.

\bibitem{Schramm-percolation}
O. Schramm, A percolation formula, Elec. Comm. in Probab. {\bf 6},
115 (2001).

\bibitem{Cardy3}
J. Cardy, Critical percolation in finite geometries, J. Phys. A:
Math. Gen. {\bf 25}, L201 (1992).

\bibitem{Cardy-SLEkr}
J. Cardy, SLE($\kappa, \rho$) and conformal field theory, arXiv:
math-ph/0412033;

\bibitem{Bauer-Friedrich-toolbox}
R. O. Bauer, R. Friedrich, The correlator toolbox, metrics and
moduli, Nucl. Phys. B {\bf 733}, 91 (2006); arXiv: hep-th/0506046.

\bibitem{MRR-boundary-Coulomb}
S. Moghimi-Araghi, M. A. Rajabpour, and S. Rouhani,
SLE($\kappa,\rho$) and boundary Coulomb gas, Nucl. Phys. B {\bf
740}, 348 (2006); arXiv: hep-th/0508047.

\bibitem{Schulze}
J. Schulze, Coulomb gas on the half plane, Nucl.\ Phys.\ {\bf B489},
580 (1997); {\bf arxiv}

\bibitem{kawai}
S.~Kawai, Coulomb-gas approach for boundary conformal field theory,
Nucl. Phys. B {\bf 630}, 203 (2002); {\bf arxiv}

\bibitem{kondev}
J. Kondev, Loop models, marginally rough interfaces, and the Coulomb
gas, Int. J. Mod. Phys. {\bf B11}, 153 (1997).

\bibitem{Kondev-Henley}
J. Kondev, C. L. Henley, Kac-Moody symmetries of critical ground
states, Nucl. Phys. B {\textbf{464}}, 540 (1996).

\bibitem{cardyoperators-1}
J. L. Cardy, Conformal invariance and surface critical behavior,
Nucl. Phys. B {\textbf{240}}, 514 (1984).

\bibitem{cardyoperators-2}
J. L. Cardy, Boundary conditions, fusion rules and the verlinde
formula, Nucl. Phys. B {\textbf{324}}, 581 (1989).

\bibitem{BB-zigzag}
M. Bauer, D. Bernard, SLE, CFT and zig-zag probabilities, arXiv:
math-ph/0401019.

\bibitem{harmonic measure-1}
J. B. Garnett, D. E. Marshall, {\it Harmonic measure}, Cambridge
University Press, 2005.

\bibitem{harmonic measure-2}
C. Pommerenke, {\it Boundary behaviour of conformal maps}, Springer,
1992.

\bibitem{Makarov-1}
N. G. Makarov, On the distortion of boundary sets under conformal
mappings, Proc. London Math. Soc. {\bf 51}, 369 (1985).

\bibitem{Makarov-2}
N. G. Makarov, Fine structure of harmonic measure, St. Petersburg
Math. J. {\bf 10}, 217 (1999).

\bibitem{DiFrancesco-QG-review}
P. Di Francesco, P. Ginsparg, and J. Zinn-Justin, 2D gravity and
random matrices, Phys. Rep. {\bf 254}, 1 (1995); arXiv:
hep-th/9306153.

\bibitem{Kontsevich}
M. Kontsevich, CFT, SLE and phase boundaries, preprint
MPI-2003-60-a, available online at URL
\verb|http://www.mpim-bonn.mpg.de/html/|
\verb|preprints/preprints.html|

\bibitem{radial SLE}
M. Bauer, D. Bernard, CFTs of SLEs: the radial case, Phys. Lett. B
{\bf 583}, 324--330 (2004); arXiv: math-ph/0310032.

\bibitem{dipolar SLE}
M. Bauer, D. Bernard, J. Houdayer, Dipolar stochastic Loewner
evolutions, J. Stat. Mech. (2005) P03001; arXiv: math-ph/0411038.

\bibitem{Schramm-Wilson}
O. Schramm and D. B. Wilson, SLE coordinate changes, New York J.
Math. {\bf 11}, 659 (2005); arXiv: math.PR/0505368.

\bibitem{multiply-connected SLE-1}
R. O. Bauer, Restricting SLE(8/3) to an annulus, arXiv:
math.PR/0602391.

\bibitem{multiply-connected SLE-2}
R. O. Bauer and R. Friedrich, Stochastic Loewner evolution in
multiply connected domain, C. R. Acad. Sci. Paris, Ser. I {\bf 339},
579 (2004); arXiv: math.PR/0408157.

\bibitem{multiply-connected SLE-3}
R. O. Bauer and R. Friedrich, On radial stochastic Loewner evolution
in multiply connected domains, arXiv: math.PR/0412060;

\bibitem{multiply-connected SLE-4}
R. O. Bauer and R. Friedrich, On chordal and bilateral SLE in
multiply connected domains, arXiv: math.PR/0503178;

\bibitem{multiply-connected SLE-5}
D. Zhan, Stochastic Loewner evolution in doubly connected domains,
Probab. Theory Related Fields {\bf 129}, 340--380 (2004); arXiv:
math.PR/0310350;

\bibitem{Lawler-EMP}
G. F. Lawler, Stochastic Loewner evolution, a draft of a
contribution on SLE to Encyclopedia of Mathematical Physics to be
published by Elsevier, available online at URL
\verb|http://www.math.cornell.edu/~lawler/papers.html|

\bibitem{SLEkr-1}
J. Dubedat, SLE$(\kappa,\rho)$ martingales and duality, Ann. Probab.
{\bf 33},  223 (2005); arXiv: math.PR/0303128.

\bibitem{SLEkr-2}
W. Werner, Girsanov's transformation for SLE$(\kappa,\rho)$
processes, intersection exponents and hiding exponents,  Ann. Fac.
Sci. Toulouse Math. (6) {\bf 13}, 121 (2004); arXiv:
math.PR/0302115.

\bibitem{SLEkr-3}
R. O. Bauer and R. Friedrich, Diffusing polygons and
SLE($\kappa,\rho$), arXiv: math.PR/0506062.

\bibitem{SLEkr-4}
K. Kyt\"ol\"a, On conformal field theory of SLE($\kappa,\rho$),
arXiv: math-ph/0504057.

\bibitem{multiple SLE-1}
J. Cardy, Stochastic Loewner evolution and Dyson's circular
ensembles, J. Phys. A: Math. Gen. {\bf 36}, L379 (2003); arXiv:
math-ph/0301039.

\bibitem{multiple SLE-2}
J. Dubedat, Commutation relations for SLE, arXiv: math.PR/0411299.

\bibitem{multiple SLE-3}
J. Dubedat, Euler integrals for commuting SLEs, arXiv:
math.PR/0507276.

\bibitem{multiple SLE-4}
M. Bauer, D. Bernard, K. Kyt\"ol\"a, Multiple Schramm-Loewner
evolutions and statistical mechanics martingales, J. Stat. Phys.
{\bf 120}, 1125 (2005); arXiv: math-ph/0503024.

\bibitem{multiple SLE-5}
K. Kyt\"ol\"a, Virasoro module structure of local martingales for
multiple SLEs, arXiv: math-ph/0604047.

\bibitem{multiple SLE-6}
M. J. Kozdron and G. F. Lawler,  The configurational measure on
mutually avoiding SLE paths, arXiv: math.PR/0605159.

\bibitem{Levy-SLE}
Qing-Yang Guan, M. Winkel, SLE and alpha-SLE driven by L\'evy
processes, arXiv: math.PR/0606685.

\bibitem{WZW-SLE-1}
J. Rasmussen, On SU(2) Wess-Zumino-Witten models and stochastic
evolutions, arXiv: hep-th/0409026.

\bibitem{WZW-SLE-2}
E. Bettelheim, I. Gruzberg, A. W. W. Ludwig, P. Wiegmann, Stochastic
Loewner evolution for conformal field theories with Lie-group
symmetries, Phys. Rev. Lett. {\bf 95}, 251601 (2005); arXiv:
hep-th/0503013.

\bibitem{SLE-turbulence}
D. Bernard, G. Boffetta, A. Celani, G. Falkovich, Conformal
invariance in two-dimensional turbulence, Nature Physics {\bf 2},
124 (2006).

\bibitem{SLE-spin-glass}
C. Amoruso, A. K. Hartmann, M. B. Hastings, and M. A. Moore,
Conformal invariance and SLE in two-dimensional Ising spin
glasses, arXiv: cond-mat/0601711.

\bibitem{SLE-chaos}
J. P. Keating, J. Marklof, I. G. Williams, Nodal domain statistics
for quantum maps, percolation and SLE, arXiv: nlin.CD/0603068.

\end{thebibliography}
\end{document}